\documentclass[11pt,twocolumn,tighten]{aastex63}
\turnoffedit

\usepackage{apjfonts}
\usepackage{url}
\usepackage{hyperref}
\usepackage{natbib}
\usepackage{amsmath,amstext,amssymb}
\usepackage[caption=false]{subfig} 
\usepackage{xcolor, fontawesome}
\usepackage{color}
\usepackage{enumitem}

\newcommand{\kms}{{km\,s$^{-1}$}}

\newcommand{\caltech}{Department of Astronomy, California Institute of Technology, Pasadena, CA 91125, USA}
\newcommand{\carnegie}{Observatories of the Carnegie Institution for Science, Pasadena, CA 91101, USA}

\newcommand{\berkeley}{Astronomy Department, University of California,
Berkeley, CA 94720, USA}
\newcommand{\cmu}{Department of Statistics \& Data Science, Carnegie
Mellon University, Pittsburgh, PA 15213, USA}
\newcommand{\osaka}{Department of Earth and Space Science, Osaka University, Osaka 560-0043, Japan}

\newcommand{\nuniqstarsantosrot}{53{,}663}
\newcommand{\nuniqstarsantosrotgyroappl}{23{,}813}

\newcommand{\nplwgyroagenograzing}{795}

\newcommand{\nuniqstarsantosrotteffcut}{45{,}229}

\newcommand{\ratioobtoybstars}{2.1}

\newcommand{\ratioobtoybplanets}{2.6}

\newcommand{\nplyounggyro}{109}

\newcommand{\nplmidgyro}{201}

\newcommand{\nploldgyro}{281}

\newcommand{\nplyounggyrotwosigma}{63}
\newcommand{\nplhostsyounggyrotwosigma}{50}

\newcommand{\nlithiumstars}{1{,}464}

\newcommand{\nkoisnofp}{4{,}716}
\newcommand{\nhireshours}{304}
\newcommand{\nnonfopkoissomeageinfo}{2{,}461}

\newcommand{\nplyounggyrotwosigmanograzingnoruwe}{46}

\newcommand{\nuniqstarfinitegyroage}{37{,}892}
\newcommand{\fracconsistentallages}{95}

\newcommand{\allagesyesconsistent}{697}
\newcommand{\allagesmaybeconsistent}{12}
\newcommand{\allagesnoconsistent}{22}

\newcommand{\fracpotentiallyconsistentallages}{97}

\newcommand{\nnewdavidtwentyone}{178}
\newcommand{\kepsixsixtgyro}{$1431^{+675}_{-364}$}
\newcommand{\kepsixseventgyro}{$878^{+108}_{-121}$}
\newcommand{\ratiosfr}{2.81}
\newcommand{\uncratiosfr}{0.12}
\newcommand{\trestwotli}{$785^{+845}_{-419}$}
\newcommand{\kepseveneightsix}{$228^{+168}_{-87}$}
\newcommand{\kepsixteenfourfour}{$77^{+144}_{-53}$}
\newcommand{\kepsixteenninenine}{$85^{+106}_{-58}$}
\newcommand{\kepnineteenfourthree}{$409^{+520}_{-254}$}

\newcommand{\ltonegyrhighqconfirmedtwosided}{18}
\newcommand{\ltonegyrhighqconfirmedonesided}{37}
\newcommand{\kepthirteentwelve}{$357^{+75}_{-109}$}
\newcommand{\kepfifteensixone}{$426^{+74}_{-78}$}
\newcommand{\kepsixteentwonine}{$529^{+62}_{-62}$}

\newcommand{\nplwithspecandprot}{1{,}170}
\newcommand{\nstwithspecandprot}{797}

\newcommand{\nconfirmedplyounggyrotwosigmanograzingnoruwe}{37}
\newcommand{\ncandidateplyounggyrotwosigmanograzingnoruwe}{nine}

\newcommand{\nminineptuneshighq}{28}
\newcommand{\nsubsaturnshighq}{seven}
\newcommand{\nsuperearthshighq}{18}
\newcommand{\nearthshighq}{four}
\newcommand{\njupitershighq}{six}
\newcommand{\nlongperiodhighq}{four}

\newcommand{\mcquillanonlyratioobtoybstars}{1.9}
\newcommand{\mcquillanonlyratiosfr}{2.24}
\newcommand{\mcquillanonlyuncratiosfr}{0.11}

\newcommand{\fivepctnuniqstarsantosrotgyroappl}{1{,}190}
\newcommand{\ngyroappllttendays}{2{,}706}
\newcommand{\nplyounggrazingorhighruwe}{17}

\newcommand{\nkeplerstars}{$\approx$160{,}000}
\newcommand{\fracstarswithprotwithbtwenty}{{$\approx$94\%}}
\newcommand{\fracstarswithprotwithoutbtwenty}{{$\approx$6\%}}
\newcommand{\frackoisnofpwithprotwithbtwenty}{{$\approx$92\%}}

\defcitealias{2013ApJ...775L..11M}{M13}
\defcitealias{McQuillan_2014}{M14}
\defcitealias{Mazeh_2015}{M15}
\defcitealias{Santos_2019}{S19}
\defcitealias{Santos_2021}{S21}
\defcitealias{Fulton_2018}{F18}
\defcitealias{Berger_2020a_catalog}{B20}

\begin{document}

\title{Ages of Stars and Planets in the Kepler Field Younger Than Four Billion Years}

\correspondingauthor{Luke G. Bouma}
\email{lbouma@carnegiescience.edu}

\received{2024 June 21}
\revised{2024 September 4}
\accepted{2024 October 7}
\shorttitle{Kepler's Demographic Cliff} 

\shortauthors{Bouma et al.}

\author[0000-0002-0514-5538]{Luke~G.~Bouma}
\altaffiliation{51 Pegasi b Fellow;  Carnegie Fellow}
\affiliation{\caltech}
\affiliation{\carnegie}

\author{Lynne~A.~Hillenbrand}
\affiliation{\caltech}

\author[0000-0001-8638-0320]{Andrew W. Howard}
\affiliation{\caltech}

\author[0000-0002-0531-1073]{Howard Isaacson}
\affiliation{\berkeley}

\author[0000-0003-1298-9699]{Kento Masuda}
\affiliation{\osaka}

\author[0000-0001-7967-1795]{Elsa~K.~Palumbo}
\affiliation{\caltech}
\affiliation{\cmu}

\begin{abstract}
  Recent analyses of FGK stars in open clusters have helped clarify
  the precision with which a star's rotation rate and lithium content
  can be used as empirical indicators for its age.
  Here we apply this knowledge to stars observed by Kepler.
  Rotation periods are drawn from previous work; lithium is measured
  from new and archival Keck/HIRES spectra.
  We report rotation-based ages for \nuniqstarsantosrotgyroappl\ stars
  (harboring \nplwgyroagenograzing\ known planets) for which our method is
  applicable.
  We find that our rotational ages \deleted{accurately }recover the ages of
  stars in open clusters spanning 0.04-2.5\,Gyr; they also agree with
  $\gtrsim$90\% of the independent lithium ages.
  The resulting yield includes \nplyounggyrotwosigma\ planets younger
  than 1\,Gyr at 2$\sigma$, and \nplyounggyro\ with median ages below
  1\,Gyr.
  This is about half the number expected under the classic assumption
  of a uniform star formation history.
  \deleted{We find that the scarcity of sub-gigayear systems can be
  attributed to the star formation rate in the Kepler field dropping
  by a factor of \ratiosfr$\pm$\uncratiosfr over the past 3\,Gyr.  We
  observe this trend both for known planet hosts and for the parent
  stellar sample.  This ``demographic cliff'' in the Galaxy's star
  formation history has been previously reported, and its confirmation
  helps clarify the age distribution of the known transiting
  exoplanets.}
  \added{The age distribution that we observe, rather than being uniform,
  shows that the youngest stars in the Kepler field are 3-5 times
  rarer than stars 3\,Gyr old.
  This trend holds for both known planet hosts and for the parent
  stellar sample.
  We attribute this ``demographic cliff'' to a
  combination of kinematic heating and a declining star formation rate
  in the Galaxy's thin disk, and highlight its impact on the age
  distribution of known transiting exoplanets.
  }
\end{abstract}

\keywords{Stellar ages (1581), Planet hosting stars (1242), Field
stars (2103), Exoplanet evolution (491), Milky Way evolution (1052)}

\section{Introduction}
\label{sec:intro}

Exoplanet science is thriving, fueled by the discovery of thousands of
worlds orbiting close to their host stars
\citep{Borucki10,2015JATIS...1a4003R}.  However, most known exoplanets
are billions of years old.  This fact leaves many gaps in our
knowledge of the exact physical and dynamical origins of these
objects.  The reason is that processes such as thermal cooling
\citep{2007ApJ...659.1661F}, atmospheric loss
\citep{2019AREPS..47...67O}, giant impacts
\citep{2014prpl.conf..595R}, and dynamical instabilities
\citep{2017MNRAS.470.1750I} are expected to be most efficient over
timescales of much less than 1\,Gyr.  For most known exoplanets, these
processes have run their course.

Young ($<$1\,Gyr) exoplanets represent one means of building the
timeline for exoplanet evolution.  Informative individual exemplars
include HIP~67522\,b, a Jupiter-sized planet with a sub-Neptunian mass
(\citealt{Rizzuto_2020}; \citealt{Thao2024}), and V1298~Tau, a
resonant chain of puffy planets that is a likely precusor to the
compact multiplanet systems \citep{David_2019}.  Population-level
analyses have similarly suggested differences in the size distribution
of young exoplanets relative to their older counterparts
\citep{Berger_2020b_rpage,David_2021,Sandoval_2021,2023AJ....166..248C,Vach2024}.

Discovering a young planet requires solving two problems: finding the
planet and measuring the star's age.  Each problem admits a range of
solutions
\citep[e.g.][]{2008Sci...322.1348M,2012ApJ...756L..33Q,2024AJ....167..193T}.
In this article we will consider planets whose existence has been
previously established using transits, and infer new stellar ages
using rotation and lithium.

To begin, imagine that the ages of nearby stars in the Galaxy are
uniformly distributed from 0--10\,Gyr
\citep[][]{2000MNRAS.318..658B,Nordstrom_2004}.  This approximation
suggests that $\approx$1\% of nearby stars have ages $t$$<$100\,Myr, and
$\approx$10\% have $t$$<$1\,Gyr.  Studies of currently forming
protoplanets \citep{2018A&A...617A..44K}, and of exoplanets evolving
just after disk dispersal \citep[e.g.][]{2022MNRAS.512.5067K}, are
thereby capped in their maximum achievable sample sizes by the tyranny
of the galactic star formation rate.

Despite this and other observational challenges, young close-in planet
discovery has matured over the past decade, primarily due to Kepler,
K2, TESS, and Gaia
\citep[e.g.][]{Meibom_2013,Mann_K2_25_2016,Curtis_2018,Livingston_2018,Bouma_2020_toi837,Plavchan_2020,Newton_2021,Nardiello_2022,Barber_2022,Zhou_2022,Zakhozhay_2022,Wood_2023}.
The strategy pursued by most groups during the 2010s was to focus on
stars with known ages---obvious members of open clusters---and to
search them for transiting planets.  The resulting stellar, and
assumed planetary, ages are precise at the $\approx$10\% level.  
An alternative strategy, facilitated by Gaia, is to select \deleted{transiting }planets
based on empirical youth indicators, and to then search their local
neighborhoods for comoving \added{stellar }companions \citep[e.g.][]{Tofflemire_2021}.

While stellar ensembles provide the gold standard for astronomical ages,
this ``cluster-first'' approach has a major limitation: most 
stars are in the field.  Only $\lesssim$1\% of stars within
$\approx$500\,pc have been associated with their birth cluster
\citep[e.g.][]{Zari_2018,CantatGaudin_2020,Kounkel_2020,Kerr_2021}.
The implication for known planetary ages can be appreciated by querying the NASA Exoplanet Archive
\citep[NEA;][]{2013PASP..125..989A} for transiting planets younger
than 1\,Gyr.  Requiring $t$$<$1\,Gyr at 2$\sigma$ precision gives
$\approx$50 \deleted{such }planets at the time of writing.  Most of these young
planets are in clusters, and were found by Kepler, K2, or TESS.
However, these surveys have cumulatively discovered \deleted{a total of }$\approx$5{,}000 planets.
Assuming a constant star formation history, we would expect an
order of magnitude more sub-gigayear transiting planets\added{ than the 50 currently known}.

This study aims to resolve two questions.  First, how wrong is it to
assume a uniform age distribution for transiting planet host stars?
Second, where are the missing young planets?  We will find that
uniform is wrong by a factor of a few, and that stellar activity may
be a less significant bias in young transiting planet detection than
the challenge of precisely determining ages for field stars.

Despite Kepler's main mission ending quite some time ago, the ages of
many Kepler planets remain uncertain.  While isochrone ages have been
calculated for Kepler stars \citep{Berger_2020b_rpage} and Kepler
Objects of Interest \citep[KOIs;][]{Petigura_2022}, such ages are most
precise for stars whose luminosities and temperatures separate them
from the main sequence.  For sub-gigayear stars on the main sequence,
isochrone ages are therefore limited.

Stellar rotation periods offer a promising alternative.  The idea of
using a star's spin-down as a clock has a rich history
\citep{Skumanich_1972,Noyes_1984,Kawaler_1989,Barnes03,Mamajek_2008,Angus_2015,2023ApJ...954L..50E}.
Empirical models now yield ages precise to $\lesssim$30\% for FGK
stars between 1-4\,Gyr, and constraining age posteriors for younger
ages \citep{Bouma_2023}.  Physics-based models
\citep{Matt_2015,Gallet_Bouvier_2015,Spada_2020} can connect these
empirical relations to the evolution of stellar winds, internal
structure, and angular momentum transport.

Rotational ages have been reported for various subsets of Kepler
stars since the early data releases
\citep[e.g.][]{Walkowicz_2013,McQuillan_2014,Reinhold_2015,Angus_2018}.
More recent work has further explored incorporating information from
stellar kinematics \citep{2021AJ....161..189L,2024AJ....167..159L},
and from stellar colors, luminosities, and starspot amplitudes
\citep{2023ApJ...952..131M}.  Our analysis is motivated by a few
factors that can yield improvements, particularly for sub-gigayear
stars.   These factors are as follows.

\begin{enumerate}[label={\it \roman*)},leftmargin=12pt,topsep=0pt,itemsep=-1ex,partopsep=1ex,parsep=1ex]
  \item The Kepler Object of Interest catalog, and our vetting of
    false positives within it, has now reached maturity
    \citep[e.g.][]{Thompson_2018}.
  \item Measured rotation periods of FGK stars in open clusters now
    show not only the average evolution of $P_{\rm rot}(T_{\rm
    eff},t)$, but also how the astrophysical dispersion of stars
    around this average converges by the $\approx$700\,Myr age of
    Praesepe
    \citep[e.g.][]{Curtis_2019_ngc6811,Gillen_2020,Rampalli_2021,Fritzewski_2021,Rebull_2022,Dungee_2022,2023AJ....166...14B}.
  \item Using open cluster data, we can marginalize over the range of
    ages that might explain any one star's rotation period
    \citep{Bouma_2023}.  This represents an improvement in the
    accuracy of uncertainty propagation relative to previous
    calibrations.
  \item We can identify \replaced{binary stars}{stars that are
    binaries and that have high metallicity} with greater fidelity\deleted{ than
    in the past}, which can clarify otherwise problematic estimates of
    rotation-based ages.
  \item New open clusters in various stages of dissolution have been
    found in the Kepler field
    \cite[e.g.][]{2019AJ....158..122K,Barber_2022,Kerr2024}.
    These clusters offer stars that we can use to test
    the reliability of \replaced{the available single-star}{our} age-dating methods.
\end{enumerate}

Recent years have also yielded improvements in the
lithium age scale.  Lithium ages include depletion boundary ages for M
dwarfs and brown dwarfs in star clusters, and decline-based ages for
individual field FGK stars \citep{Soderblom_2010}.  The latter
approach relies on the observed decrease of Li abundances in
partially-convective stars as they age
\citep[e.g.][]{2005A&A...442..615S}.  The theoretical explanation for
these observations is debated
\citep[e.g.][]{1995ApJ...441..865C,2010ApJ...716.1269D,2019MNRAS.485.4052C}.
Empirical understanding however has improved due to work by
\citet{Jeffries_2023}, who modeled the time evolution of the
\ion{Li}{1} 6708\,\AA\ equivalent width (EW) using a set of 6{,}200
stars in 52 open clusters.  Two-sided lithium ages are useful for
Kepler (FGK) stars between $\approx$0.03-0.5\,Gyr, though with a
strong dependence on spectral type.  The precision of lithium ages in
this regime are now in the range of 0.3-0.5~dex.

We discuss our method for selecting the star and planet samples in
Section~\ref{sec:selection}, and describe the origin of our adopted
stellar parameters other than ages in Section~\ref{sec:stellarprops}.
We present our age-dating methods in Sections~\ref{sec:rotage}
and~\ref{sec:liage}, and test them in
Section~\ref{sec:litagecomparison} using clusters in the Kepler field.
We discuss population-level trends in Sections~\ref{sec:results}
and~\ref{sec:disc}, and offer a few conclusions in
Section~\ref{sec:conclusions}.

\section{Selecting the Stars and Planets}
\label{sec:selection}

\begin{figure}[!t]
	\begin{center}
		\subfloat{
			\includegraphics[width=0.42\textwidth]{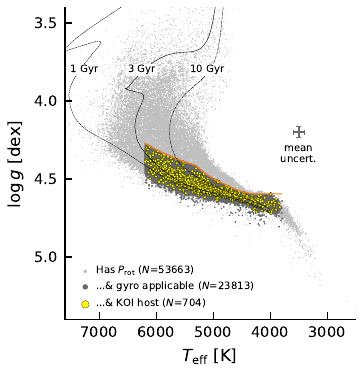}
		}
	
		\vspace{-0.3cm}
		\subfloat{
			\includegraphics[width=0.42\textwidth]{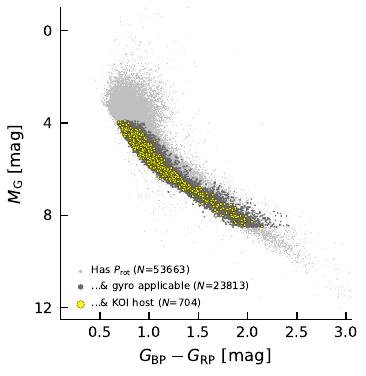}
		}
	\end{center}
	\vspace{-0.5cm}
  \caption{{\bf The stars.}  Our analysis focuses on stars observed by
  Kepler with previously reported rotation periods (gray points).  The
  rotation periods are primarily drawn from
  \citet{Santos_2019,Santos_2021}.  About half of the stars with
  rotation periods are suitable for gyrochronology (dark gray\deleted{ points}),
  based on factors including their \added{apparent }non-binarity and proximity to the
  main sequence (orange line in \replaced{lower}{upper} panel; see
  Section~\ref{subsec:flags}).  Some host ``confirmed'' or
  ``candidate'' KOIs that meet additional planetary quality criteria
  (yellow points; Section~\ref{subsec:plflags}).  Surface gravities
  and effective temperatures were derived photometrically by
  \citet{Berger_2020a_catalog}.  Isochrones in the \replaced{lower}{upper} panel are
  from MIST \citep{Choi_2016}.
	}
	\label{fig:stellarprops}
\end{figure}

This work is focused on Kepler stars for which ages can be inferred
using either rotation, lithium, or both.  Such stars are a minority of
the \nkeplerstars\ Kepler targets.  Rotation periods have been
reported for roughly one in three Kepler targets
\citep[e.g.][]{McQuillan_2014,Santos_2021}.  High-resolution spectra
suitable for measurement of the \ion{Li}{1} 6708\,\AA\ doublet have
only been acquired for the Kepler objects of interest (KOIs), which
comprise a few percent of Kepler's targets.  In the following, we will
describe the set of stars that we adopt with measured rotation periods
(Section~\ref{subsec:rotsel}), the set of objects we adopt as hosting
planets (Section~\ref{subsec:planetsel}), and the subset of these with
high-resolution spectra suitable for lithium analysis
(Section~\ref{subsec:lithiumsel}).  Figure~\ref{fig:stellarprops}
provides a visualization of the various samples.

\subsection{Stellar Rotation Periods}
\label{subsec:rotsel}

\begin{figure*}[!t]
	\begin{center}
		\subfloat{
			\includegraphics[width=0.46\textwidth]{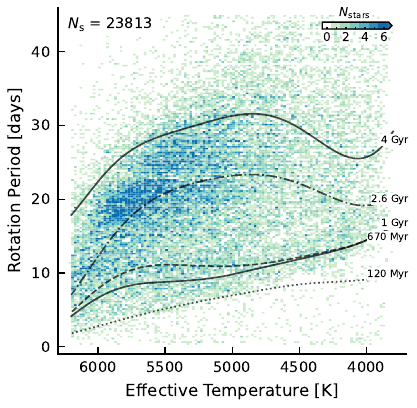}
		}
		\subfloat{
			\includegraphics[width=0.46\textwidth]{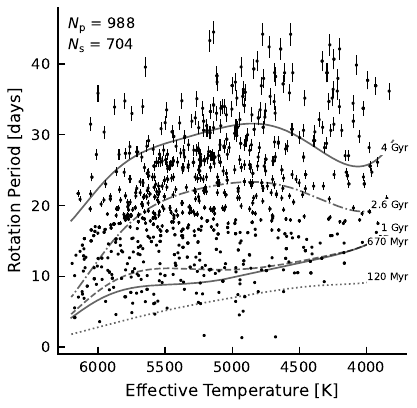}
		}
	\end{center}
	\vspace{-0.5cm}
	\caption{
    {\bf Rotation periods for Kepler target stars (left) and known
    planet hosts (right)}.  The gray lines are \replaced{``mean fits'' to the
    rotation sequences of open clusters}{polynomials
    fitted to the slow rotation sequences of open clusters}.  The 2-D histogram of the
    stellar sample (left) shows only apparently single stars near
    the main sequence with $\log g$$>$4.2, RUWE$<$1.4, and
    temperatures of 3800--6200\,K.  The planet hosts (right) require
    the same stellar cuts, and include only the confirmed and
    candidate planets described in Section~\ref{subsec:plflags}.  The
    total number of stars $N_{\rm s}$ and planets $N_{\rm p}$ are noted in
    the relevant panels.  A simple estimate for the number of young
    planets discovered by Kepler follows by counting points below the
    \deleted{mean }rotational isochrones.
	}
	\label{fig:prot_vs_teff}
\end{figure*}

To select stars with rotation periods, we turn to previous work.  Many
investigators have derived rotation periods both specifically for KOIs
\citep{McQuillan_2013,Walkowicz_2013,Mazeh_2015,Angus_2018,David_2021},
and also for the broader \replaced{set of all}{sample of} Kepler target stars
\citep{McQuillan_2014,Reinhold_2015,Santos_2019,Santos_2021,Reinhold2023}.
These studies used a range of detection methods and selection
functions.  We are interested in understanding the age distribution of
all Kepler target stars irrespective of KOI status.  Out of these
studies, the most homogeneous \deleted{and precise }analyses of both Kepler
targets and KOIs appear to be those by
\citet{Santos_2019,Santos_2021}, hereafter \citetalias{Santos_2019}
and \citetalias{Santos_2021}.  For discussion of the work by
\citet{Reinhold2023}, see Appendix~\ref{app:reinhold23}.
\citetalias{Santos_2019} and \citetalias{Santos_2021} combined a
wavelet analysis and autocorrelation-based approach, and cumulatively
reported rotation periods for 55{,}232 main-sequence and subgiant FGKM
stars.  They included known KOIs and binaries in their analysis, and
removed transits and eclipses during the stellar rotation measurement
process. 

We therefore adopt the results of \citetalias{Santos_2019} and
\citetalias{Santos_2021} as our default rotation periods.
\citetalias{Santos_2021} provided a comparison against
\citet{McQuillan_2014} (hereafter \citetalias{McQuillan_2014}); the brief summary is that the periods
agree for 99.0\% of the 31{,}038 period detections in common between
the two studies.  \citetalias{Santos_2021} classified the 2{,}992
remaining stars from \citetalias{McQuillan_2014} as not showing
rotation periods based on updated knowledge of contaminants
(such as giant stars and eclipsing binaries) and visual inspection.  In
addition, \citetalias{Santos_2021} reported rotation periods for
24{,}182 main-sequence and subgiant FGKM stars that were not reported
as periodic by \citetalias{McQuillan_2014}.  Many of these reported
signals have lower variability amplitudes and longer periods than
those reported by \citetalias{McQuillan_2014}. 

Analyzing the compilation of \citetalias{Santos_2019} and
\citetalias{Santos_2021} rotation periods for the KOI hosts, we
noticed that some known KOIs with rotation periods were missing.  This
is not surprising, since the rotation periods of KOIs have received
more scrutiny than those of ordinary Kepler stars.  We therefore
decided to split our subsequent analysis into a homogeneous portion
that used only the \citetalias{Santos_2019} and
\citetalias{Santos_2021} data, and an inhomogeneous portion that also
considered a broader set of available KOI rotation periods.  For the
latter portion, we first included 32 KOIs with orbital and rotation
periods within $\approx$20\% that had been
excluded 
from the \citetalias{Santos_2019} and \citetalias{Santos_2021}
catalogs (A.~Santos, private communication).  We then incorporated an
additional \nnewdavidtwentyone\ rotation periods for KOI hosts that
\citet{David_2021} described as either ``reliable'' or ``highly
reliable'' in their visual analysis of previously reported KOI
rotation periods from \citet{McQuillan_2013}, \citet{Walkowicz_2013},
\citet{Mazeh_2015} and \citet{Angus_2018}.  Inclusion of these
additional KOI rotation periods is a supplementary measure aimed at
completeness in our final KOI age catalog; the provenances of the
individual adopted periods are noted in the relevant tables.

Finally, since our scope is focused on rotation-based ages, we
restricted our attention to stars with reported $P_{\rm
rot}$$<$45\,days.  The slowest-rotating FGK stars in the open clusters
used to calibrate our gyrochronology model have $P_{\rm
rot}$$\approx$35\,days.  Figure~\ref{fig:prot_vs_teff} shows a subset
of the resulting \nuniqstarsantosrot\ Kepler stars with rotation
periods compiled from \citetalias{Santos_2019},
\citetalias{Santos_2021}, and our extended KOI list.\added{  The gray
lines in this figure are the seventh-order polynomials described by \citet{Bouma_2023}
that 
fit the slow rotation sequences of the Pleiades \citep{Rebull_2016},
Praesepe \citep{Rampalli_2021}, NGC-6811 \citep{Curtis_2019_ngc6811}, NGC-6819 \citep{Meibom_2015},
Ruprecht-147 \citep{Curtis_2020}, and M67 \citep{Barnes2016,Dungee_2022,Gruner_2023}.}

To assess the statistical uncertainties of \replaced{these}{our adopted} rotation periods, we
compared our \deleted{adopted} periods with those reported by
\citet{McQuillan_2014}.  The details are in
Appendix~\ref{app:reinhold23}.  We found that for $P_{\rm
rot}\lesssim$15\,days, the two datasets agree at a precision of
$\lesssim$0.01$P_{\rm rot}$.  At longer periods of $P_{\rm
rot}\approx$30\,days, the agreement was typically at the
$\lesssim$0.03$P_{\rm rot}$ level, and the envelope of the period
difference increased roughly linearly with period.  Based on this
comparison, we adopted a simple prescription for the period
uncertainties, such that there are 1\% relative uncertainties below
$P_{\rm rot}=15$\,days, and a linear increase thereafter, with slope
set to require 3\% $P_{\rm rot}$ uncertainties at rotation periods of
30 days.

\subsection{Kepler Objects of Interest}
\label{subsec:planetsel}

We considered planets in the NEA cumulative KOI table as of 2023 June
6, which included the best knowledge available on any given planet
candidate while also incorporating human-based vetting.  These planets
represent a superset of those in the fully automated Q1-Q17 DR25 KOI
Table \citep{Thompson_2018}, which could be adopted in future work for
planet occurrence rate calculations.  This version of the cumulative
KOI table included \nkoisnofp\ objects that are either ``confirmed''
or ``candidate'' planets, after excluding known false positives. 

\subsection{High Resolution Spectra}
\label{subsec:lithiumsel}

The final piece of our analysis involves assessing ages based on the
\ion{Li}{1} 6708\,\AA\ doublet.  We analyzed spectra from the High
Resolution Echelle Spectrometer (HIRES; \citealt{vogt_hires_1994}) on
the Keck I 10m telescope.  These spectra were primarily collected
through the California Kepler Survey
\citep{2017AJ....154..107P,2017AJ....154..108J,2017AJ....154..109F}.
We supplemented the existing archive with $\approx$10~hours of new
observations for 22 stars between Fall 2022 and Spring 2024.  These
stars were chosen to ensure that confirmed planets with rotational
evidence for ages below 1\,Gyr had spectra, since this is the age
range in which lithium is most likely to yield useful age constraints.

Lithium equivalent widths and abundances for the Kepler Objects of
Interest were already analyzed by \citet{2018ApJ...855..115B} for
roughly three quarters of the spectra in our sample.  However, new
spectra have since been acquired, and our approach and selection
function are different.  We therefore performed our own line width
measurements on the reduced HIRES spectra.

We collected all blaze-corrected HIRES spectra from our group's
observations of non-false positive Kepler Objects of Interest with a
``multiple event statistic'' (MES, {\it koi\_max\_mult\_ev}) of at
least 10.  This yielded at least one spectrum for 1{,}464 stars
hosting 2{,}174 planets.  About half of these stars have measured
rotation periods (\nstwithspecandprot\ stars and \nplwithspecandprot\
planets, respectively).   For stars with multiple spectra available,
we analyzed only the spectrum with the highest number of counts.  The
resulting spectra were acquired between 2009 September 6 and 2024 May
16, and cumulatively comprise \nhireshours\,hours of open-shutter
time.

\begin{figure}[!t]
	\begin{center}
		\leavevmode
		\includegraphics[width=0.47\textwidth]{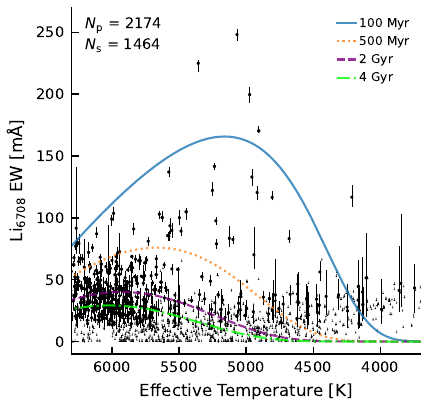}
	\end{center}
	\vspace{-0.5cm}
	\caption{{\bf Equivalent widths (EW) of the \ion{Li}{1} 6708\,\AA\ doublet
    for planet-hosting stars.} These measurements were made from
    Keck/HIRES spectra collected from 2009--2024.  Lines are the ``mean''
    isochrones from \citet{Jeffries_2023}.  The intrinsic dispersion
    around these isochrones becomes much larger than changes in the
    mean at $\gtrsim$1\,Gyr (see Figure~\ref{fig:models}).  Some stars
    with lithium detections do not have detected rotation periods, and
    vice-versa.
		\label{fig:li_vs_teff}
	}
\end{figure}

We measured the lithium equivalent widths using a procedure adapted
from previous work \citep{Bouma_2021}.  Our stars of interest are FGK
stars, and so the continuum in the vicinity of the \ion{Li}{1}
6708\,\AA\ doublet is well-defined.  We Doppler-corrected the spectra
to a common reference wavelength by cross-correlating against a high
S/N template for Kepler-1698, chosen because $|\gamma|$$<$10\kms,
$T_{\rm eff}$$\approx$5000\,K and $v\sin i$$\approx$5\,\kms, which
puts it in the middle of our sample's temperature range and gives it
mild line-broadening.  We then trimmed the Doppler-corrected spectra
to a local window centered on the lithium line, using a window width
of 15\,\AA\ (we also considered 10\,\AA\ and 20\,\AA; the results were
consistent).  We continuum-normalized by fitting a third-degree
Chebyshev series, while excluding regions with absorption lines.  We
then numerically integrated the resulting spectrum using a
one-component Gaussian with free amplitude, width, and mean, and
estimated uncertainties on the line width through a Monte Carlo
procedure that bootstrapped against the local scatter in the
continuum.  The resulting EWs are shown in
Figure~\ref{fig:li_vs_teff}.

Our EW measurement approach did not correct for the neighboring
\ion{Fe}{1} 6707.44\,\AA\ blend.  To evaluate the accuracy and
precision of our method, after applying an initial iteration on the
\nlithiumstars\ stars with spectra, we compared our lithium equivalent
widths with those reported by \citet{2018ApJ...855..115B}.  For the
stars in both samples, we found broad agreement at $>$30\,m\AA, and
significant differences at $<$30\,m\AA\ because
\citet{2018ApJ...855..115B} required positive EWs, while we allowed
for statistically negative ones.  At $>$30\,m\AA, there is however a
small offset in our respective scales caused by our non-treatment of
the iron blend, such that (B24-B18)$=$7.50$\pm$8.55\,m\AA.  We
therefore directly subtracted this constant mean value (7.50\,m\AA)
when calculating lithium-based ages.  This offset is at worst five
times smaller than the astrophysical scatter present in lithium EWs
for calibration clusters at any given age
\citep[see][]{Jeffries_2023}.

\section{Stellar Properties}
\label{sec:stellarprops}


Our default source for stellar temperatures and surface gravities was
the Gaia-Kepler Stellar Properties Catalog (GKSP;
\citealt{Berger_2020a_catalog}).  The GKSP parameters were reported
for stars with ``AAA'' 2MASS photometry, measured parallaxes in Gaia
DR2,  and $g$-band photometry available from either SDSS or else
Kepler-INT \citep{2012AJ....144...24G}.  The parameters themselves
were derived using \texttt{isoclassify} \citep{2017ApJ...844..102H} to
interpolate over the MIST isochrone grids
\citep{Choi_2016,2016ApJS..222....8D}, given the SDSS $g$ and 2MASS
$K_{\rm s}$ photometry, the Gaia DR2 parallaxes, and $V$-band
extinction from the \citet{2019ApJ...887...93G} reddening map.  The
resulting stellar parameters are available for
\fracstarswithprotwithbtwenty\ of the \nuniqstarsantosrot\ Kepler
stars with rotation periods.  For the remaining
\fracstarswithprotwithoutbtwenty\ of stars that lack temperatures and
surface gravities from \citetalias{Berger_2020a_catalog}, we adopted
the values reported by \citet{Santos_2019} and \citet{Santos_2021},
which are primarily derived from the photometric \citet{Mathur_2017}
DR25 Kepler Stellar Properties Catalog.  In the planet sample,
\frackoisnofpwithprotwithbtwenty\ of the non-false-positive KOIs with
rotation periods have parameters from \citet{Berger_2020a_catalog},
and the remainder are drawn from DR25. 

\citet{David_2021} compared the photometric
\citetalias{Berger_2020a_catalog} stellar parameters ($T_{\rm eff}$,
$R_\star$, [Fe/H]) against the spectroscopic parameters from
\citet{Fulton_2018}.  The temperature scales showed a few-percent
systematic difference, with \citetalias{Fulton_2018} quoting higher
temperatures than \citetalias{Berger_2020a_catalog} for mid K dwarfs,
and lower temperatures for early F dwarfs.  Our age analysis,
described below, adopts the maximum of the two-sided uncertainties
reported by \citetalias{Berger_2020a_catalog} as a symmetric Gaussian
temperature uncertainty.  Systematic uncertainties in the temperature
scale generally influence our age uncertainties at a smaller level
than the statistical intrinsic scatter in the open cluster rotation
sequences.

\begin{figure*}[!t]
	\begin{center}
		\leavevmode
		\subfloat{
			\includegraphics[width=0.49\textwidth]{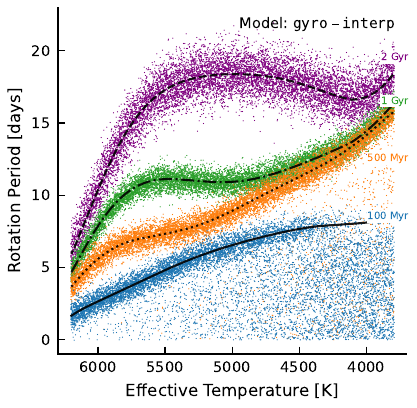}
		}
		\subfloat{
			\includegraphics[width=0.49\textwidth]{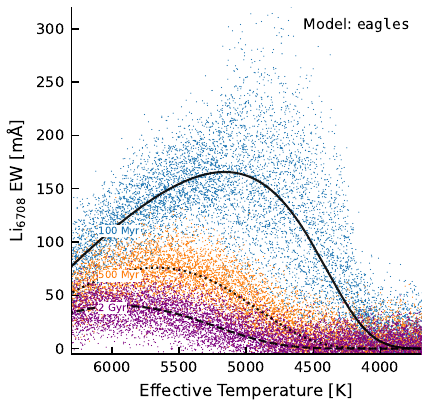}
		}
	\end{center}
	\vspace{-0.6cm}
  \caption{{\bf The models.}
    Points represent $10^4$ draws from models that have been fitted to
    rotation periods \citep{Bouma_2023} and lithium equivalent widths
    \citep[EWs;][]{Jeffries_2023} of stars in open clusters.  Lines
    are the ``mean models'' at various ages.  The intrinsic dispersion
    around the mean, which is what the models fit, sets the
    theoretical precision floor for the age-dating methods.
    Additional sources of uncertainty, including measurement
    uncertainty, impose further limits\deleted{ on achievable precision}.  These
    models are calibrated using data from open clusters.  The sizes of
    the points are the same in each panel, so that low apparent
    density signifies greater dispersion around the mean.
		\label{fig:models}
	}
\end{figure*}

\section{Ages From Rotation}
\label{sec:rotage}

We calculated gyrochrone ages using \texttt{gyro-interp}
\citep[\added{\texttt{v0.6};}][]{Bouma_2023}, which is a method designed to address the fact
that stars with the same mass and same rotation period can have a wide
range of ages \citep[e.g.][]{Gallet_Bouvier_2015}.  The gyrochrone age
posterior should therefore incorporate the intrinsic population-level
scatter into its statistical precision.  Figure~\ref{fig:models}
highlights this problem, particularly in regions where the 0.1--1\,Gyr
stars overlap.

To estimate the stellar spin-down rate as a function of time and
stellar temperature, \texttt{gyro-interp} uses measured rotation
periods and effective temperatures from reference open clusters, and
interpolates between them using cubic Hermite polynomials.  We
calculated the probability of the rotation-based age, $t_{\rm gyro}$,
given the observed periods and temperatures (and their uncertainties)
by integrating Equation~1 of \citet{Bouma_2023}.  This procedure
marginalizes over the astrophysical scatter that is observed in the
open cluster sequences.  We adopted a linear age grid spanning
0--5\,Gyr with 500 grid points, and interpolated the resulting
posteriors to calculate any summary statistics.  The oldest cluster
for which \texttt{gyro-interp} is calibrated\added{ in \texttt{v0.6}} is the 4\,Gyr M67 cluster
\citep[][]{Dungee_2022,Gruner_2023}.  For \deleted{Kepler }stars with rotation
periods above the M67 sequence, our resulting age constraints are\added{ therefore} quoted as lower limits.

\added{Finally, we opted to modify \texttt{gyro-interp} in \texttt{v0.6} to
incorporate an age-dependent population-level Gaussian scatter, as detailed in the change log at \url{gyro-interp.readthedocs.io}.
While age precisions at $\leq$1\,Gyr are unaffected, this update
accounts for the observed intrinsic scatter in
Ruprecht-147 and M67 being larger than in Praesepe or NGC-6811,
potentially due to differential rotation \citep[e.g.,][]{Epstein_2014}.  
For $\approx$2-4\,Gyr old stars with rotation periods measured at $\approx$5\% precision, 
this change yields a $\lesssim$50\% increase in age uncertainties.}
	
\subsection{Stellar Quality Flags}
\label{subsec:flags}
We calculated gyrochrone ages for all \nuniqstarsantosrotteffcut\
stars with reported rotation periods that had effective temperatures
of 3800--6200\,K.  To identify stars for which we suspect these ages
may not be valid, we then built a set of quality flags which we
condensed into a single binary number: $Q_{\rm star}$.  When and how
this bitmask should be applied depends on the question being asked.
If the goal is to construct a false-positive free sample, all the
quality flags could be applied.  If the goal is to construct a
complete sample, then consider the examples of Kepler~1627Ab
($\approx$40\,Myr) and Kepler~51c ($\approx$625\,Myr).  The former has
a high RUWE due to a resolved binary companion \citep{Bouma_2022a};
the latter is on a grazing orbit \citep{2014ApJ...783...53M}.  We
leave selection for or against such cases as a decision to the user.
For our own analysis, we assume that a star is suitable for
gyrochronology if none of bits zero through \replaced{nine}{ten} (inclusive) are
raised.  For analyses that require all stellar rotation periods to
come from the same detection pipeline, we further require bit \replaced{11}{12} to be
zero. 

Three assumptions must hold for a rotation-based age dating method
like \texttt{gyro-interp} to be valid.  {\it 1)} The evolutionary
state of the star must be well-specified by its temperature and its
age, {\it 2)} the star's spin-down must not be influenced by binary
companions, and {\it 3)} the rotation period distribution for field
stars of a given temperature and age must be identical to that of
equally aged open clusters (metallicity differences, for instance, are
ignored).  A rephrasing of the first condition is that the star must
be ``near the main sequence'' because during the post main sequence
stellar temperatures change.  From $\approx$0.08 to $\approx$3\,Gyr,
the stars of interest in this work (3800--6200\,K,
0.5--1.2\,$M_\odot$) have temperatures constant to $\approx$1\%
\citep{Choi_2016}.  From $\approx$3--4\,Gyr, a 1.2\,$M_\odot$ star's
temperature drops from $\approx$6200\,K to $\approx$6000\,K
($\approx$3\%), because its outer layers expand as it begins hydrogen
shell burning as a subgiant.  We treat this issue in a manner
discussed in ``bit 9'' below.

{\it Temperature range (bit 0)}---We require stars to have effective
temperatures $T_{\rm eff}$ between 3800 and 6200\,K in order to report
gyrochrone ages \citep{Bouma_2023}.   Stars hotter than 6200\,K spin
down very slowly, if at all.  Stars cooler than 3800\,K do spin down
over gigayear timescales
\citep{2016ApJ...821...93N,2023ApJ...954L..50E,2024arXiv240312129C},
but the currently available open cluster data have yet to clarify when
the intrinsic scatter in the population decreases.

{\it Surface gravity (bit 1)}---We flagged stars as possible subgiants
if they had $\log g < 4.2$. 

{\it Absolute luminosity (bit 2)}---We calculated the absolute Gaia
DR3 $G$-band luminosity, ignoring reddening, using the reported
apparent $G$-band magnitude and parallaxes.  We flagged stars with
$M_{\rm G} < 3.9$ or $M_{\rm G} > 8.5$, corresponding to main-sequence
spectral types earlier than $\approx$F8V or later than $\approx$M0.5V
\citep{Pecaut_2013}.

{\it Known eclipsing binaries (bit 3)}---We flagged any stars reported
to be in the final Kepler eclipsing binary catalog (KEBC;
\citealt{2016AJ....151...68K}).

{\it Kepler-Gaia crossmatch quality (bits 4 and 5)}--- To leverage 
Gaia DR3 data, we used M.~Bedell's 4$''$
Kepler-to-Gaia crossmatch 
of the NEA \texttt{q1\_q17\_dr25\_stellar} catalog with Gaia DR3
(available at \url{https://gaia-kepler.fun}).  The
separation distribution of the Kepler-Gaia DR3 crossmatches is such
that 99.2\% of candidate matches are within 1$''$.   We nonetheless
noted an upturn in the candidate match rate from 3-4$''$; such sources
are flagged using bit 4.  For KIC stars with multiple potential Gaia
matches within the 4$''$ radius, we adopted the brightest star as the
default match.  In most such cases this was unambiguous because there
is a large brightness difference between the primary and any apparent
neighbors.  However cases with multiple stars within 4$''$ within
$\Delta G$$<$$0.5$\,mag are noted using bit flag 5.  

{\it Gaia DR3 non-single-stars (bit 6)}---The Gaia DR3
\texttt{non\_single\_star} column in the \texttt{gaia\_source} table
flags known eclipsing, astrometric, and spectroscopic binaries.  We
directly included this column.

{\it RUWE (bit 7)}---We inspected diagrams of the Gaia DR3
renormalized unit weight error (RUWE) as a function of other stellar
parameters, and flagged stars with RUWE$>$1.4 as possible binaries.  Such
astrometric outliers can be either bona fide astrometric binaries, or
more often are marginally resolved point sources for which a
single-source PSF model provides a poor fit.

{\it Crowding (bit 8)}---We searched the Gaia DR3 point source catalog
for stars within 1 Kepler pixel (4$''$) of every target star.  While
such companions may not not be physically associated with the target
star, their presence can confuse rotation period measurements.  We
therefore flagged any stars with neighbors down to 1/10$^{\rm th}$ the
brightness of the target star within this region ($\Delta G$$<$2.5).\added{ We also
considered a deeper cut ($\Delta G$$<$5), and found it had negligible impact
on our conclusions.}

{\it Near the main sequence (bit 9)}---Figure~\ref{fig:stellarprops} shows
that many stars observed by Kepler
are far from the main sequence.  Some of the challenges this introduces
for rotational ages include unresolved binaries, metallicity,
reddening, and drivers of rotation other than magnetic braking.
After exploring various options, we settled on the orange locus in the
$\log g$--$T_{\rm eff}$ plane shown in Figure~\ref{fig:stellarprops}
as a way of flagging unresolved binaries, as well as evolved
late-F and early-G stars.  While the exact details of how this locus
is constructed are arbitrary (see Appendix~\ref{app:linemethod}), the
general aim is to flag stars for which there is evidence based on
their location in the \replaced{HR}{HR or Kiel} diagram\added{s} that our gyrochronology model may not
be reliable.  While the $\log g$--$T_{\rm eff}$ cut flags about half
of Kepler stars with rotation periods as potentially not being
suitable for gyrochronology, a few outliers in $M_{\rm G}$ vs.~$G_{\rm
BP}$-$G_{\rm RP}$ remained after applying it.  Such outliers are
likely also questionable.  We therefore also fitted a polynomial to
the KOI main sequence in $M_{\rm G}$ vs.~$G_{\rm
	BP}$-$G_{\rm RP}$, and flagged stars more than 1\,magnitude from
this locus as part of the same bitflag.

\added{{\it Metallicity (bit 10)}---Models
	suggest that stars with non-solar metallicities
	spin down at different rates than solar-metallicity stars 
	\citep{Amard2020}.   This expectation has been hard
	to verify due to a dearth of nearby non-solar metallicity open clusters.
	However, if true, this could cause systematic
	biases for gyrochrone ages \citep[e.g.~Figure~9 of][]{Claytor2020}.
	We therefore flagged stars with spectroscopic $|[{\rm Fe/H}]|$$>$0.3
	from either LAMOST DR5 \citep{Zong2018} or CKS \citep{Petigura_2022}
	as being outside the range of our gyrochronology calibration.
	We discuss limitations of this approach in Section~\ref{sec:disc}.
}

{\it Candidate pulsators and close-in binaries (bit
\replaced{10}{11})}---\citeauthor{Santos_2021} included in a flag for candidate
``classical pulsators'' (e.g.\ Cepheids) and close-in binaries.
Visual inspection shows that although this flag selects many bona fide
objects in these classes, it also selects known young planets
with $P_{\rm rot}$$\lesssim$5\,day.  These objects are
neither classical pulsators nor close binaries.  While we propagated
this flag into our set of quality flags, we therefore generally ignore
it.

{\it Not in the homogeneous stellar rotation sample (bit \replaced{11}{12})}--- This
flag was set for the KOIs with rotation periods drawn from a source
other than the \citeauthor{Santos_2019} pipeline.

\subsection{Planet Quality Flags} \label{subsec:plflags} Some planets
are more reliably identified than others.  We used the following
additional quality indicators to assess the reliability and utility of
a planet, and assembled them into a separate bitmask, $Q_{\rm
planet}$.

{\it Candidate reliability (bit 0)}---The NEA's Kepler Objects of
Interest Table includes both an overall planet-candidate disposition
status ({\it koi\_disposition}), as well as a disposition based only
on the Kepler data ({\it koi\_pdisposition}).  We required both to
include only ``planet candidates'' and ``confirmed planets.'' 

{\it Candidate S/N (bit 1)}---In any transit survey, the false
positive rate increases greatly toward the noise floor for planet
detection \citep[e.g.][]{2002ApJ...564..495J}.  We required a S/N in
excess of Kepler's usual 7.1$\sigma$ floor, through a cut on the
maximum ``multiple event statistic'' (MES, {\it koi\_max\_mult\_ev}):
we required ${\rm MES}$$>$10.

{\it Grazing planets (bit 2)}---Grazing objects, for which the impact
parameter $b$ is greater than $1-R_{\rm p}/R_\star$, often yield
biased planetary parameters \citep[e.g.][]{2022AJ....163..111G}.  For
large planets, they also include astrophysical false positives at
higher rates \citep{2016ApJ...822...86M}, in part due to the
size-impact parameter degeneracy.  We flagged planet candidates as
potentially grazing if $b$$>$0.8, using the impact parameters reported
by \citet{Thompson_2018}.

\section{Ages From Lithium}
\label{sec:liage} 

Figure~\ref{fig:li_vs_teff} shows our measured equivalent widths for
the lithium 6708\,\AA\ absorption doublet, plotted over the mean
isochrone models from \texttt{EAGLES} \citep{Jeffries_2023}.  We
show an upper-limit for plotting purposes if the 1$\sigma$ lower
limit on the equivalent width is below 10\,m\AA.  Our measured EWs
span -40 to 250\,m\AA; all negative EWs have uncertainties that are
statistically consistent with zero.

To calculate lithium ages, we used \texttt{EAGLES} (git commit
\texttt{ac09637}), which, similar to \texttt{gyro-interp}, is based on
an empirical interpolation approach.  \texttt{EAGLES} 
was calibrated on lithium measurements of stars observed by the
Gaia-ESO spectroscopic survey in 52 open clusters with ages spanning 2
to 6{,}000\,Myr \citep{Jeffries_2023}.  We adopted a linear prior in
age, spanning $10^6$ to $10^{10}$\,yr, and symmetrize our EW
uncertainties for purposes of interfacing with \texttt{EAGLES} by
taking the maximum of the measured positive and negative 1$\sigma$
uncertainty intervals.

In part because many of our EWs are upper limits, many of the lithium
ages are lower limits.  In such cases,  the inferred age posterior is
strongly influenced by the assumed intrinsic dispersion in
the lithium model, and by the prior.  In these instances, we
quote the 3$\sigma$ (99.7$^{\rm th}$ percentile) limits.

\section{Cluster Age Comparison}
\label{sec:litagecomparison}

\begin{figure*}[!t]
	\begin{center}
		\subfloat{
			\includegraphics[width=0.95\textwidth]{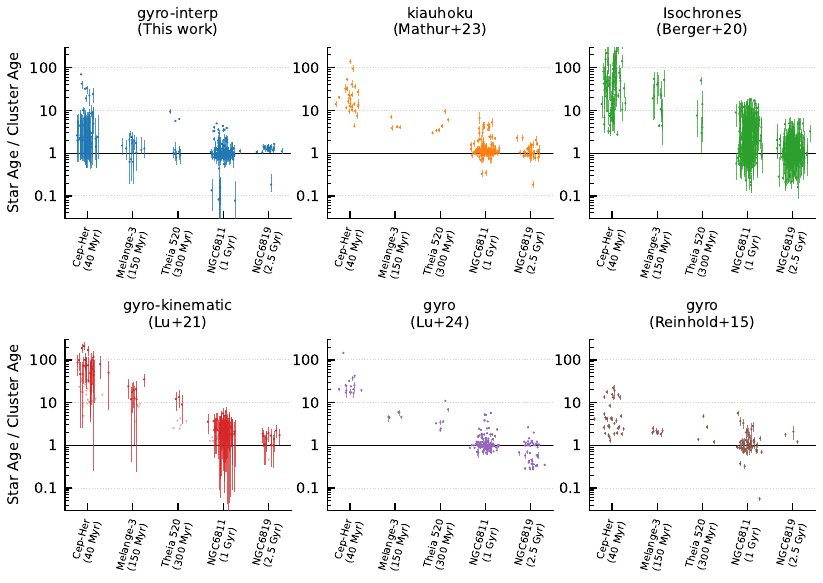}
		}
	\end{center}
	\vspace{-0.5cm}
	\caption{
		{\bf Reported ages of individual stars in benchmark open clusters.}
    Each point denotes a star's reported age, normalized by the age of
    a stellar ensemble in which the star is a candidate member.
    Cluster membership was evaluated without knowledge of rotation\added{, and for this plot,
    no attempt was made to exclude known binaries}.
    Although some field interlopers may be present in the membership
    lists, outliers can be compared between different methods on a
    relative basis.  Each study was cross-matched against the same
    cluster list; certain methods report ages for more stars than
    others.  Horizontal scatter is added to visually clarify the
    statistical age uncertainties.  Below 1\,Gyr, our ages are
    generally accurate, and their statistical uncertainties match
    their dispersion.
	}
	\label{fig:agescalecompone}
\end{figure*}

Stars that formed in the same birth cluster are the gold standard for
the astronomical age scale \citep{Soderblom_2010}.  Before Gaia, the
few open clusters known in the Kepler field had been cataloged by
\citet{1864RSPT..154....1H}.  Gaia has enabled the discovery of
new stellar ensembles that are more diffuse, but which nonetheless share a
common age based on isochrone, rotation, and lithium dating.
Specifically in the Kepler field,
NGC6811 (1\,Gyr; \citealt{Curtis_2019_ngc6811}) and NGC6819 (2.5\,Gyr; \citealt{Meibom_2015}) 
have been known for centuries,
while 
Theia~520 ($\approx$300\,Myr; \citealt{2019AJ....158..122K}), Melange-3 ($\approx$150\,Myr; \citealt{Barber_2022}), and Cep-Her ($\approx$40\,Myr; \citealt{Bouma_2022b} and \citealt{Kerr2024}) are more recent discoveries.

In Figure~\ref{fig:agescalecompone}, we compare our rotational
ages, and available ages from the literature, against the ages of
stars in these open clusters. 
From the literature, we
drew ages from \citet{Reinhold_2015}, \citet{Berger_2020a_catalog},
\citet{2021AJ....161..189L}, \citet{2023ApJ...952..131M}, and
\citet{2024AJ....167..159L}.  We followed any guidance available from
each study for removing ages that were unreliable, and plotted stars
within $1\sigma$ of zero as upper limits.  For the
\citeauthor{Reinhold_2015} ages, we used those calculated using the
\citet{Mamajek_2008} calibration; for the
\citeauthor{2023ApJ...952..131M} ages, we show those from
\texttt{kiauhoku} \citep{Claytor2020} rather than \texttt{STAREVOL}
\citep{Amard2019}, since the former showed better agreement with the
cluster age scale.

Cluster membership is a nuanced subject.
Figure~\ref{fig:agescalecompone} is showing reported ages for a set of
spatially and kinematically selected stars that could be cluster
members, or they could be field interlopers.  For NGC6811 and NGC6819,
we adopted candidate members from
\citet{2018A&A...618A..93C,CantatGaudin_2020} and
\citet{Kounkel_2020}.  For Theia~520, we used candidate members from
\citet{Kounkel_2020}.  For Melange-3, we used the candidates reported
by \citet{Barber_2022} and required ``offset'' tangential velocities
below 2\,\kms.  For Cep-Her, we used candidate members from
\citet{Kerr2024} with $P_{\rm fin}$$>$0.8.  Even with
contaminants, we can compare the relative age distributions derived by
the different studies because in all instances we are comparing
reported ages against a fixed list of stars.

There are two main metrics for success in this test. {\it 1)} Do the
reported ages agree with the cluster age? {\it 2)} Do the reported age
uncertainties agree with their dispersion around the cluster age?

Figure~\ref{fig:agescalecompone} show that although previous studies
reported ages that agree with the cluster scale for $\gtrsim$1\,Gyr
stars, sub-gigayear stars have historically had their ages
overestimated by 0.3--2\,dex, with severely understated uncertainties.
For isochrone and kinematic ages, this is because these methods rely
on parameters that do not appreciably change at $t$$\lesssim$1\,Gyr.
For the rotational ages in \citet{2023ApJ...952..131M} and
\citet{2024AJ....167..159L}, the discrepancy is caused by details of
the calibration methodologies.  The \texttt{kiauhoku} spin-down model
for instance has known ``stretches'' and ``compressions'' in its age
scale relative to observed open clusters \citep[see][Sec.~7.3]{2023ApJ...952..131M}.  The \citet{Mamajek_2008} calibration used
by \citet{Reinhold_2015} appears more accurate than the former two
models, because it was fitted to reproduce the open cluster sequences
known at the time.  However, the uncertainties from all of these
methods appear to be underestimated.  This is probably because they do
not marginalize over the range of rotation periods accessible to
sub-gigayear stars of a fixed mass and age.

\section{Results}
\label{sec:results}

\begin{figure*}[!t]
  \begin{center}
    \leavevmode
    \subfloat{
        \includegraphics[width=.98\textwidth]{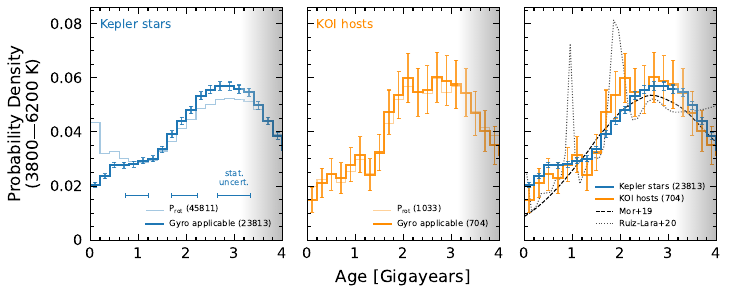}
    }

	\vspace{-0.35cm}
    \subfloat{
        \includegraphics[width=.98\textwidth]{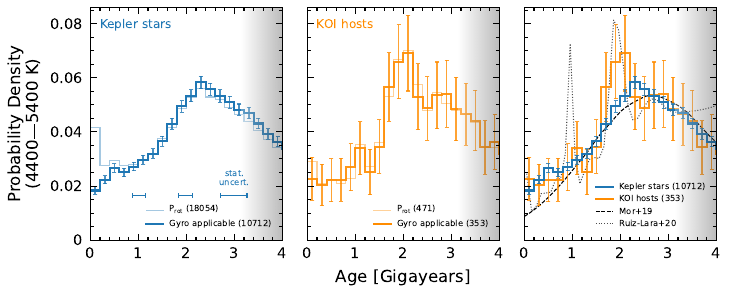}
    }
  \end{center}
  \vspace{-0.6cm}
  \caption{{\bf Kepler's demographic cliff,} visible in the
  rotation-derived age distributions of its stars (left) and planet
  hosts (middle).  The top row shows \replaced{all stars with temperatures of
  3800--6200\,K for which we calculated rotational ages}{stars with temperatures of 3800--6200\,K}.  Opaque lines
  impose quality cuts on binarity, \added{metallicity, }crowding, and the star's
  evolutionary state; transparent lines do not\added{ (see
  Section~\ref{subsec:flags})}.  The bottom row shows
  stars with temperatures of 4400--5400\,K, which \replaced{have more precise
  ages due to their fast spin-down}{have more precise ages}.  The
  statistical uncertainty for an average star at 1, 2, and 3\,Gyr is
  shown, and is identical across each
  row.  
  Finally, the right panel compares the \replaced{rotation-derived}{rotational} ages
  against star formation histories derived \replaced{using CMD
  fitting}{from CMD fits to other regions of the Galaxy} (dashed
  and dotted lines).  Completeness in Kepler's $P$$_{\rm rot}$
  detection sensitivity is near unity until $t$$\lesssim$3\,Gyr\added{, and decreases rapidly at older ages}
  \citep{2022ApJ...937...94M}.
  \label{fig:hist_tgyro}
  }
\end{figure*}

\subsection{Rotation-Based Age Distribution}

\subsubsection{Kepler's Demographic Cliff}

The stellar ages are given in Table~\ref{tab:planets} for the known
planet hosts, and in Table~\ref{tab:stars} for all Kepler stars.
Figure~\ref{fig:hist_tgyro} shows the rotation-based age distributions
of all the stars (left column) and the KOI hosts (middle column).
These ``histograms'' are sampled from the posterior probability
distributions by drawing ten random samples from each posterior, and
then computing the normalized histogram of the resulting samples.
We also considered an alternative approach for constructing these
plots using hierarchical Bayesian deconvolution
\citep{Masuda2022infer}, and found similar results.  The plotted
uncertainties \replaced{assume Poisson statistics}{are Poissonian}.  We truncated the plots
at $4$\,Gyr, which is the upper bound for our
rotational age calibration.  The overall slope \replaced{yields}{shows} a paucity of
young stars relative to the expectation of a uniform age distribution.

The age distributions of Kepler stars and KOI hosts in
Figure~\ref{fig:hist_tgyro} appear similar.  One diagnostic for
whether the two distributions are drawn from the same underlying
distribution is the Kolmogorov-Smirnov test;  we calculated this
statistic in a manner that accounts for the Poisson uncertainties by
performing 1000 random draws of 70\% of the stars from each sample
when requiring $t_{\rm gyro}$$<$3\,Gyr.  The resulting $\log_{\rm
10}p$ values spanned (2.5$^{\rm th}$ to 97.5$^{\rm th}$ percentiles)
-4.5 to -1.8 for the K dwarfs, and -4.5 to -1.9 for all Kepler stars.
This agrees with the visual impression that while small differences
may be present, they are not drastic.

The similarity of the KOI host and Kepler star age distributions adds
some nuance to the argument that young transiting planets are hard to
detect due to the photometric variability of their host stars.  If
true, this statement \replaced{seems to}{might} hold only for the very youngest
($\lesssim$0.4\,Gyr) stars, where there is a marginal deficit in the
KOI host age distribution relative to the parent stellar sample.
Examining scatter plots of the planet properties as a function of
age, we similarly find that fewer systems at $t_{\rm
gyro}$$<$0.4\,Gyr are detected with orbital periods beyond 30\,days
than at $t_{\rm gyro}$$>$0.4\,Gyr.

We can quantify the relative counts of stars as a function of age by
labelling stars between 0-1\,Gyr, 1-2\,Gyr, and 2-3\,Gyr as ``young'',
``intermediate-age'', and ``old''.  A simple counting exercise from
Figure~\ref{fig:hist_tgyro} tells us that there are
\ratioobtoybstars\ times as many old stars in the Kepler field as young stars.
Similarly, there are \ratioobtoybplanets\ times as many old planet
hosts as young planet hosts.  Focusing only on the tails of the
distributions (0-0.3\,Gyr and 2.7-3\,Gyr), the implication is that the
formation rate of stars in the Kepler field decreased by a factor of
\ratiosfr$\pm$\uncratiosfr\ over the past three billion years.  If
this decrease continues linearly into the future, then a simple
extrapolation would imply that in $\approx$1.4$\pm$0.3\,Gyr the
abundance of newly-formed stars in this region of the Galaxy will
reach zero.

In terms of detected planet counts, our rotation-based ages for the
Kepler sample yield \{\nplyounggyro, \nplmidgyro, \nploldgyro\}
detected planets in the 0-1\,Gyr, 1-2\,Gyr, and 2-3\,Gyr bins.

\subsubsection{Nuances in the Age Distribution}

\added{{\it Completeness}---Figure~\ref{fig:hist_tgyro} has a gray
overlay beyond $\gtrsim$3\,Gyr because Kepler's sensitivity to
rotation signals diminishes at old ages.} \citet{2022ApJ...937...94M}
studied \deleted{this issue}\added{Kepler's completeness to rotation
signals using MIST-derived isochrone ages}, and found for Sun-like
stars that the fraction of rotation signals \added{that are
}detectable \deleted{by Kepler }is near unity up to $\approx$3\,Gyr,
and that it drops to almost zero by $\approx$5\,Gyr.  \deleted{This trend is
opposite in functional form to our derived age
distributions.}\added{This estimate suggests that while aspects of
Figure~\ref{fig:hist_tgyro} might be interpretable in terms of the Galaxy's star
formation history at $\lesssim$3\,Gyr, at older ages incompleteness
precludes any such interpretation.}

\added{{\it Temperature sensitivity}---There are quantitative differences
in Figure~\ref{fig:hist_tgyro} between the
K dwarfs (4400--5400\,K) and the FGK stars (3800--6200\,K).  For instance,
the distribution peaks at 2.3\,Gyr for K dwarfs, and at 2.9\,Gyr for
the FGK sample.  The peak is also sharper for the K dwarfs;
its decrease from 2.3-3\,Gyr is in an age range in which we believe the
rotation period catalog completeness to be near unity.  We
show further details of the $t_{\rm gyro}$--$T_{\rm eff}$
distributions in Appendix~\ref{app:age_diagnostic}, and propose possible
explanations for these differences in Section~\ref{subsec:caveats}. }

\subsection{Planets Younger Than One Billion Years}

\deleted{Sub-gigayear planets
can be particularly informative for studies of planet
evolution.}
\added{This analysis provides an important path toward expanding the population
of young and evolving exoplanets.}
Our catalog has \nplyounggyro\ confirmed
and candidate planets with median ages below 1\,Gyr.  Requiring
$t_{\rm gyro}$$<$1\,Gyr at 2$\sigma$ yields \nplyounggyrotwosigma\
planets orbiting \nplhostsyounggyrotwosigma\ stars.  The youth claim
is most secure for the systems that are either in clusters
(Section~\ref{subsec:clusterplanets}), or that have independent
rotation and lithium-based ages
(Section~\ref{subsec:rotnlithiumcomp}).  These samples provide broader
context for the planets around field stars that push the boundaries of
the current young exoplanet census (Section~\ref{subsec:notables}).

\subsubsection{Planets in Clusters}
\label{subsec:clusterplanets}

Four stellar ensembles in the Kepler field have to date yielded
\deleted{a total of }fourteen transiting planets.  Our analysis
blindly recovers the youth of all of these planets.

{\it Cep-Her}---The Cep-Her complex ($\approx$40\,Myr) contains
Kepler-1627Ab, Kepler-1643b, Kepler-1974b and Kepler-1975\,Ab
\citep{Bouma_2022a,Bouma_2022b}.  Different sub-groups of the complex
vary in age by $\approx$50\% \citep{Kerr2024}.  The planets
themselves are all on close-in orbits (5-25\,days), with sizes from
2-4\,$R_\oplus$.  Our rotation and lithium ages agree with the cluster
age for Kepler-1627A, Kepler-1974, and Kepler-1975A.  For Kepler-1643,
the rotation and cluster ages agree, and the lithium age is
2.0$\sigma$ above the cluster age \replaced{reported by}{from} \citet{Bouma_2022b} for
RSG-5.

{\it MELANGE-3}---\citet{Barber_2022} reported a $105$$\pm$$10$\,Myr
association in the Kepler field containing two transiting
mini-Neptunes: Kepler-970 and Kepler-1928.  We find rotational
ages of $t_{\rm gyro}=176_{-40}^{+120}$\,Myr and
$144_{-88}^{+104}$\,Myr for Kepler-970 and Kepler-1928 respectively.
The former \deleted{rotation-based }age being slightly older than the suggested
Pleiades age for the association agrees with Figure~4 of
\citet{Barber_2022}.

{\it Theia~520}---We find rotational ages of $\approx$350\,Myr
for Kepler-968 and Kepler-52.  Based on spatial and kinematic
clustering, these stars are candidate 
members of Theia~520 \citep{2019AJ....158..122K}.  The core of this cluster is also known
as UBC-1 \citep{2018A&A...618A..59C}.  Based on isochrone and
rotational age-dating, Theia~520 seems to be
$\approx$230$\pm$70\,Myr old
\citep{2024A&A...681A..13F}, which is within
1$\sigma$ of our $t_{\rm gyro}$ measurements.  Ongoing work by
\citet{Curtis2024} broadly finds that the two planet-hosting stars are
indeed members of this diffuse population.  Given the cluster-level
and individual-star evidence, this makes Kepler-968 and Kepler-52 the
youngest multiplanet systems currently known from the main Kepler
mission.

{\it NGC~6811}---\citet{Meibom_2013} reported the discovery of
Kepler-66 and -67b, two mini-Neptunes in NGC~6811 ($\approx$1\,Gyr).
We derive rotational ages for these systems of
\kepsixsixtgyro\,Myr and \kepsixseventgyro\,Myr for Kepler-66 and
-67b, respectively.  Kepler-66 ($T_{\rm eff}$$\approx$5900\,K, $P_{\rm
rot}$$\approx$10.4\,days) has a lower precision and a more asymmetric
posterior because of its slow spin-down rate.  Kepler-67 ($T_{\rm
eff}$$\approx$5100\,K, $P_{\rm rot}$$\approx$10.3\,days) is marginally
hotter than the later K dwarfs that are in ``stalled'' spin-down at
this time, enabling its rotation period to be diagnostic of its age.

\begin{figure}[!t]
  \begin{center}
    \includegraphics[width=0.48\textwidth]{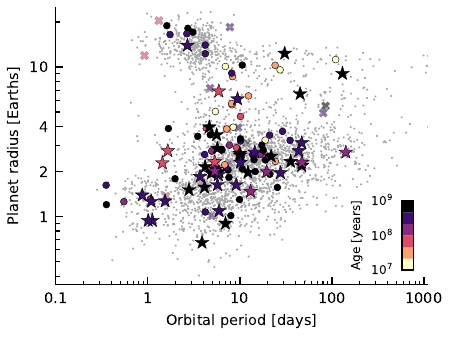}
  \end{center}
  \vspace{-0.5cm}
  \caption{
    {\bf Sizes, orbital periods, and rotation-based ages of transiting
    exoplanets younger than one billion years.}
    \replaced{Star symbols denote \nplyounggyrotwosigmanograzingnoruwe\ planets
    that are characterized in this work to have $t_{\rm gyro}$<1\,Gyr
    at $2$$\sigma$; circles are planets from the literature meeting
    the same age requirement, but with a heterogeneous set of age
    provenances.  Gray points are transiting planets from the NASA
    Exoplanet Archive older than 1\,Gyr.  If one were to instead
    select for precise ages ($t/\sigma_{\rm t}>3$), this would hide
    $\approx$ten $<$0.5\,Gyr planets, and add $\approx$fifteen
    0.5-1\,Gyr planets.  For this plot, we require our Kepler stellar
    hosts to not have $Q_{\rm star}$ bits 0--9 raised,
    and for the planets to similarly no have quality flags
    raised.}{Systems are selected to have $t$<1\,Gyr at $2$$\sigma$.
    Star symbols denote \nplyounggyrotwosigmanograzingnoruwe\ planets
    characterized in this work to have such ages at the highest
    confidence, meaning that none of $Q_{\rm star}$ bits 0--10 are raised,
    and the planets have no quality flags
    raised.  Transparent crosses denote \nplyounggrazingorhighruwe\
    planets for which either the planet is grazing, or the star has
    RUWE$>$1.4.  Gray points are from the NASA Exoplanet Archive. }
    \label{fig:rp_period_age_results}
  }
\end{figure}

\subsubsection{Rotation vs.~Lithium Ages}
\label{subsec:rotnlithiumcomp}

We compared the rotation and lithium ages for the known planet hosts
in the ``Consistent?'' column of Table~\ref{tab:planets}.  There were
three cases of interest.  {\it i)} If two-sided posteriors for both
$t_{\rm gyro}$ and $t_{\rm Li}$ existed, and their median values were
consistent within 2$\sigma$, we listed them as consistent; if they
were consistent within 2-3$\sigma$, we listed them as ``maybe''
consistent.  Otherwise, they disagreed.  {\it ii)} If $t_{\rm Li}$ was
a lower limit, we compared this lower limit with the $1\sigma$ upper
limit from $t_{\rm gyro}$; if the two overlapped, we judged the age
estimates to be consistent.  {\it iii)} If $t_{\rm Li}$ provided a
two-sided posterior, and no rotation period was found, we judged the
age estimates to be inconsistent.  This is because a two-sided lithium
constraint can only be provided for G and K dwarfs $\lesssim$2\,Gyr
old, and Kepler should have been sensitive to the rotation periods of
such stars.

In the entire planet sample (i.e.,~without imposing any quality cuts)
this yielded \allagesyesconsistent\ consistent cases,
\allagesmaybeconsistent\ maybe consistent cases, and
\allagesnoconsistent\ inconsistent cases.  Rephrased, in the overall
sample, \fracconsistentallages\% of stars have consistent rotation and
lithium-based ages (\fracpotentiallyconsistentallages\% potentially
consistent).

Appendix~\ref{app:inconsistent} describes all cases of ``discrepant''
confirmed planets for which a sub-gigayear age is reported by at least
one age indicator.  All but two systems are either evolved stars or
else unresolved binaries, and are automatically flagged as such.  The
more interesting of the two is Kepler-786, an early K dwarf with
$t_{\rm Li}$=\kepseveneightsix\,Myr, but with a $\approx$33\,day
rotation period that implies $t_{\rm
gyro}$$\approx$4.4\,Gyr.  Other age indicators in the
spectrum similarly suggest that the star is not young.  It 
therefore appears to be anomalously lithium-rich.

\subsubsection{Notable New Young Planets}
\label{subsec:notables}

Let us focus on blemish-free stars with blemish-free planets: $Q_{\rm
star}$ must not have any of bits zero through \replaced{nine}{ten} raised, and $Q_{\rm
planet}$ must similarly have no quality flags raised.  Under this
constraint, our results include \ltonegyrhighqconfirmedtwosided\
confirmed planets younger than 1\,Gyr with two-sided $t_{\rm gyro}$
and $t_{\rm Li}$, and \ltonegyrhighqconfirmedonesided\ confirmed
planets with comparable ages for which $t_{\rm gyro}$ is two-sided
while $t_{\rm Li}$ is one-sided.

Figure~\ref{fig:rp_period_age_results} shows the planets and planet
candidates with $t_{\rm gyro}$$<$1\,Gyr at 2$\sigma$.  \replaced{This}{The stars in this} figure
include\deleted{s} \nconfirmedplyounggyrotwosigmanograzingnoruwe\ confirmed
Kepler planets, and \ncandidateplyounggyrotwosigmanograzingnoruwe\
candidate planets, all \deleted{of which are }listed in Table~\ref{tab:planets}.
These include \nearthshighq\ objects Earth-sized or smaller,
\njupitershighq\ Jovian-sized planet candidates, \nsubsaturnshighq\
super-Neptunes (4--10\,$R_\oplus$), \nminineptuneshighq\ mini-Neptunes
above the radius valley as parametrized by
\citealt{2018MNRAS.479.4786V}, and \nsuperearthshighq\ super-Earth
sized planets below the radius valley.  While most of these planets
have orbital periods below 50\,days, \nlongperiodhighq\ are on more
distant orbits.

Among these systems, four highlights are Kepler-1529, Kepler-1565,
Kepler-1312, and Kepler-1629.  Kepler-1529\,b is a
$\approx$100$\pm$50\,Myr mini-Neptune with well-measured rotation and
lithium ages.  Kepler-1565\,b is an analogous $\approx$170-230\,Myr
super-Earth.  Kepler-1312 is a $t_{\rm gyro}$=\kepthirteentwelve\,Myr
near-Solar analog with an Earth-sized planet on a one-day orbit, and a
mini-Neptune on a five-day orbit.  The Earth-sized planet,
Kepler-1312\,c, along with the near-identical Kepler-1561\,b ($t_{\rm
gyro}$=\kepfifteensixone\,Myr), \replaced{rank among}{could be} the youngest Earth-sized
planets currently known, \added{and are }comparable to planets such as HD~63433\,d
\citep[1.1\,$R_\oplus$, $414$$\pm$$23$\,Myr;][]{2024AJ....167...54C}
and TOI-1807\,b \citep[1.3\,$R_\oplus$,
180$\pm$40\,Myr;][]{2021AJ....162...54H}.  Kepler-1629\,b, with a size
just two-thirds that of Earth (0.67$\pm$0.05\,$R_\oplus$), is
marginally older, also based on rotation (\kepsixteentwonine\,Myr).
The lithium age from a reconnaissance TRES spectrum confirms the
point.

\section{Discussion}
\label{sec:disc}

\begin{figure*}[!t]
	\begin{center}
		\leavevmode
		\subfloat{
			\includegraphics[width=.99\textwidth]{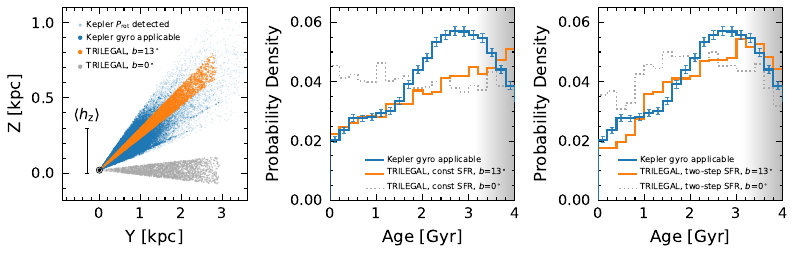}
		}
	\end{center}
	\vspace{-0.53cm}
	\caption{{\bf Impact of star formation rate (SFR) and kinematic heating on age distribution.}   
		Most Kepler stars are a few hundred parsecs above the Galactic plane ($\hat{Z}$ points to the north galactic pole;  $\hat{Y}$ toward the direction of
		galactic rotation).  This is comparable to the $\approx$300\,pc mass-weighted scale height.
		The dotted gray and solid orange lines show synthetic age distributions
		from TRILEGAL assuming a constant SFR (middle) and a two-step
		SFR (right), described in the text.
		A combination of both kinematic heating and
		a decreasing SFR are needed to qualitatively match the observed
    paucity of the youngest stars (blue line; as in
    Figure~\ref{fig:hist_tgyro}).
		\label{fig:trilegal}
	}
\end{figure*}

\subsection{Kinematic Heating and Star Formation in the Thin Disk}
\label{subsec:trilegal}

\added{Two separate astrophysical effects might explain our observed 
age distribution (Figure~\ref{fig:hist_tgyro}): a declining star formation
rate (SFR), and kinematic heating.   Kinematic heating is associated with an
increase in both stellar velocity dispersions and the disk's vertical 
scale height over time \citep[e.g.][]{Villumsen1983,Schmidt2024}.
The Kepler field points toward $(l,b)$=(76.3$^\circ$,13.5$^\circ$).
A typical Kepler star at 1\,kpc is therefore $\approx$230\,pc
above the galactic plane, a significant fraction of the thin disk's 
mass-weighted $\approx$300\,pc scale height \citep{Bland-Hawthorn2016}.
}

\added{We assessed the relative importance of these two effects
	using the TRILEGAL Monte Carlo population synthesis tool }\citep[v1.6;][]{Girardi2005}.
\added{TRILEGAL treats the Galaxy as a linear combination of a thin disk, thick disk,
	halo, and bulge, and allows for two possible SFRs: constant, and a ``two-step''
	SFR in which the thin disk SFR is 50\% larger from 1-4\,Gyr than at other ages.
	This two-step SFR was introduced to improve the model's ability to
	match star counts from Hipparcos and 2MASS }\citep[see Section~5.1 of][]{Girardi2005}.
\added{For kinematic heating, TRILEGAL assumes that
	the disk scale height grows as
$h_{\rm z} = z_0 (1 + t/t_0)^\alpha$, for $\alpha$=5/3, $z_0$=94.7\,pc,
	and $t_0$=5.55\,Gyr, similar to \citet{Rana1991}.
	We queried the web interface toward the
	Kepler field, and toward the galactic plane (at $l$=76.3$^\circ$) to
	negate the effects of kinematic heating.
	We limited the output to stars with apparent Kepler magnitudes brighter than 16,
	and set the field of view to 10\,${\rm deg}^2$, the maximum
	available through the web interface.
}

\added{Figure~\ref{fig:trilegal} compares the results against
	our observed age distribution.
	The middle panel shows that assuming a constant SFR,
	kinematic heating alone can yield a two-fold difference
	in the abundance of the youngest stars relative to those $\approx$3\,Gyr old.
	The right-most panel
	shows that additionally including 
	the time-variable SFR yields the best agreement with the slope of
	our observed age distribution.
}

\deleted{Most stars observed by Kepler are in the thin disk.
This can be verified by following \citet{Gaia_2018}, and labeling
stars with 2D tangential velocities $v_{\rm T}$$<$40\,\kms\ as thin
disk members, and those with 60$<$$v_{\rm T}$$<$150\,\kms\ as thick
disk members.  These
criteria yield 95{,}694 thin and 54{,}039 thick disk stars
observed by Kepler.  The thin disk stars have
a threefold larger detected rotation period fraction: 35{,}675
rotators are in the thin disk, while 7{,}312 are in the thick disk.
Further imposing the quality flags discussed
in Section~\ref{subsec:flags} yields 17{,}755 thin disk stars for
which gyrochronology is applicable, and 2{,}462 thick disk stars.}

\deleted{Classifying Kepler stars as thin vs.\ thick disk members helps connect
them to previous work on the Galaxy's star formation history.  On
extragalactic scales, the star formation rate in spiral galaxies with
mass similar to the Milky Way peaked $\approx$10\,Gyr ago, and has
since decreased by an order of magnitude. }
\added{Despite the expected differences between pencil-beam surveys
and volume-complete samples, our observed age distribution qualitatively
agrees with a number of previous studies.}
\replaced{In our own
galaxy, previous studies have}{Previous work has} measured star formation histories
through CMD fitting of resolved stellar populations
\citep[][]{2019A&A...624L...1M,2020NatAs...4..965R,2021MNRAS.501..302A,2022Natur.603..599X},
and by modeling the local white dwarf luminosity function
\citep[e.g.][]{2019ApJ...878L..11I}.  \citet{2019A&A...624L...1M} \replaced{and
\citet{2019ApJ...878L..11I} for instance focused on stars (and white
dwarfs)}{and \citet{Mazzi2024} for instance focused on stars} within a few hundred parsecs, and both found a peak in the
local star formation rate 2-3\,Gyr ago.  \citet{2020NatAs...4..965R}
focused on stars in a wider 2\,kpc bubble, and additionally reported
three local maxima in the SFR, which they
associated with pericenter passages of the Sagittarius satellite
galaxy.

The right column of Figure~\ref{fig:hist_tgyro} compares our derived
age distribution against star formation histories (SFHs) reported by
\citet{2019A&A...624L...1M} and \citet{2020NatAs...4..965R} based on
CMD fitting.  The overall slope of both SFHs broadly agrees with what
we find from rotation-based ages.  The SFH from
\citet{2019A&A...624L...1M} seems entirely consistent, particularly
after accounting for the statistical uncertainties of that study.
Regarding the episodic star formation bursts reported by
\citet{2020NatAs...4..965R}, they are not apparent in our overall FGK
star sample (top row).  However, rotational ages are most precise
for $\approx$G8V-K4V dwarfs older than $\approx$1.5\,Gyr
\citep{Bouma_2023}.  The lower row of Figure~\ref{fig:hist_tgyro}
selects these stars with a temperature cut, and may show a \added{weak }hint of the
\added{$\approx$1\,Gyr and }$\approx$2\,Gyr spike\added{s} reported by \citet{2020NatAs...4..965R}.  \deleted{The
$\approx$1\,Gyr spike however is not recovered. } This
$\approx$2.5\,Gyr local maximum is similarly present in the isochrone
ages derived by \citet{Berger_2020a_catalog} for the Kepler field, and
in the red giant asteroseismic ages derived by
\citet{SilvaAguirre2018}.  The main novelty of our ages relative to
\replaced{those studies is their improved accuracy at $<$}{these previous studies is their accuracy below} 1\,Gyr\added{ (see Figure~\ref{fig:agescalecompone})}.

\subsection{Caveats \& Limitations}
\label{subsec:caveats}

\begin{figure*}[!t]
	\begin{center}
		\leavevmode
		\subfloat{
			\includegraphics[width=.98\textwidth]{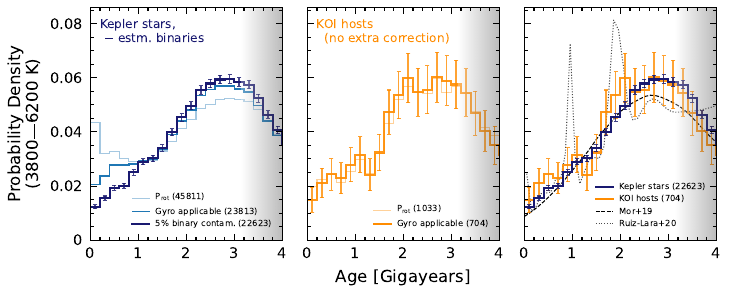}
		}
	\end{center}
	\vspace{-0.6cm}
	\caption{{\bf Potential impact of unresolved binaries.}   This figure shows
		the same raw (thinnest line) and quality-flag cleaned (``gyro applicable'') distributions
		as in Figure~\ref{fig:hist_tgyro}.
		However, even after applying our quality cuts, up to
		\fivepctnuniqstarsantosrotgyroappl\ of \ngyroappllttendays\ apparently clean rotators with
		$P_{\rm rot}$$<$10\,days could be unresolved $\lesssim$0.5\,AU  binaries (see Section~\ref{subsec:caveats}).
		The dark blue ``5\% binary contamination'' distribution corrects for this possibility by randomly removing
		what we estimate (estm.) to be the maximum possible number of such systems; the true
		correction might be smaller.  We expect such a correction to be more important in
		the sample of all Kepler stars than in the KOI host sample.
		\label{fig:hist_tgyro_corr}
	}
\end{figure*}

Our most constraining ages tend to come from one source of
information: rotation.  Factors other than \deleted{stellar }age and mass can
influence \deleted{stellar }rotation rates.  \added{Astrophysical factors include
metallicity, latitudinal differential rotation, mergers, and tidal spin-up.  Observational
limitations include the difficulty of measuring small-amplitude,
long-period signals that are only quasi-coherent due to the starspot
lifetime approaching the stellar rotation period.  There is also the issue that
our age-dating approach is not calibrated beyond 4\,Gyr.
In the following, we attempt to
quantify the importance of these effects in our analysis, ordered by
what we perceive to be their descending importance.}

{\it Binaries}:
Stars in binaries are often biased
toward rapid rotation, even beyond the typical
realm of tidal synchronization \citep[see][]{Meibom_2007,Gruner_2023}.
\added{\citet{2019ApJ...871..174S} for instance found that 59\% of
Kepler stars with rotation periods between 1.5 and 7\,days lie
0.3\,mag above the main sequence, compared with 28\% of their full
rotation sample.
We attempted to mitigate the effects of binarity with five distinct
quality flags (Section~\ref{subsec:flags}).   This attempt was at least
somewhat effective, based on the peak near zero in the raw
distributions of Figure~\ref{fig:hist_tgyro} being flattened in the
cleaned distributions.
However, the completeness of our quality flags is hard to assess.  One specific concern is that these
flags might have minimal sensitivity to binaries with
separations of $\approx$0.05-0.5\,AU, for all but the highest mass ratios.  This inner limit is set by the onset
of detectable ellipsoidal variations in \citet{2016AJ....151...68K}; the outer limit is set by the typical 
orbital periods of astrometric binaries in Gaia DR3 }\citep[see Figure~3 of][]{GaiaCollaboration2023}.

\added{If we were not at all sensitive to
binaries with separations of 0.05-0.5\,AU, then based on
the binary fraction and separation distribution from
\citet{Raghavan2010}, as much as $\approx$5\% of our ``gyro
applicable'' sample
(\fivepctnuniqstarsantosrotgyroappl/\nuniqstarsantosrotgyroappl\
stars) might be contaminated.  In a worst case scenario, all of
these unidentified binaries might induce $P_{\rm rot}$$<$10
days, which would then be mixed in with the bona fide young, single stars.
This would constitute a major source of contamination,
given that our nominal sample of apparently clean stars had
\ngyroappllttendays\ stars with $P_{\rm
rot}$$<$10\,days. }

\added{Figure~\ref{fig:hist_tgyro_corr} shows the rotation-based age distribution
calculated by randomly dropping \fivepctnuniqstarsantosrotgyroappl\ of the
$P_{\rm rot}$$<$10\,day stars.
Relative to the nominal age distribution, there are three important differences.
There is a much larger drop in the star formation rate over the past three gigayears,
by a factor of 4.81$\pm$0.26, rather than \ratiosfr$\pm$\uncratiosfr.
The rate of decline in the age distribution is also constant, rather than having a ``kink'' at 1.5\,Gyr.
And finally, the corrected distributions show closer agreement with both the SFHs
derived from CMD fitting, and also with the KOI host distribution,
which is almost devoid of close binaries within 1\,AU \citep[e.g.][and references therein]{Moe2021}.
We caution however that this correction assumes that {\it none} of our
quality flags were able to identify any of these unresolved binaries.
Since we did in fact include flags based on e.g.,~location in the HR diagram,
this correction is best interpreted as an upper limit on the possible impact of binarity.}

\added{{\it Rotational outliers:} }
We were interested in not only the overall age distribution of the
Kepler field, but also in the reliability of individual ages for \replaced{young
stars known to host planets}{each star}.  We attempted to assess this reliability using
quality bitmasks (Table~\ref{tab:planets}), a comparison against \deleted{open }cluster ages (Section~\ref{sec:litagecomparison}), and a comparison
between rotation and lithium ages.  The \deleted{open }cluster comparison
suggested that our ages were mostly accurate at $\ll$0.5\,dex
across 0.04--2.5\,Gyr, similar to their quoted precision.
Nonetheless, the upper-left panel of Figure~\ref{fig:agescalecompone}
does show outliers, typically $\lesssim$10\% of the population
in any cluster.  \added{While most of these outliers have known bad quality flags, some do not.} 
These stars could be either field star contaminants,
or else anomalous stars whose physical rotation histories were altered
by processes not captured by our statistical uncertainties.
Similarly, although we found consistent $t_{\rm gyro}$ and $t_{\rm Li}$
for \ltonegyrhighqconfirmedtwosided\ planets orbiting ``high-quality''
stars, we did find one system, Kepler-786, with radically different
lithium (\kepseveneightsix\,Myr) and rotational
($\approx$4.4\,Gyr) ages.  Planetary mergers
are one process that could produce such a signal, but
testing this would require a method for measuring differential abundances
across a large number of elements.

\added{A separate plausible origin for rotational outliers is that they come from
	erroneously reported rotation periods.  While
	the large surveys do show good internal consistency (Appendix~\ref{app:mcqonly}),
	the referee correctly pointed out that this is only a comparative test of methods,
	rather than an ability to recover ground truth \citep[e.g.][]{Aigrain2015}.
	One could imagine an adversarial scenario in which, say, 10-20\%\ of stars
	with $P_{\rm rot}$$>$20\,days are erroneously reported, and in fact have much
	longer rotation periods.  In this scenario, a similar analysis to that in Figure~\ref{fig:hist_tgyro_corr} showed
	that our analysis would overestimate the
	degree by which the SFR in the Kepler field has decreased, by 7-14\%. }

\added{{\it Weakened magnetic braking:} 
If stellar rotation rates and spot-induced photometric amplitudes
remained {\it constant} after $\approx$3\,Gyr, then older stars could
contaminate the peaks in Figure~\ref{fig:hist_tgyro}.  This
possibility warrants consideration, given asteroseismic evidence
suggesting slowed spin-down rates after $\approx$4\,Gyr for Sun-like
stars \citep[][]{vanSaders_2016,Hall2021,2024ApJ...962..138S}, and the
``long-period edge'' in the $P_{\rm rot}$-$T_{\rm eff}$ distribution
which may pile up over similar timescales \citep{David_2022}.
However, this stalling is thought to happen at $>$5\,Gyr for late-G
dwarfs and early K dwarfs \citep[e.g.~Figure~4
by][]{2024ApJ...962..138S}, which means that it cannot account for the
2.3\,Gyr peak in the 4400-5400\,K age distribution shown in
Figure~\ref{fig:hist_tgyro}.}

\added{Additionally, \citet{2022ApJ...937...94M} found that the
\citet{McQuillan_2014} sample, discussed in Appendix~\ref{app:mcqonly},
includes few stars older than $\approx$4\,Gyr, based on
isochrone-fitting to a high-quality subset of Kepler planet hosts.
Some solar twins have also been shown to have rotation periods of
30-50 days
\citep{LorenzoOliveira2019,LorenzoOliveira2020,doNascimento2020},
indicating that a completely stalled spin-down may be unlikely. These
arguments suggest that while the peak in the age distribution is not
primarily caused by an accumulation of much older stars with slowed
spin-down rates, contamination by such stars beyond $\approx$3\,Gyr
remains possible, particularly for stars hotter than the Sun.}

\added{{\it Metallicity:}  
Stars with higher metallicity are predicted to have longer convective turnover
times, which may slow their rate of spin-down
	\citep{Amard2020}.
While recent observational work has empirically confirmed that such an effect is
present in the Kepler data
\citep[][]{See2024},
the anti-correlation between close,
unresolved binaries and stellar metallicity \citep{Moe2019} could be a complicating factor.
Nonetheless, based on these studies, we flagged stars with known
	spectroscopic $|[{\rm Fe/H}]|$$>$0.3, using either LAMOST or Keck/HIRES
	spectra \citep{Zong2018,Petigura_2022}.
The main limitation of this approach is that only 53.3\% of the stars in Table~\ref{tab:stars} with
$t_{\rm gyro}$$<$4\,Gyr had such spectra available.
For the remainder of our sample,
our quality flags were not sensitive to metallicity outliers.
The metallicity distribution of stars for which we do have spectra suggests that
at most $\approx$5\% of stars in our apparently clean sample might have $|[{\rm Fe/H}]|$$>$0.3 and not
be flagged,
with a slight bias toward lower metallicity stars over higher metallicity.
Based on work by \citet{Claytor2020},  this will be most important
for stars near the Kraft break, which could have ages over-estimated
by $\approx$50-100\% if they are in fact metal-poor, and under-estimated in the converse case.
}

\added{{\it Differential rotation:}
The Sun's surface rotates once per $\approx$25-30\,days at latitudes between the equator
and $\approx$$\pm$60$^\circ$ \citep[e.g.][]{Snodgrass1990}.
This latitudinal surface differential rotation yields two separate limitations on rotational ages.
The first is
on the accuracy with which any one star's rotation period can be measured over a few-year mission like Kepler;
the second is on the degree to which the intrinsic scatter in open cluster rotation sequences can be measured \citep[e.g.][]{Epstein_2014}.
In this study, we accounted for the latter intrinsic scatter based on available measurements
from Ruprecht-147 \citep{Curtis_2020} and M67 data
\citep{Barnes2016,Dungee_2022,Gruner_2023}.
This means that irrespective of the assumed $P_{\rm rot}$ measurement precision,
our model imposed a $\approx$5-10\% level of intrinsic $P_{\rm rot}$ scatter
at old ages.
While this accounts for the statistical impact of differential rotation on the inferred
age distribution,
the accuracy issue remains;
any one individual star's observed rotation period might be faster
or slower than average stars of the same mass and age.  
This effect can yield a $\approx$20\% age error for individual Sun-like stars with ages
of a few billion years; the statistical uncertainties in Table~\ref{tab:stars} should be
treated with appropriate caution.}

\deleted{Metallicity may also be relevant
\citep{2020MNRAS.499.3481A,See2024}.  An additional caveat
is that our rotation-based ages used photometric effective
temperatures (Section~\ref{sec:stellarprops}), even though
spectroscopic temperatures are available for the planet-hosting
subset.  This decision was driven by a desire for homogeneity 
irrespective of planet-hosting status.  However it implies that the
estimated planet ages might shift if one were to account for the
spectroscopic information.}

\deleted{Other age indicators may help in verifying the ages of the
$\lesssim$3\,Gyr stars that are the main focus of this work.  Specifically,
chromospheric emission in the X-ray, \ion{Ca}{2} HK line, and the UV
can serve as an age tracer
\citep{Mamajek_2008,2014MNRAS.441.2361V,2024ApJ...960...62E}.
However, these age indicators are all in a sense ``rotation-powered''.
The dynamo converts kinetic energy into magnetic energy, which is
emitted through these chromospheric pathways.  We did not attempt to
incorporate these indicators into this study due to concerns regarding
sensitivity, homogeneity, and the question of whether they in fact
provide age information that is truly independent from rotation.}

\subsection{Future Directions}

\added{{\it Ages of stars older than the Sun}: }\deleted{A broader question, beyond
the scope of this work, concerns the ages of stars $\gtrsim$4\,Gyr
old.}\added{This work was limited to stars younger than 4\,Gyr}. For
age-dating methods based only on rotation rates, an important challenge
is that the detectability of the rotation signals for 3-10\,Gyr stars,
especially Sun-like stars, is at the limits of the Kepler data
\citep{2022ApJ...937...94M}.  In the ``old star'' regime, methods that
leverage either asteroseismology
\citep{vanSaders_2016,2024ApJ...962..138S}, kinematics \citep{2021AJ....161..189L,See2024},
or else that combine both
the evolution of stellar luminosity and stellar rotation
\citep{Angus_2019,Claytor2020,2023ApJ...952..131M} seem the most
capable of providing useful age constraints for single stars over the
full span of their main-sequence lifetimes.

{\it \added{Planet} occurrence rates}:
The age-dependent trends predicted for exoplanet populations, such as
the Kelvin-Helmholtz cooling of mini-Neptunes \citep{Gupta_2019},
time-dependent carving of the photoevaporation desert
\citep{Owen2018}, and the time evolution of the radius valley
\citep{Rogers_2021} can be explored using our data.  We defer this
analysis to a separate publication.  Some care is required since
beyond age, exoplanet demographics also depend on stellar metallicity
and mass \citep[e.g.][]{Petigura_2018,Miyazaki2023}.  

{\it Field star ages from other surveys}:
While this study focused on Kepler, other photometric surveys
(e.g.~K2, TESS, HAT, WASP, NGTS, ZTF, ATLAS) open opportunities for
age-dating a far broader set of stars and planets.  Future prospects
also include PLATO \citep{Rauer14}, Earth 2.0
\citep{2022arXiv220606693G}, and the Roman Galactic Bulge Time Domain
Survey \citep{Wilson2023}.  Rotational ages from these surveys could
yield new insight
into \replaced{whether the star formation history that we see in the Kepler field is
universal across the thin disk.}{the relative importance of star formation and
kinematic heating across different components of the Galaxy.}

{\it Searches for new associations}:
Some of the youngest stars in Tables~\ref{tab:planets}
and~\ref{tab:stars} have been linked to their birth clusters.
Others have not. A systematic search for the birth clusters of the
youngest ``field stars'' could combine positions and velocities from
Gaia with rotation measurements from TESS. Given the $\sim$100\,Myr
decoherence time of young clusters, \replaced{the youngest ``field stars''}{these young stars} may
have easily identifiable young neighbors.

\section{Conclusions}
\label{sec:conclusions}

We began \deleted{this work }with two questions: how wrong is the assumption of
a uniform age distribution for stars in the galactic thin disk?  And
why are only $\approx$50 sub-gigayear transiting planets known, rather
than the $\approx$500 that would be expected under the assumption of a
uniform star formation \replaced{history}{rate} for the $\approx$5{,}000 known planets?

Our approach to answering these questions was to curate a sample of
stellar rotation periods, lithium equivalent widths, and temperatures
using archival and new data from the Kepler field.  We derived new
ages using empirical interpolation-based methods, and assessed the
reliability of these ages by comparing them against benchmark open
clusters (Figure~\ref{fig:agescalecompone}).

Tables~\ref{tab:planets} and~\ref{tab:stars} summarize the results for
planets and stars, respectively.  Our recovered ages are accurate for
all 14 known Kepler planets in clusters.  While lithium provided
minimal added information for most of the sample\added{ due to its astrophysical
depletion rate}, $t_{\rm Li}$ and
$t_{\rm gyro}$ agreed in over 90\% of cases for which comparison
was possible.  Our \replaced{resulting ages}{results} included two-sided $t_{\rm gyro}$
and $t_{\rm Li}$ for \ltonegyrhighqconfirmedtwosided\ sub-gigayear
confirmed planets, and two-sided $t_{\rm gyro}$ with one-sided $t_{\rm
Li}$ for \ltonegyrhighqconfirmedonesided\ \replaced{sub-gigayear confirmed}{such}
planets.  Allowing for ``candidate'' planets, grazing transit
geometries, and stars with RUWE far from unity expands the counts by a
factor of two.

The sizes and periods for the most secure set of planets are shown in
Figure~\ref{fig:rp_period_age_results}.   While the new young planets
are mostly mini-Neptunes, some are near the lower boundary of the
``sub-Jovian desert'' \citep{Owen2018}, which could have an
evolutionary connection to their youth.  Other discoveries include
Earth-sized planets \added{(e.g.~Kepler-1312\,c and Kepler-1561\,b) }with \deleted{new }ages of only a few hundred million years.  

\added{Significant care is 
needed not only when assessing the reliability of individual rotational ages, but also when interpreting distributions of rotational ages inferred from many stars.
Despite a slew of stellar quality flags, including on metallicity, evolutionary state,
and binarity, up to a third of our ``apparently clean'' 
Kepler stars with $P_{\rm rot}$$<$10\,days
could potentially be influenced by unresolved binaries (Figure~\ref{fig:hist_tgyro_corr}).
An additional nuance is that even after correcting for such effects,
the star formation history does not directly trace
the observed age distribution: 
particularly for pencil-beam surveys such as Kepler,
kinematic heating of stars in the
thin disk significantly sculpts the observed stellar (and therefore planetary) age distribution.}

Our \deleted{main }conclusions with respect to our original \added{two }questions\added{ -- what is the age distribution in the thin disk, and where are the missing young transiting exoplanets -- }are as
follows.

\begin{enumerate}[leftmargin=*,topsep=0pt,itemsep=-0.5ex,partopsep=1ex,parsep=1ex]
  \item Rather than being uniform, the age distributions of both the
    Kepler target stars and the known Kepler planet hosts show a
    demographic cliff.  There are \added{at least }twice as many ``old'' (2-3\,Gyr)
    stars in the Kepler field as ``young'' (0-1\,Gyr) stars.  \deleted{The star
    formation rate today is \ratiosfr$\pm$\uncratiosfr\ times lower
    than it was three billion years ago.}
\item \added{The youngest Kepler stars (0-0.3\,Gyr)
    are at least  \ratiosfr$\pm$\uncratiosfr\ times rarer than stars 2.7-3\,Gyr old. The statistical
    uncertainty on this value is dwarfed by the systematic
    uncertainty in the contamination rate of unresolved
    photometric binaries, which could push it as high as 4.81$\pm$0.26 (Section~\ref{subsec:caveats})}.
     \replaced{This result}{This range of values} \replaced{from
    rotation-based ages broadly agrees}{is broadly consistent} with \replaced{recent reports of a
    declining star-formation rate}{modeling expectations (Section~\ref{subsec:trilegal}), and with age distributions derived from} CMD fitting and white-dwarf
    chronology\replaced{, though with the}{.  The }advantage of \added{rotation-based ages however is their }\replaced{sensitivity}{accuracy} for
    individual FGK stars at $t$$<$1\,Gyr (e.g.~Figure~\ref{fig:agescalecompone}).
  \item Rather than expecting $\approx$500 cumulative exoplanets younger than one
    billion years, the age distribution of the Kepler field
    \replaced{implies}{suggests} that we \replaced{should}{might} instead expect $\approx$250.
  \item We have \replaced{derived}{reported} rotational ages for \nplyounggyrotwosigma\
    Kepler planets younger than 1\,Gyr at 2$\sigma$, and
    \nplyounggyro\ planets with median ages below 1\,Gyr.
    Concatenating these planets against the existing literature yields
    $\approx$170 known sub-gigayear planets.  This lessens the
    original factor of ten discrepancy to a factor of at most two.
\end{enumerate}

\acknowledgements
We thank Kevin Schlaufman and the reviewer for suggestions which significantly improved the 
manuscript.
This work was supported by the Heising-Simons 51~Pegasi~b Fellowship
(LGB, EKP), the Carnegie Institution for Science's Carnegie Fellowship (LGB),
and the Arthur R.~Adams SURF Fellowship (EKP).
The HIRES data were obtained at the Keck Observatory.
We recognize the importance that the summit of Maunakea has always had
within the indigenous Hawaiian community, and we are deeply grateful 
for the opportunity to conduct observations from this mountain.

{\it \large Contributions}: Per \url{https://credit.niso.org/}:
Conceptualization: LGB.
Data curation: LGB, AWH, HI.
Formal analysis: LGB, KM.
Funding acquisition: LGB.
Investigation: LGB, EKP.
Methodology: LGB, EKP, KM.
Project administration: LGB, LAH, AWH.
Resources: LGB, AWH.
Software: LGB.
Supervision: LGB, LAH.
Validation: LGB, KM.
Visualization: LGB.
Writing – original draft: LGB.
Writing – review \& editing: all authors.

\facilities{
  Gaia \citep{GaiaCollaboration2023},
  Kepler \citep{Borucki10},
  Keck:I (HIRES) \citep{vogt_hires_1994},
	2MASS \citep{Skrutskie06},
	SDSS \citep{2000AJ....120.1579Y}.
}

\software{
  astropy \citep{astropy:2013,astropy:2018,astropy:2022},
  claude \citep{claude2024anthropic},
  eagles \citep{Jeffries_2023},
  gyro-interp (\citealt{Bouma_2023}; \texttt{v0.6} available at doi:~\href{https://doi.org/10.5281/zenodo.13733242}{10.5281/zenodo.13733242}),
  matplotlib \citep{matplotlib},
  numpy \citep{numpy},
  scipy \citep{scipy}.
}

\clearpage 

\startlongtable
\begin{deluxetable*}{llllllllrrrrrrl}
  \tabletypesize{\scriptsize}
  \tablecaption{Ages of Kepler planets and planet candidates.  This
  version of the table is truncated to include the youngest
  systems, sorted by the minimum of either the
  rotation or lithium-based age.  The full machine-readable table
  contains ages and age limits for \nnonfopkoissomeageinfo\ non-false
  positive KOIs with MES$>$10.  
  A bash script to decode the $Q_{\rm star}$ quality flag is
  {\bf
  \href{https://gist.github.com/lgbouma/20368253f1a98da1b39cf32fdda0be13}{available
  online}}.  A python script to select stars with specific bit flags
  is {\bf \href{https://gist.github.com/lgbouma/87ad8bde42625e766a5a8857cc5a183a}{also available}}.
  All quoted age uncertainties are statistical.
  \label{tab:planets}}
  \startdata
  KOI & Kepler &  $T_{\rm eff}$ & $P_{\rm rot}$ & EW$_{\rm Li}^{\rm \ast}$ &
  $t_{\rm gyro}$ & $t_{\rm Li} $ & Consistent? & $R_{\rm p}$ & $P_{\rm orb}$ &
  $Q_{\rm planet}$ & $Q_{\rm star}$ &  Spec? & Comment \\
  -- &   -- & K & days & m\AA & Myr & Myr &  str &   Earths &    days &       int  & int & bool & -- \\
  \hline
  K05245.01 & Kepler-1627 b & 5357 & 2.62 & $225\pm7$ & $81^{+158}_{-55}$ & $51^{+38}_{-27}$ & Yes & 3.79 & 7.2 & 0 & 2176 & 1 & Cep-Her \\
K07368.01 & Kepler-1974 b & 5068 & 2.56 & $248\pm4$ & $88^{+183}_{-60}$ & $54^{+47}_{-25}$ & Yes & 2.22 & 6.84 & 0 & 2560 & 1 & Cep-Her \\
K06228.01 & Kepler-1644 b & 5521 & 1.43 & $-2\pm13$ & $77^{+144}_{-53}$ & $> 767$ & No & 1.88 & 21.09 & 4 & 2690 & 1 & Unres. Binary \\
K06186.01 & Kepler-1643 b & 4918 & 5.05 & $120\pm6$ & $79^{+182}_{-54}$ & $191^{+92}_{-76}$ & Yes & 2.11 & 5.34 & 0 & 0 & 1 & Cep-Her \\
K03933.01 & Kepler-1699 b & 5496 & 4.16 & $-11\pm7$ & $85^{+106}_{-58}$ & $> 889$ & No & 1.32 & 3.49 & 0 & 2688 & 1 & Unres. Binary \\
K03916.01 & Kepler-1529 b & 4974 & 6.43 & $200\pm6$ & $109^{+117}_{-71}$ & $90^{+53}_{-39}$ & Yes & 2.01 & 5.34 & 0 & 0 & 1 & \checkmark \checkmark \\
K01804.01 & Kepler-957 b & 4947 & 4.52 & $24\pm9$ & $96^{+196}_{-66}$ & $> 241$ & Yes & 6.9 & 5.91 & 0 & 0 & 1 & \checkmark \\
K07913.01 & Kepler-1975 b & 4450 & 3.36 & $56\pm9$ & $96^{+223}_{-66}$ & $> 57$ & Yes & 2.03 & 24.28 & 0 & 2816 & 1 & Cep-Her \\
K03936.02 & Kepler-1930 b & 4906 & 7.1 & $170\pm4$ & $174^{+106}_{-65}$ & $115^{+55}_{-49}$ & Yes & 1.52 & 13.03 & 4 & 0 & 1 &  \\
K03876.01 & Kepler-1928 b & 5577 & 4.64 & $137\pm4$ & $148^{+102}_{-87}$ & $189^{+150}_{-94}$ & Yes & 1.86 & 19.58 & 0 & 2048 & 1 & MELANGE-3 \\
K04069.01 & Kepler-1938 b & 4617 & 7.82 & $6\pm16$ & $152^{+112}_{-42}$ & $> 208$ & Yes & 1.47 & 13.06 & 0 & 2048 & 1 & \checkmark \\
K02678.01 & Kepler-1313 b & 5236 & 6.13 & $142\pm3$ & $197^{+112}_{-91}$ & $174^{+96}_{-72}$ & Yes & 1.71 & 3.83 & 4 & 2048 & 1 &  \\
K04194.01 & Kepler-1565 b & 4958 & 7.4 & $133\pm7$ & $230^{+111}_{-85}$ & $174^{+81}_{-72}$ & Yes & 1.27 & 1.54 & 0 & 0 & 1 & \checkmark \checkmark \\
K03835.01 & Kepler-1521 b & 4806 & 7.82 & $117\pm5$ & $207^{+98}_{-68}$ & $176^{+79}_{-69}$ & Yes & 2.3 & 47.15 & 0 & 2048 & 1 & \checkmark \checkmark \\
K01838.01 & Kepler-970 b & 4314 & 9.23 & $36\pm14$ & $177^{+124}_{-39}$ & $> 92$ & Yes & 2.15 & 16.74 & 4 & 0 & 1 & MELANGE-3 \\
K00063.01 & Kepler-63 b & 5486 & 5.49 & $89\pm4$ & $223^{+98}_{-92}$ & $542^{+475}_{-256}$ & Yes & 5.64 & 9.43 & 0 & 4096 & 1 & \checkmark \checkmark \\
K01199.01 & Kepler-786 b & 4680 & 33.06 & $83\pm6$ & $4364^{+359}_{-374}$ & $228^{+168}_{-87}$ & No & 2.31 & 53.53 & 0 & 0 & 1 & Mystery \\
K03316.01 & Kepler-1467 b & 5252 & 6.31 & $122\pm6$ & $230^{+112}_{-98}$ & $236^{+151}_{-95}$ & Yes & 3.11 & 47.06 & 0 & 0 & 1 & \checkmark \checkmark \\
K01074.01 & Kepler-762 b & 5921 & 4.01 & $-27\pm25$ & $245^{+114}_{-100}$ & $> 548$ & Maybe & 15.19 & 3.77 & 0 & 512 & 1 &  \\
K01839.01 & Kepler-971 b & 5447 & 6.22 & $105\pm6$ & $306^{+95}_{-115}$ & $366^{+290}_{-164}$ & Yes & 3.93 & 9.59 & 0 & 128 & 1 &  \\
K01833.01 & Kepler-968 b & 4413 & 10.46 & $10\pm18$ & $328^{+108}_{-87}$ & $> 159$ & Yes & 1.85 & 3.69 & 0 & 0 & 1 & Theia-520 \\
K01833.03 & Kepler-968 c & 4413 & 10.46 & $10\pm18$ & $328^{+108}_{-87}$ & $> 159$ & Yes & 1.63 & 5.71 & 0 & 0 & 1 & Theia-520 \\
K01833.02 & Kepler-968 d & 4413 & 10.46 & $10\pm18$ & $328^{+108}_{-87}$ & $> 159$ & Yes & 2.28 & 7.68 & 4 & 0 & 1 & Theia-520 \\
K02675.01 & Kepler-1312 b & 5584 & 6.13 & $86\pm4$ & $357^{+75}_{-109}$ & $642^{+617}_{-318}$ & Yes & 2.07 & 5.45 & 0 & 4096 & 1 & \checkmark \checkmark \\
K02675.02 & Kepler-1312 c & 5584 & 6.13 & $86\pm4$ & $357^{+75}_{-109}$ & $642^{+617}_{-318}$ & Yes & 0.94 & 1.12 & 0 & 4096 & 1 & \checkmark \checkmark \\
K00775.02 & Kepler-52 b & 4164 & 11.85 & $22\pm18$ & $360^{+208}_{-97}$ & $> 100$ & Yes & 2.19 & 7.88 & 0 & 0 & 1 & Theia-520 \\
K00775.01 & Kepler-52 c & 4164 & 11.85 & $22\pm18$ & $360^{+208}_{-97}$ & $> 100$ & Yes & 2.04 & 16.38 & 0 & 0 & 1 & Theia-520 \\
K00775.03 & Kepler-52 d & 4164 & 11.85 & $22\pm18$ & $360^{+208}_{-97}$ & $> 100$ & Yes & 2.03 & 36.45 & 0 & 0 & 1 & Theia-520 \\
K04004.01 & Kepler-1933 b & 5576 & 6.21 & $85\pm3$ & $366^{+74}_{-109}$ & $642^{+603}_{-318}$ & Yes & 1.01 & 4.94 & 4 & 0 & 1 &  \\
K02174.03 & Kepler-1802 b & 4245 & 11.45 & -- & $379^{+201}_{-109}$ & -- & -- & 1.71 & 7.73 & 4 & 820 & 0 &  \\
K02174.02 & Kepler-1802 c & 4245 & 11.45 & -- & $379^{+201}_{-109}$ & -- & -- & 2.05 & 33.14 & 0 & 820 & 0 &  \\
K03935.01 & Kepler-1532 b & 5554 & 6.48 & $90\pm6$ & $397^{+71}_{-102}$ & $567^{+545}_{-278}$ & Yes & 1.26 & 1.09 & 0 & 0 & 1 & \checkmark \checkmark \\
K01801.01 & Kepler-955 b & 5221 & 7.5 & $79\pm4$ & $397^{+99}_{-131}$ & $536^{+481}_{-237}$ & Yes & 2.69 & 14.53 & 0 & 0 & 1 & \checkmark \checkmark \\
K01800.01 & Kepler-447 b & 5648 & 6.4 & $103\pm3$ & $420^{+64}_{-78}$ & $405^{+355}_{-203}$ & Yes & 18.49 & 7.79 & 4 & 2048 & 1 &  \\
K03370.02 & Kepler-1481 b & 4832 & 9.11 & $22\pm7$ & $407^{+126}_{-118}$ & $> 210$ & Yes & 1.09 & 5.94 & 0 & 0 & 1 & \checkmark \\
K04156.01 & Kepler-1943 b & 6002 & -- & $99\pm5$ & -- & $409^{+520}_{-254}$ & No & 1.29 & 4.85 & 4 & 518 & 1 &  \\
K00448.01 & Kepler-159 b & 4511 & 10.5 & $19\pm10$ & $415^{+160}_{-110}$ & $> 161$ & Yes & 2.3 & 10.14 & 0 & 4096 & 1 & \checkmark \\
K00448.02 & Kepler-159 c & 4511 & 10.5 & $19\pm10$ & $415^{+160}_{-110}$ & $> 161$ & Yes & 2.75 & 43.59 & 0 & 4096 & 1 & \checkmark \\
K00046.01 & Kepler-101 b & 5498 & -- & $100\pm5$ & -- & $419^{+349}_{-196}$ & No & 5.9 & 3.49 & 0 & 1542 & 1 &  \\
K04169.01 & Kepler-1561 b & 5742 & 6.18 & $66\pm3$ & $426^{+74}_{-78}$ & $1409^{+1718}_{-788}$ & Maybe & 0.94 & 1.01 & 0 & 2048 & 1 & \checkmark \checkmark \\
K02708.01 & Kepler-1320 b & 4536 & 10.46 & $25\pm32$ & $430^{+173}_{-113}$ & $> 140$ & Yes & 1.39 & 0.87 & 0 & 0 & 1 & \checkmark \\
K00119.01 & Kepler-108 b & 5626 & -- & $100\pm5$ & -- & $438^{+402}_{-220}$ & No & 8.2 & 49.18 & 0 & 418 & 1 &  \\
K00119.02 & Kepler-108 c & 5626 & -- & $100\pm5$ & -- & $438^{+402}_{-220}$ & No & 7.78 & 190.32 & 4 & 418 & 1 &  \\
K00323.01 & Kepler-523 b & 5267 & 7.6 & $49\pm4$ & $444^{+87}_{-114}$ & $1824^{+2706}_{-1047}$ & Maybe & 1.9 & 5.84 & 0 & 0 & 1 & \checkmark \checkmark \\
K02115.01 & Kepler-67 b & 5126 & 10.39 & $83\pm10$ & $878^{+108}_{-121}$ & $458^{+451}_{-206}$ & Yes & 2.96 & 15.73 & 0 & 0 & 1 & \checkmark \checkmark \\
K00002.01 & Kepler-2 b & 6436 & -- & $83\pm4$ & -- & $485^{+924}_{-353}$ & -- & 16.42 & 2.2 & 0 & 7 & 1 &  \\
K03371.02 & Kepler-1482 b & 5330 & 7.7 & $52\pm3$ & $491^{+76}_{-91}$ & $1630^{+2285}_{-904}$ & Maybe & 1.0 & 12.25 & 0 & 640 & 1 &  \\
K03497.01 & Kepler-1512 b & 4894 & 9.33 & $14\pm8$ & $505^{+135}_{-115}$ & $> 295$ & Yes & 0.8 & 20.36 & 4 & 4738 & 1 &  \\
K03864.01 & Kepler-1698 b & 4866 & 9.49 & $2\pm8$ & $518^{+151}_{-123}$ & $> 358$ & Yes & 0.9 & 1.21 & 0 & 1024 & 1 &  \\
K03010.01 & Kepler-1410 b & 3808 & 14.19 & $-21\pm24$ & $523^{+612}_{-196}$ & $> 80$ & Yes & 1.39 & 60.87 & 0 & 512 & 1 &  \\
K05447.02 & Kepler-1629 b & 5585 & 7.45 & -- & $529^{+62}_{-62}$ & -- & -- & 0.67 & 3.88 & 0 & 0 & 0 &  \\
K04246.01 & Kepler-1576 b & 5794 & 7.09 & $17\pm8$ & $559^{+103}_{-71}$ & $> 613$ & Yes & 0.9 & 6.98 & 0 & 0 & 1 & \checkmark \\
K03324.01 & Kepler-1469 b & 5356 & 8.15 & $-5\pm25$ & $562^{+75}_{-82}$ & $> 535$ & Yes & 2.53 & 21.86 & 0 & 0 & 1 & \checkmark \\
K01779.01 & Kepler-318 b & 5799 & 7.09 & $65\pm3$ & $562^{+106}_{-72}$ & $1507^{+1876}_{-858}$ & Yes & 3.97 & 4.66 & 0 & 0 & 1 & \checkmark \checkmark \\
K01779.02 & Kepler-318 c & 5799 & 7.09 & $65\pm3$ & $562^{+106}_{-72}$ & $1507^{+1876}_{-858}$ & Yes & 3.1 & 11.82 & 4 & 0 & 1 &  \\
K02084.01 & Kepler-1792 b & 4942 & 9.49 & $13\pm11$ & $587^{+155}_{-144}$ & $> 312$ & Yes & 2.15 & 4.2 & 0 & 0 & 1 & \checkmark \\
K02035.01 & Kepler-1066 b & 5847 & 7.0 & $60\pm4$ & $588^{+158}_{-85}$ & $2018^{+2616}_{-1187}$ & Maybe & 1.96 & 1.93 & 4 & 1024 & 1 &  \\
K03274.01 & Kepler-1451 b & 5675 & 7.82 & $42\pm4$ & $597^{+79}_{-65}$ & $> 316$ & Yes & 2.33 & 35.62 & 0 & 0 & 1 & \checkmark \\
K01615.01 & Kepler-908 b & 5670 & 7.88 & $67\pm4$ & $602^{+77}_{-65}$ & $1317^{+1607}_{-724}$ & Yes & 1.36 & 1.34 & 0 & 4608 & 1 &  \\
K02022.01 & Kepler-349 b & 5756 & 7.71 & $64\pm6$ & $617^{+103}_{-72}$ & $1686^{+2185}_{-976}$ & Yes & 1.99 & 5.93 & 0 & 0 & 1 & \checkmark \checkmark \\
K02022.02 & Kepler-349 c & 5756 & 7.71 & $64\pm6$ & $617^{+103}_{-72}$ & $1686^{+2185}_{-976}$ & Yes & 1.97 & 12.25 & 0 & 0 & 1 & \checkmark \checkmark \\
K00620.01 & Kepler-51 b & 5635 & 8.14 & $48\pm8$ & $623^{+75}_{-65}$ & $> 258$ & Yes & 6.62 & 45.16 & 0 & 0 & 1 & \checkmark \\
K00620.03 & Kepler-51 c & 5635 & 8.14 & $48\pm8$ & $623^{+75}_{-65}$ & $> 258$ & Yes & 5.49 & 85.32 & 4 & 0 & 1 &  \\
K00620.02 & Kepler-51 d & 5635 & 8.14 & $48\pm8$ & $623^{+75}_{-65}$ & $> 258$ & Yes & 9.04 & 130.18 & 0 & 0 & 1 & \checkmark \\
K02803.01 & Kepler-1877 b & 5506 & 8.37 & $3\pm12$ & $624^{+69}_{-65}$ & $> 726$ & Maybe & 0.55 & 2.38 & 0 & 1024 & 1 &  \\
K00720.04 & Kepler-221 b & 5070 & 9.3 & $21\pm5$ & $636^{+120}_{-115}$ & $> 341$ & Yes & 1.51 & 2.8 & 0 & 0 & 1 & \checkmark \\
K00720.01 & Kepler-221 c & 5070 & 9.3 & $21\pm5$ & $636^{+120}_{-115}$ & $> 341$ & Yes & 2.86 & 5.69 & 0 & 0 & 1 & \checkmark \\
K00720.02 & Kepler-221 d & 5070 & 9.3 & $21\pm5$ & $636^{+120}_{-115}$ & $> 341$ & Yes & 2.57 & 10.04 & 0 & 0 & 1 & \checkmark \\
K00720.03 & Kepler-221 e & 5070 & 9.3 & $21\pm5$ & $636^{+120}_{-115}$ & $> 341$ & Yes & 2.58 & 18.37 & 4 & 0 & 1 &  \\
K03097.02 & Kepler-431 b & 6259 & 16.16 & $80\pm3$ & -- & $671^{+1092}_{-458}$ & -- & 0.93 & 6.8 & 4 & 4103 & 1 &  \\
K03097.03 & Kepler-431 c & 6259 & 16.16 & $80\pm3$ & -- & $671^{+1092}_{-458}$ & -- & 0.93 & 8.7 & 4 & 4103 & 1 &  \\
K03097.01 & Kepler-431 d & 6259 & 16.16 & $80\pm3$ & -- & $671^{+1092}_{-458}$ & -- & 1.08 & 11.92 & 4 & 4103 & 1 &  \\
K01982.01 & Kepler-1781 b & 5363 & 9.17 & -- & $708^{+80}_{-78}$ & -- & -- & 1.95 & 4.89 & 0 & 0 & 0 &  \\
K03375.01 & Kepler-1918 b & 5522 & 9.11 & -- & $722^{+75}_{-71}$ & -- & -- & 2.19 & 47.06 & 0 & 0 & 0 &  \\
K01835.02 & Kepler-326 b & 5142 & 9.56 & $8\pm8$ & $724^{+108}_{-103}$ & $> 506$ & Yes & 1.25 & 2.25 & 0 & 1664 & 1 &  \\
K01835.01 & Kepler-326 c & 5142 & 9.56 & $8\pm8$ & $724^{+108}_{-103}$ & $> 506$ & Yes & 1.38 & 4.58 & 0 & 1664 & 1 &  \\
K01835.03 & Kepler-326 d & 5142 & 9.56 & $8\pm8$ & $724^{+108}_{-103}$ & $> 506$ & Yes & 1.31 & 6.77 & 0 & 1664 & 1 &  \\
K01797.01 & Kepler-954 b & 4736 & 10.68 & $4\pm11$ & $727^{+222}_{-184}$ & $> 270$ & Yes & 2.16 & 16.78 & 0 & 1024 & 1 &  \\
K01821.01 & Kepler-963 b & 5383 & 9.42 & -- & $748^{+80}_{-78}$ & -- & -- & 2.64 & 9.98 & 0 & 0 & 0 &  \\
K03681.01 & Kepler-1514 b & 5852 & 7.87 & $69\pm2$ & $758^{+416}_{-142}$ & $1302^{+1622}_{-754}$ & Yes & 11.94 & 217.83 & 0 & 1540 & 1 &  \\
K03681.02 & Kepler-1514 c & 5852 & 7.87 & $69\pm2$ & $758^{+416}_{-142}$ & $1302^{+1622}_{-754}$ & Yes & 1.17 & 10.51 & 4 & 1540 & 1 &  \\
K00647.01 & Kepler-634 b & 6272 & -- & $77\pm3$ & -- & $768^{+1250}_{-527}$ & -- & 2.13 & 5.17 & 0 & 7 & 1 &  \\
K02037.01 & Kepler-1995 b & 4746 & 10.8 & $9\pm21$ & $772^{+213}_{-194}$ & $> 223$ & Yes & 3.46 & 73.76 & 0 & 4096 & 1 & \checkmark \\
K01781.02 & Kepler-411 b & 4920 & 10.32 & $4\pm3$ & $775^{+164}_{-166}$ & $> 396$ & Yes & 2.2 & 3.01 & 4 & 0 & 1 &  \\
K01781.01 & Kepler-411 c & 4920 & 10.32 & $4\pm3$ & $775^{+164}_{-166}$ & $> 396$ & Yes & 3.47 & 7.83 & 0 & 0 & 1 & \checkmark \\
  K07375.01 & -- & 4212 & 3.88 & $117\pm9$ & $104^{+235}_{-72}$ & $70^{+37}_{-21}$ & Yes & 1.73 & 4.85 & 0 & 2560 & 1 &  \\
K03991.01 & -- & 5226 & 5.22 & $98\pm3$ & $71^{+91}_{-48}$ & $354^{+253}_{-145}$ & Maybe & 1.37 & 1.57 & 0 & 2176 & 1 &  \\
K01546.01 & -- & 5639 & 0.9 & $33\pm24$ & $75^{+134}_{-51}$ & $> 316$ & Maybe & 11.92 & 0.92 & 0 & 2176 & 1 & $|P_\mathrm{rot}$-$P_\mathrm{orb}|$$<$0.1$\,\mathrm{days}$ \\
K05482.01 & -- & 5519 & 0.81 & -- & $77^{+144}_{-53}$ & -- & -- & 2.96 & 31.71 & 4 & 2690 & 0 &  \\
K00064.01 & -- & 5306 & 2.23 & $3\pm4$ & $81^{+129}_{-55}$ & $> 567$ & No & 10.49 & 1.95 & 4 & 4614 & 1 &  \\
K06188.01 & -- & 5209 & 1.62 & $6\pm11$ & $84^{+171}_{-58}$ & $> 554$ & Maybe & 2.75 & 1.65 & 0 & 2048 & 1 & $|P_\mathrm{rot}$-$P_\mathrm{orb}|$$<$0.1$\,\mathrm{days}$ \\
K02695.01 & -- & 5174 & 2.88 & -- & $86^{+176}_{-59}$ & -- & -- & 20.23 & 2.5 & 4 & 640 & 0 &  \\
K07449.01 & -- & 4928 & 1.31 & -- & $92^{+194}_{-63}$ & -- & -- & 20.43 & 1.32 & 4 & 2048 & 0 & $|P_\mathrm{rot}$-$P_\mathrm{orb}|$$<$0.1$\,\mathrm{days}$ \\
K06130.01 & -- & 4560 & 3.02 & $-1\pm25$ & $94^{+217}_{-65}$ & $> 186$ & Yes & 1.45 & 1.54 & 0 & 2976 & 1 &  \\
K06195.01 & -- & 4677 & 1.42 & $-8\pm30$ & $94^{+210}_{-65}$ & $> 208$ & Yes & 2.28 & 1.44 & 0 & 2048 & 1 & $|P_\mathrm{rot}$-$P_\mathrm{orb}|$$<$0.1$\,\mathrm{days}$ \\

  \enddata
  \tablecomments{
  	EW$_{\rm Li}^{\rm \ast}$ is the lithium equivalent width {\it after} subtracting a constant 7.5\,m\AA\ to account for the \ion{Fe}{1} 6707.44\,\AA\ blend (see Section~\ref{subsec:lithiumsel}).
  	Two checkmarks ($\checkmark\checkmark$) denote confirmed planets with two-sided $t_{\rm
  gyro}$ and $t_{\rm Li}$ for which both the age and the planet are
  expected to be reliable.  One checkmark ($\checkmark$) denotes
  confirmed planets with two-sided $t_{\rm gyro}$ only. 
  ``Confirmed'' planets appear in the machine-readable
  version before ``candidate'' planets.  
  Planetary sizes are mostly drawn from (in order of precedence)
  \citet{Petigura_2022}, \citet{Berger_2020b_rpage}, and \citet{Thompson_2018}, and might be unphysical
  for grazing planets.
  The bit quality flags for the
  rotation-based ages, $Q_{\rm star}$, are described in
  Section~\ref{subsec:flags}.  Concisely summarized, they are:
  Bit 0: $T_{\rm eff}/{\rm K} \in [ 3800-6200]$ ?
  Bit 1: $\log g$$<$4.2?
  Bit 2: $M_{\rm G}$<3.9 or $M_{\rm G}>8.5$?
  Bit 3: In KEBC?
  Bit 4: Large $d_{\rm xm,Kep-Gaia}$?
  Bit 5: Confused Kep-Gaia crossmatch?
  Bit 6: Gaia DR3 non-single star?
  Bit 7: RUWE$>$1.4?
  Bit 8: Crowded?
  Bit 9: Far from main sequence?
  Bit 10: Spectroscopic $|[{\rm Fe/H}]|$$>$0.3?
  Bit 11: \citetalias{Santos_2021} CP/CB?
  Bit 12: $P_{\rm rot}$ not in the homogeneous
  \citetalias{Santos_2019} \citetalias{Santos_2021} sample?
  As an example, Kepler-1627 has $Q_{\rm star}$ flagged with bit 11 and bit 7.
  The analogous planet quality bitmask, $Q_{\rm planet}$, has the
  following meaning.
  Bit 0: Candidate disposition not reliable?
  Bit 1: MES$<$10?
  Bit 2: Grazing?
  For a star to have a high likelihood of being ``reliable for
  gyrochronology'' we suggest a $Q_{\rm star}$ bitmask with bits 0--10
  not raised, and for a planet to be ``reliable'', we suggest
  $Q_{\rm planet}$ to be zero.
  }
\end{deluxetable*}

\startlongtable
\begin{deluxetable*}{lllllllrrrrrr}
  \tabletypesize{\scriptsize}
  \tablecaption{
  Ages of Kepler target stars derived from rotation periods.  The
  full machine-readable table includes
  \nuniqstarfinitegyroage\ stars with
  finite reported gyrochrone ages.  The quality
  bitmask is as in Table~\ref{tab:planets}; requiring bits zero to
	ten to be null (e.g.~using the suggested {\bf
\href{https://gist.github.com/lgbouma/87ad8bde42625e766a5a8857cc5a183a}{python script}}) yields \nuniqstarsantosrotgyroappl\ stars for
  which gyrochronology is likely to be valid.  \label{tab:stars}}
  \startdata
  KIC & Gaia DR3 &  $T_{\rm eff}$ & $P_{\rm rot}$ & $t_{\rm gyro}$ & $Q_{\rm star}$ \\
  -- &   -- & K & days &  Myr &    int  \\
  \hline
  8367679 & 2126373303927600000 & 5980 & 7.44 & $1062^{+815}_{-369}$ & 0 \\
11954016 & 2133663925009149952 & 5423 & 24.88 & $3230^{+330}_{-301}$ & 0 \\
1865152 & 2051034045638265984 & 5878 & 7.67 & $879^{+891}_{-274}$ & 0 \\
4447913 & 2100652256617096832 & 4654 & 35.67 & $> 4000$ & 0 \\
11085139 & 2129657064120722688 & 5507 & 20.49 & $2596^{+322}_{-267}$ & 0 \\
11072100 & 2131603612018238976 & 4814 & 23.87 & $2726^{+253}_{-216}$ & 0 \\
2860187 & 2052239591421076352 & 5546 & 26.29 & $3639^{+354}_{-345}$ & 518 \\
8737920 & 2106733414911973504 & 4844 & 25.94 & $3035^{+286}_{-255}$ & 640 \\
11962034 & 2130137688140649984 & 4925 & 27.61 & $3327^{+326}_{-287}$ & 0 \\
7958926 & 2126140894655659264 & 4323 & 17.0 & $1990^{+214}_{-161}$ & 0 \\
  \enddata
\end{deluxetable*}



\bibliographystyle{aasjournal}
\bibliography{bibliography}

\begin{thebibliography}{}
\expandafter\ifx\csname natexlab\endcsname\relax\def\natexlab#1{#1}\fi
\providecommand{\url}[1]{\href{#1}{#1}}

\bibitem[{{Aigrain} {et~al.}(2015){Aigrain}, {Llama}, {Ceillier}, {Chagas},
  {Davenport}, {Garc{\'\i}a}, {Hay}, {Lanza}, {McQuillan}, {Mazeh}, {de
  Medeiros}, {Nielsen}, \& {Reinhold}}]{Aigrain2015}
{Aigrain}, S., {Llama}, J., {Ceillier}, T., {et~al.} 2015, \mnras, 450, 3211

\bibitem[{{Akeson} {et~al.}(2013){Akeson}, {Chen}, {Ciardi}, {Crane}, {Good},
  {Harbut}, {Jackson}, {Kane}, {Laity}, {Leifer}, {Lynn}, {McElroy}, {Papin},
  {Plavchan}, {Ram{\'\i}rez}, {Rey}, {von Braun}, {Wittman}, {Abajian}, {Ali},
  {Beichman}, {Beekley}, {Berriman}, {Berukoff}, {Bryden}, {Chan}, {Groom},
  {Lau}, {Payne}, {Regelson}, {Saucedo}, {Schmitz}, {Stauffer}, {Wyatt}, \&
  {Zhang}}]{2013PASP..125..989A}
{Akeson}, R.~L., {Chen}, X., {Ciardi}, D., {et~al.} 2013, \pasp, 125, 989

\bibitem[{{Alzate} {et~al.}(2021){Alzate}, {Bruzual}, \&
  {D{\'\i}az-Gonz{\'a}lez}}]{2021MNRAS.501..302A}
{Alzate}, J.~A., {Bruzual}, G., \& {D{\'\i}az-Gonz{\'a}lez}, D.~J. 2021,
  \mnras, 501, 302

\bibitem[{{Amard} \& {Matt}(2020)}]{Amard2020}
{Amard}, L., \& {Matt}, S.~P. 2020, \apj, 889, 108

\bibitem[{{Amard} {et~al.}(2019){Amard}, {Palacios}, {Charbonnel}, {Gallet},
  {Georgy}, {Lagarde}, \& {Siess}}]{Amard2019}
{Amard}, L., {Palacios}, A., {Charbonnel}, C., {et~al.} 2019, \aap, 631, A77

\bibitem[{{Angus} {et~al.}(2015){Angus}, {Aigrain}, {Foreman-Mackey}, \&
  {McQuillan}}]{Angus_2015}
{Angus}, R., {Aigrain}, S., {Foreman-Mackey}, D., \& {McQuillan}, A. 2015,
  \mnras, 450, 1787

\bibitem[{{Angus} {et~al.}(2018){Angus}, {Morton}, {Aigrain}, {Foreman-Mackey},
  \& {Rajpaul}}]{Angus_2018}
{Angus}, R., {Morton}, T., {Aigrain}, S., {Foreman-Mackey}, D., \& {Rajpaul},
  V. 2018, \mnras, 474, 2094

\bibitem[{{Angus} {et~al.}(2019){Angus}, {Morton}, {Foreman-Mackey}, {van
  Saders}, {Curtis}, {Kane}, {Bedell}, {Kiman}, {Hogg}, \&
  {Brewer}}]{Angus_2019}
{Angus}, R., {Morton}, T.~D., {Foreman-Mackey}, D., {et~al.} 2019, \aj, 158,
  173

\bibitem[{{Anthropic}({2024})}]{claude2024anthropic}
{Anthropic}. {2024}, {Claude}, v{3},  {Anthropic}, {Conversational AI model
  used for editing manuscript and generating testable code; the authors wrote
  all of the original manuscript text.}
\newblock \url{{https://claude.ai/}}

\bibitem[{{Astropy Collaboration} {et~al.}(2013){Astropy Collaboration},
  {Robitaille}, {Tollerud}, {Greenfield}, {Droettboom}, {Bray}, {Aldcroft},
  {Davis}, {Ginsburg}, {Price-Whelan}, {Kerzendorf}, {Conley}, {Crighton},
  {Barbary}, {Muna}, {Ferguson}, {Grollier}, {Parikh}, {Nair}, {Unther},
  {Deil}, {Woillez}, {Conseil}, {Kramer}, {Turner}, {Singer}, {Fox}, {Weaver},
  {Zabalza}, {Edwards}, {Azalee Bostroem}, {Burke}, {Casey}, {Crawford},
  {Dencheva}, {Ely}, {Jenness}, {Labrie}, {Lim}, {Pierfederici}, {Pontzen},
  {Ptak}, {Refsdal}, {Servillat}, \& {Streicher}}]{astropy:2013}
{Astropy Collaboration}, {Robitaille}, T.~P., {Tollerud}, E.~J., {et~al.} 2013,
  \aap, 558, A33

\bibitem[{{Astropy Collaboration} {et~al.}(2018){Astropy Collaboration},
  {Price-Whelan}, {Sip{\H{o}}cz}, {G{\"u}nther}, {Lim}, {Crawford}, {Conseil},
  {Shupe}, {Craig}, {Dencheva}, {Ginsburg}, {Vand erPlas}, {Bradley},
  {P{\'e}rez-Su{\'a}rez}, {de Val-Borro}, {Aldcroft}, {Cruz}, {Robitaille},
  {Tollerud}, {Ardelean}, {Babej}, {Bach}, {Bachetti}, {Bakanov}, {Bamford},
  {Barentsen}, {Barmby}, {Baumbach}, {Berry}, {Biscani}, {Boquien}, {Bostroem},
  {Bouma}, {Brammer}, {Bray}, {Breytenbach}, {Buddelmeijer}, {Burke},
  {Calderone}, {Cano Rodr{\'\i}guez}, {Cara}, {Cardoso}, {Cheedella}, {Copin},
  {Corrales}, {Crichton}, {D'Avella}, {Deil}, {Depagne}, {Dietrich}, {Donath},
  {Droettboom}, {Earl}, {Erben}, {Fabbro}, {Ferreira}, {Finethy}, {Fox},
  {Garrison}, {Gibbons}, {Goldstein}, {Gommers}, {Greco}, {Greenfield},
  {Groener}, {Grollier}, {Hagen}, {Hirst}, {Homeier}, {Horton}, {Hosseinzadeh},
  {Hu}, {Hunkeler}, {Ivezi{\'c}}, {Jain}, {Jenness}, {Kanarek}, {Kendrew},
  {Kern}, {Kerzendorf}, {Khvalko}, {King}, {Kirkby}, {Kulkarni}, {Kumar},
  {Lee}, {Lenz}, {Littlefair}, {Ma}, {Macleod}, {Mastropietro}, {McCully},
  {Montagnac}, {Morris}, {Mueller}, {Mumford}, {Muna}, {Murphy}, {Nelson},
  {Nguyen}, {Ninan}, {N{\"o}the}, {Ogaz}, {Oh}, {Parejko}, {Parley}, {Pascual},
  {Patil}, {Patil}, {Plunkett}, {Prochaska}, {Rastogi}, {Reddy Janga},
  {Sabater}, {Sakurikar}, {Seifert}, {Sherbert}, {Sherwood-Taylor}, {Shih},
  {Sick}, {Silbiger}, {Singanamalla}, {Singer}, {Sladen}, {Sooley},
  {Sornarajah}, {Streicher}, {Teuben}, {Thomas}, {Tremblay}, {Turner},
  {Terr{\'o}n}, {van Kerkwijk}, {de la Vega}, {Watkins}, {Weaver}, {Whitmore},
  {Woillez}, {Zabalza}, \& {Astropy Contributors}}]{astropy:2018}
{Astropy Collaboration}, {Price-Whelan}, A.~M., {Sip{\H{o}}cz}, B.~M., {et~al.}
  2018, \aj, 156, 123

\bibitem[{{Astropy Collaboration} {et~al.}(2022){Astropy Collaboration},
  {Price-Whelan}, {Lim}, {Earl}, {Starkman}, {Bradley}, {Shupe}, {Patil},
  {Corrales}, {Brasseur}, {N{"o}the}, {Donath}, {Tollerud}, {Morris},
  {Ginsburg}, {Vaher}, {Weaver}, {Tocknell}, {Jamieson}, {van Kerkwijk},
  {Robitaille}, {Merry}, {Bachetti}, {G{"u}nther}, {Aldcroft},
  {Alvarado-Montes}, {Archibald}, {B{'o}di}, {Bapat}, {Barentsen}, {Baz{'a}n},
  {Biswas}, {Boquien}, {Burke}, {Cara}, {Cara}, {Conroy}, {Conseil}, {Craig},
  {Cross}, {Cruz}, {D'Eugenio}, {Dencheva}, {Devillepoix}, {Dietrich},
  {Eigenbrot}, {Erben}, {Ferreira}, {Foreman-Mackey}, {Fox}, {Freij}, {Garg},
  {Geda}, {Glattly}, {Gondhalekar}, {Gordon}, {Grant}, {Greenfield}, {Groener},
  {Guest}, {Gurovich}, {Handberg}, {Hart}, {Hatfield-Dodds}, {Homeier},
  {Hosseinzadeh}, {Jenness}, {Jones}, {Joseph}, {Kalmbach}, {Karamehmetoglu},
  {Ka{l}uszy{'n}ski}, {Kelley}, {Kern}, {Kerzendorf}, {Koch}, {Kulumani},
  {Lee}, {Ly}, {Ma}, {MacBride}, {Maljaars}, {Muna}, {Murphy}, {Norman},
  {O'Steen}, {Oman}, {Pacifici}, {Pascual}, {Pascual-Granado}, {Patil},
  {Perren}, {Pickering}, {Rastogi}, {Roulston}, {Ryan}, {Rykoff}, {Sabater},
  {Sakurikar}, {Salgado}, {Sanghi}, {Saunders}, {Savchenko}, {Schwardt},
  {Seifert-Eckert}, {Shih}, {Jain}, {Shukla}, {Sick}, {Simpson},
  {Singanamalla}, {Singer}, {Singhal}, {Sinha}, {Sip{H{o}}cz}, {Spitler},
  {Stansby}, {Streicher}, {{{S}}umak}, {Swinbank}, {Taranu}, {Tewary},
  {Tremblay}, {Val-Borro}, {Van Kooten}, {Vasovi{'c}}, {Verma}, {de Miranda
  Cardoso}, {Williams}, {Wilson}, {Winkel}, {Wood-Vasey}, {Xue}, {Yoachim},
  {Zhang}, {Zonca}, \& {Astropy Project Contributors}}]{astropy:2022}
{Astropy Collaboration}, {Price-Whelan}, A.~M., {Lim}, P.~L., {et~al.} 2022,
  \apj, 935, 167

\bibitem[{{Barber} {et~al.}(2022){Barber}, {Mann}, {Bush}, {Tofflemire},
  {Kraus}, {Krolikowski}, {Vanderburg}, {Fields}, {Newton}, {Owens}, \&
  {Thao}}]{Barber_2022}
{Barber}, M.~G., {Mann}, A.~W., {Bush}, J.~L., {et~al.} 2022, \aj, 164, 88

\bibitem[{{Barnes}(2003)}]{Barnes03}
{Barnes}, S.~A. 2003, \apj, 586, 464

\bibitem[{{Barnes} {et~al.}(2016){Barnes}, {Weingrill}, {Fritzewski},
  {Strassmeier}, \& {Platais}}]{Barnes2016}
{Barnes}, S.~A., {Weingrill}, J., {Fritzewski}, D., {Strassmeier}, K.~G., \&
  {Platais}, I. 2016, \apj, 823, 16

\bibitem[{{Berger} {et~al.}(2018){Berger}, {Howard}, \&
  {Boesgaard}}]{2018ApJ...855..115B}
{Berger}, T.~A., {Howard}, A.~W., \& {Boesgaard}, A.~M. 2018, \apj, 855, 115

\bibitem[{{Berger} {et~al.}(2020{\natexlab{a}}){Berger}, {Huber}, {Gaidos},
  {van Saders}, \& {Weiss}}]{Berger_2020b_rpage}
{Berger}, T.~A., {Huber}, D., {Gaidos}, E., {van Saders}, J.~L., \& {Weiss},
  L.~M. 2020{\natexlab{a}}, \aj, 160, 108

\bibitem[{{Berger} {et~al.}(2020{\natexlab{b}}){Berger}, {Huber}, {van Saders},
  {Gaidos}, {Tayar}, \& {Kraus}}]{Berger_2020a_catalog}
{Berger}, T.~A., {Huber}, D., {van Saders}, J.~L., {et~al.} 2020{\natexlab{b}},
  \aj, 159, 280

\bibitem[{{Binney} {et~al.}(2000){Binney}, {Dehnen}, \&
  {Bertelli}}]{2000MNRAS.318..658B}
{Binney}, J., {Dehnen}, W., \& {Bertelli}, G. 2000, \mnras, 318, 658

\bibitem[{{Bland-Hawthorn} \& {Gerhard}(2016)}]{Bland-Hawthorn2016}
{Bland-Hawthorn}, J., \& {Gerhard}, O. 2016, \araa, 54, 529

\bibitem[{{Bonomo} {et~al.}(2014){Bonomo}, {Sozzetti}, {Lovis}, {Malavolta},
  {Rice}, {Buchhave}, {Sasselov}, {Cameron}, {Latham}, {Molinari}, {Pepe},
  {Udry}, {Affer}, {Charbonneau}, {Cosentino}, {Dressing}, {Dumusque},
  {Figueira}, {Fiorenzano}, {Gettel}, {Harutyunyan}, {Haywood}, {Horne},
  {Lopez-Morales}, {Mayor}, {Micela}, {Motalebi}, {Nascimbeni}, {Phillips},
  {Piotto}, {Pollacco}, {Queloz}, {S{\'e}gransan}, {Szentgyorgyi}, \&
  {Watson}}]{2014A&A...572A...2B}
{Bonomo}, A.~S., {Sozzetti}, A., {Lovis}, C., {et~al.} 2014, \aap, 572, A2

\bibitem[{{Borucki} {et~al.}(2010){Borucki}, {Koch}, {Basri}, {Batalha},
  {Brown}, {Caldwell}, {Caldwell}, {Christensen-Dalsgaard}, {Cochran},
  {DeVore}, {Dunham}, {Dupree}, {Gautier}, {Geary}, {Gilliland}, {Gould},
  {Howell}, {Jenkins}, {Kondo}, {Latham}, {Marcy}, {Meibom}, {Kjeldsen},
  {Lissauer}, {Monet}, {Morrison}, {Sasselov}, {Tarter}, {Boss}, {Brownlee},
  {Owen}, {Buzasi}, {Charbonneau}, {Doyle}, {Fortney}, {Ford}, {Holman},
  {Seager}, {Steffen}, {Welsh}, {Rowe}, {Anderson}, {Buchhave}, {Ciardi},
  {Walkowicz}, {Sherry}, {Horch}, {Isaacson}, {Everett}, {Fischer}, {Torres},
  {Johnson}, {Endl}, {MacQueen}, {Bryson}, {Dotson}, {Haas}, {Kolodziejczak},
  {Van Cleve}, {Chandrasekaran}, {Twicken}, {Quintana}, {Clarke}, {Allen},
  {Li}, {Wu}, {Tenenbaum}, {Verner}, {Bruhweiler}, {Barnes}, \&
  {Prsa}}]{Borucki10}
{Borucki}, W.~J., {Koch}, D., {Basri}, G., {et~al.} 2010, Science, 327, 977

\bibitem[{{Bouma} {et~al.}(2021){Bouma}, {Curtis}, {Hartman}, {Winn}, \&
  {Bakos}}]{Bouma_2021}
{Bouma}, L.~G., {Curtis}, J.~L., {Hartman}, J.~D., {Winn}, J.~N., \& {Bakos},
  G.~{\'A}. 2021, \aj, 162, 197

\bibitem[{{Bouma} {et~al.}(2023){Bouma}, {Palumbo}, \&
  {Hillenbrand}}]{Bouma_2023}
{Bouma}, L.~G., {Palumbo}, E.~K., \& {Hillenbrand}, L.~A. 2023, \apjl, 947, L3

\bibitem[{{Bouma} {et~al.}(2020){Bouma}, {Hartman}, {Brahm}, {Evans},
  {Collins}, {Zhou}, {Sarkis}, {Quinn}, {de Leon}, {Livingston}, {Bergmann},
  {Stassun}, {Bhatti}, {Winn}, {Bakos}, {Abe}, {Crouzet}, {Dransfield},
  {Guillot}, {Marie-Sainte}, {M{\'e}karnia}, {Triaud}, {Tinney}, {Henning},
  {Espinoza}, {Jord{\'a}n}, {Barbieri}, {Nandakumar}, {Trifonov}, {Vines},
  {Vuckovic}, {Ziegler}, {Law}, {Mann}, {Ricker}, {Vanderspek}, {Seager},
  {Jenkins}, {Burke}, {Dragomir}, {Levine}, {Quintana}, {Rodriguez}, {Smith},
  \& {Wohler}}]{Bouma_2020_toi837}
{Bouma}, L.~G., {Hartman}, J.~D., {Brahm}, R., {et~al.} 2020, \aj, 160, 239

\bibitem[{{Bouma} {et~al.}(2022{\natexlab{a}}){Bouma}, {Curtis}, {Masuda},
  {Hillenbrand}, {Stefansson}, {Isaacson}, {Narita}, {Fukui}, {Ikoma},
  {Tamura}, {Kraus}, {Furlan}, {Gnilka}, {Lester}, \& {Howell}}]{Bouma_2022a}
{Bouma}, L.~G., {Curtis}, J.~L., {Masuda}, K., {et~al.} 2022{\natexlab{a}},
  \aj, 163, 121

\bibitem[{{Bouma} {et~al.}(2022{\natexlab{b}}){Bouma}, {Kerr}, {Curtis},
  {Isaacson}, {Hillenbrand}, {Howard}, {Kraus}, {Bieryla}, {Latham},
  {Petigura}, \& {Huber}}]{Bouma_2022b}
{Bouma}, L.~G., {Kerr}, R., {Curtis}, J.~L., {et~al.} 2022{\natexlab{b}}, \aj,
  164, 215

\bibitem[{{Boyle} \& {Bouma}(2023)}]{2023AJ....166...14B}
{Boyle}, A.~W., \& {Bouma}, L.~G. 2023, \aj, 166, 14

\bibitem[{{Cantat-Gaudin} {et~al.}(2018){Cantat-Gaudin}, {Jordi}, {Vallenari},
  {Bragaglia}, {Balaguer-N{\'u}{\~n}ez}, {Soubiran}, {Bossini}, {Moitinho},
  {Castro-Ginard}, {Krone-Martins}, {Casamiquela}, {Sordo}, \&
  {Carrera}}]{2018A&A...618A..93C}
{Cantat-Gaudin}, T., {Jordi}, C., {Vallenari}, A., {et~al.} 2018, \aap, 618,
  A93

\bibitem[{{Cantat-Gaudin} {et~al.}(2020){Cantat-Gaudin}, {Anders},
  {Castro-Ginard}, {Jordi}, {Romero-G{\'o}mez}, {Soubiran}, {Casamiquela},
  {Tarricq}, {Moitinho}, {Vallenari}, {Bragaglia}, {Krone-Martins}, \&
  {Kounkel}}]{CantatGaudin_2020}
{Cantat-Gaudin}, T., {Anders}, F., {Castro-Ginard}, A., {et~al.} 2020, \aap,
  640, A1

\bibitem[{{Capistrant} {et~al.}(2024){Capistrant}, {Soares-Furtado},
  {Vanderburg}, {Jankowski}, {Mann}, {Ross}, {Srdoc}, {Hinkel}, {Becker},
  {Magliano}, {Limbach}, {Stephan}, {Nine}, {Tofflemire}, {Kraus}, {Giacalone},
  {Winn}, {Bieryla}, {Bouma}, {Ciardi}, {Collins}, {Covone}, {de Beurs},
  {Huang}, {Jenkins}, {Kreidberg}, {Latham}, {Quinn}, {Seager}, {Shporer},
  {Twicken}, {Wohler}, {Vanderspek}, {Yarza}, \&
  {Ziegler}}]{2024AJ....167...54C}
{Capistrant}, B.~K., {Soares-Furtado}, M., {Vanderburg}, A., {et~al.} 2024,
  \aj, 167, 54

\bibitem[{{Carlos} {et~al.}(2019){Carlos}, {Mel{\'e}ndez}, {Spina}, {dos
  Santos}, {Bedell}, {Ramirez}, {Asplund}, {Bean}, {Yong}, {Yana Galarza}, \&
  {Alves-Brito}}]{2019MNRAS.485.4052C}
{Carlos}, M., {Mel{\'e}ndez}, J., {Spina}, L., {et~al.} 2019, \mnras, 485, 4052

\bibitem[{{Castro-Ginard} {et~al.}(2018){Castro-Ginard}, {Jordi}, {Luri},
  {Julbe}, {Morvan}, {Balaguer-N{\'u}{\~n}ez}, \&
  {Cantat-Gaudin}}]{2018A&A...618A..59C}
{Castro-Ginard}, A., {Jordi}, C., {Luri}, X., {et~al.} 2018, \aap, 618, A59

\bibitem[{{Chaboyer} {et~al.}(1995){Chaboyer}, {Demarque}, \&
  {Pinsonneault}}]{1995ApJ...441..865C}
{Chaboyer}, B., {Demarque}, P., \& {Pinsonneault}, M.~H. 1995, \apj, 441, 865

\bibitem[{{Chiti} {et~al.}(2024){Chiti}, {van Saders}, {Heintz}, {Hermes},
  {Ong}, {Hey}, {Ramirez-Weinhouse}, \& {Dugas}}]{2024arXiv240312129C}
{Chiti}, F., {van Saders}, J.~L., {Heintz}, T.~M., {et~al.} 2024, arXiv
  e-prints, arXiv:2403.12129

\bibitem[{{Choi} {et~al.}(2016){Choi}, {Dotter}, {Conroy}, {Cantiello},
  {Paxton}, \& {Johnson}}]{Choi_2016}
{Choi}, J., {Dotter}, A., {Conroy}, C., {et~al.} 2016, \apj, 823, 102

\bibitem[{{Christiansen} {et~al.}(2023){Christiansen}, {Zink},
  {Hardegree-Ullman}, {Fernandes}, {Hopkins}, {Rebull}, {Boley}, {Bergsten}, \&
  {Bhure}}]{2023AJ....166..248C}
{Christiansen}, J.~L., {Zink}, J.~K., {Hardegree-Ullman}, K.~K., {et~al.} 2023,
  \aj, 166, 248

\bibitem[{{Claytor} {et~al.}(2020){Claytor}, {van Saders}, {Santos},
  {Garc{\'\i}a}, {Mathur}, {Tayar}, {Pinsonneault}, \&
  {Shetrone}}]{Claytor2020}
{Claytor}, Z.~R., {van Saders}, J.~L., {Santos}, {\^A}. R.~G., {et~al.} 2020,
  \apj, 888, 43

\bibitem[{{Curtis}(2024)}]{Curtis2024}
{Curtis}, J. 2024, in American Astronomical Society Meeting Abstracts, Vol.
  243, American Astronomical Society Meeting Abstracts, 458.08

\bibitem[{{Curtis} {et~al.}(2019){Curtis}, {Ag{\"u}eros}, {Douglas}, \&
  {Meibom}}]{Curtis_2019_ngc6811}
{Curtis}, J.~L., {Ag{\"u}eros}, M.~A., {Douglas}, S.~T., \& {Meibom}, S. 2019,
  \apj, 879, 49

\bibitem[{{Curtis} {et~al.}(2018){Curtis}, {Vanderburg}, {Torres}, {Kraus},
  {Huber}, {Mann}, {Rizzuto}, {Isaacson}, {Howard}, {Henze}, {Fulton}, \&
  {Wright}}]{Curtis_2018}
{Curtis}, J.~L., {Vanderburg}, A., {Torres}, G., {et~al.} 2018, \aj, 155, 173

\bibitem[{{Curtis} {et~al.}(2020){Curtis}, {Ag{\"u}eros}, {Matt}, {Covey},
  {Douglas}, {Angus}, {Saar}, {Cody}, {Vanderburg}, {Law}, {Kraus}, {Latham},
  {Baranec}, {Riddle}, {Ziegler}, {Lund}, {Torres}, {Meibom}, {Aguirre}, \&
  {Wright}}]{Curtis_2020}
{Curtis}, J.~L., {Ag{\"u}eros}, M.~A., {Matt}, S.~P., {et~al.} 2020, \apj, 904,
  140

\bibitem[{{David} {et~al.}(2022){David}, {Angus}, {Curtis}, {van Saders},
  {Colman}, {Contardo}, {Lu}, \& {Zinn}}]{David_2022}
{David}, T.~J., {Angus}, R., {Curtis}, J.~L., {et~al.} 2022, \apj, 933, 114

\bibitem[{{David} {et~al.}(2019){David}, {Petigura}, {Luger}, {Foreman-Mackey},
  {Livingston}, {Mamajek}, \& {Hillenbrand}}]{David_2019}
{David}, T.~J., {Petigura}, E.~A., {Luger}, R., {et~al.} 2019, \apjl, 885, L12

\bibitem[{{David} {et~al.}(2021){David}, {Contardo}, {Sandoval}, {Angus}, {Lu},
  {Bedell}, {Curtis}, {Foreman-Mackey}, {Fulton}, {Grunblatt}, \&
  {Petigura}}]{David_2021}
{David}, T.~J., {Contardo}, G., {Sandoval}, A., {et~al.} 2021, \aj, 161, 265

\bibitem[{{Denissenkov} {et~al.}(2010){Denissenkov}, {Pinsonneault},
  {Terndrup}, \& {Newsham}}]{2010ApJ...716.1269D}
{Denissenkov}, P.~A., {Pinsonneault}, M., {Terndrup}, D.~M., \& {Newsham}, G.
  2010, \apj, 716, 1269

\bibitem[{{do Nascimento} {et~al.}(2020){do Nascimento}, {de Almeida},
  {Velloso}, {Anthony}, {Barnes}, {Saar}, {Meibom}, {da Costa}, {Castro},
  {Galarza}, {Lorenzo-Oliveira}, {Beck}, \& {Mel{\'e}ndez}}]{doNascimento2020}
{do Nascimento}, J.~D., J., {de Almeida}, L., {Velloso}, E.~N., {et~al.} 2020,
  \apj, 898, 173

\bibitem[{{Dotter}(2016)}]{2016ApJS..222....8D}
{Dotter}, A. 2016, \apjs, 222, 8

\bibitem[{{Dungee} {et~al.}(2022){Dungee}, {van Saders}, {Gaidos}, {Chun},
  {Garc{\'\i}a}, {Magnier}, {Mathur}, \& {Santos}}]{Dungee_2022}
{Dungee}, R., {van Saders}, J., {Gaidos}, E., {et~al.} 2022, \apj, 938, 118

\bibitem[{{Engle} \& {Guinan}(2023)}]{2023ApJ...954L..50E}
{Engle}, S.~G., \& {Guinan}, E.~F. 2023, \apjl, 954, L50

\bibitem[{{Epstein} \& {Pinsonneault}(2014)}]{Epstein_2014}
{Epstein}, C.~R., \& {Pinsonneault}, M.~H. 2014, \apj, 780, 159

\bibitem[{{Fortney} {et~al.}(2007){Fortney}, {Marley}, \&
  {Barnes}}]{2007ApJ...659.1661F}
{Fortney}, J.~J., {Marley}, M.~S., \& {Barnes}, J.~W. 2007, \apj, 659, 1661

\bibitem[{{Fritzewski} {et~al.}(2021){Fritzewski}, {Barnes}, {James}, \&
  {Strassmeier}}]{Fritzewski_2021}
{Fritzewski}, D.~J., {Barnes}, S.~A., {James}, D.~J., \& {Strassmeier}, K.~G.
  2021, \aap, 652, A60

\bibitem[{{Fritzewski} {et~al.}(2024){Fritzewski}, {Van Reeth}, {Aerts}, {Van
  Beeck}, {Gossage}, \& {Li}}]{2024A&A...681A..13F}
{Fritzewski}, D.~J., {Van Reeth}, T., {Aerts}, C., {et~al.} 2024, \aap, 681,
  A13

\bibitem[{{Fulton} \& {Petigura}(2018)}]{Fulton_2018}
{Fulton}, B.~J., \& {Petigura}, E.~A. 2018, \aj, 156, 264

\bibitem[{{Fulton} {et~al.}(2017){Fulton}, {Petigura}, {Howard}, {Isaacson},
  {Marcy}, {Cargile}, {Hebb}, {Weiss}, {Johnson}, {Morton}, {Sinukoff},
  {Crossfield}, \& {Hirsch}}]{2017AJ....154..109F}
{Fulton}, B.~J., {Petigura}, E.~A., {Howard}, A.~W., {et~al.} 2017, \aj, 154,
  109

\bibitem[{{Gaia Collaboration} {et~al.}(2023){Gaia Collaboration}, {Arenou},
  {Babusiaux}, {Barstow}, {Faigler}, {Jorissen}, {Kervella}, {Mazeh},
  {Mowlavi}, {Panuzzo}, \& et~al.}]{GaiaCollaboration2023}
{Gaia Collaboration}, {Arenou}, F., {Babusiaux}, C., {et~al.} 2023, \aap, 674,
  A34

\bibitem[{{Gallet} \& {Bouvier}(2015)}]{Gallet_Bouvier_2015}
{Gallet}, F., \& {Bouvier}, J. 2015, \aap, 577, A98

\bibitem[{{Ge} {et~al.}(2022){Ge}, {Zhang}, {Zang}, {Deng}, {Mao}, {Xie},
  {Liu}, {Zhou}, {Willis}, {Huang}, {Howell}, {Feng}, {Zhu}, {Yao}, {Liu},
  {Aizawa}, {Zhu}, {Li}, {Ma}, {Ye}, {Yu}, {Xiang}, {Yu}, {Liu}, {Yang},
  {Wang}, {Shi}, {Fang}, {Zong}, {Liu}, {Zhang}, {Zhang}, {El-Badry}, {Shen},
  {Tam}, {Hu}, {Yang}, {Zou}, {Wu}, {Lei}, {Wei}, {Wu}, {Sun}, {Wang}, {Zhang},
  {Xu}, {Yang}, {Li}, {Xiang}, {Wang}, {Wang}, {Zhang}, {Jia}, {Yuan}, {Zhang},
  {Xuesong Wang}, {Gan}, {Wang}, {Zhao}, {Liu}, {Wei}, {Kang}, {Yang}, {Qi},
  {Liu}, {Zhang}, {Zhu}, {Zhou}, {Zhang}, {Yu}, {Zhang}, {Li}, {Tang}, {Wang},
  {Wang}, {Li}, {Cheng}, {Shen}, {Li}, {Pan}, {Yang}, {Gao}, {Song}, {Wang},
  {Zhang}, {Chen}, {Wang}, {Zhang}, {Wang}, {Zeng}, {Zheng}, {Zhu}, {Guo},
  {Zhang}, {Li}, {Wen}, {Feng}, {Chen}, {Chen}, {Han}, {Yang}, {Wang}, {Duan},
  {Huang}, {Liang}, {Bi}, {Gai}, {Ge}, {Guo}, {Huang}, {Li}, {Li}, {Li},
  {Yuxi}, {Lu}, {Rix}, {Shi}, {Song}, {Tang}, {Ting}, {Wu}, {Wu}, {Yang},
  {Yin}, {Gould}, {Lee}, {Dong}, {Yee}, {Shvartzvald}, {Yang}, {Kuang},
  {Zhang}, {Liao}, {Qi}, {Yang}, {Zhang}, {Jiang}, {Ou}, {Li}, {Beck},
  {Bedding}, {Campante}, {Chaplin}, {Christensen-Dalsgaard}, {Garc{\'\i}a},
  {Gaulme}, {Gizon}, {Hekker}, {Huber}, {Khanna}, {Li}, {Mathur}, {Miglio},
  {Mosser}, {Ong}, {Santos}, {Stello}, {Bowman}, {Lares-Martiz}, {Murphy},
  {Niu}, {Ma}, {Moln{\'a}r}, {Fu}, {De Cat}, {Su}, \&
  {consortium}}]{2022arXiv220606693G}
{Ge}, J., {Zhang}, H., {Zang}, W., {et~al.} 2022, arXiv e-prints,
  arXiv:2206.06693

\bibitem[{{Gilbert}(2022)}]{2022AJ....163..111G}
{Gilbert}, G.~J. 2022, \aj, 163, 111

\bibitem[{{Gillen} {et~al.}(2020){Gillen}, {Briegal}, {Hodgkin},
  {Foreman-Mackey}, {Van Leeuwen}, {Jackman}, {McCormac}, {West}, {Queloz},
  {Bayliss}, {Goad}, {Watson}, {Wheatley}, {Belardi}, {Burleigh}, {Casewell},
  {Jenkins}, {Raynard}, {Smith}, {Tilbrook}, \& {Vines}}]{Gillen_2020}
{Gillen}, E., {Briegal}, J.~T., {Hodgkin}, S.~T., {et~al.} 2020, \mnras, 492,
  1008

\bibitem[{{Girardi} {et~al.}(2005){Girardi}, {Groenewegen}, {Hatziminaoglou},
  \& {da Costa}}]{Girardi2005}
{Girardi}, L., {Groenewegen}, M.~A.~T., {Hatziminaoglou}, E., \& {da Costa}, L.
  2005, \aap, 436, 895

\bibitem[{{Green} {et~al.}(2019){Green}, {Schlafly}, {Zucker}, {Speagle}, \&
  {Finkbeiner}}]{2019ApJ...887...93G}
{Green}, G.~M., {Schlafly}, E., {Zucker}, C., {Speagle}, J.~S., \&
  {Finkbeiner}, D. 2019, \apj, 887, 93

\bibitem[{{Greiss} {et~al.}(2012){Greiss}, {Steeghs}, {G{\"a}nsicke},
  {Mart{\'\i}n}, {Groot}, {Irwin}, {Gonz{\'a}lez-Solares}, {Greimel}, {Knigge},
  {{\O}stensen}, {Verbeek}, {Drew}, {Drake}, {Jonker}, {Ripepi}, {Scaringi},
  {Southworth}, {Still}, {Wright}, {Farnhill}, {van Haaften}, \&
  {Shah}}]{2012AJ....144...24G}
{Greiss}, S., {Steeghs}, D., {G{\"a}nsicke}, B.~T., {et~al.} 2012, \aj, 144, 24

\bibitem[{{Gruner} {et~al.}(2023){Gruner}, {Barnes}, \&
  {Weingrill}}]{Gruner_2023}
{Gruner}, D., {Barnes}, S.~A., \& {Weingrill}, J. 2023, \aap, 672, A159

\bibitem[{{Gupta} \& {Schlichting}(2019)}]{Gupta_2019}
{Gupta}, A., \& {Schlichting}, H.~E. 2019, \mnras, 487, 24

\bibitem[{{Hall} {et~al.}(2021){Hall}, {Davies}, {van Saders}, {Nielsen},
  {Lund}, {Chaplin}, {Garc{\'\i}a}, {Amard}, {Breimann}, {Khan}, {See}, \&
  {Tayar}}]{Hall2021}
{Hall}, O.~J., {Davies}, G.~R., {van Saders}, J., {et~al.} 2021, Nature
  Astronomy, 5, 707

\bibitem[{{Hedges} {et~al.}(2021){Hedges}, {Hughes}, {Zhou}, {David}, {Becker},
  {Giacalone}, {Vanderburg}, {Rodriguez}, {Bieryla}, {Wirth}, {Atherton},
  {Fetherolf}, {Collins}, {Price-Whelan}, {Bedell}, {Quinn}, {Gan}, {Ricker},
  {Latham}, {Vanderspek}, {Seager}, {Winn}, {Jenkins}, {Kielkopf}, {Schwarz},
  {Dressing}, {Gonzales}, {Crossfield}, {Matthews}, {Jensen}, {Furlan},
  {Gnilka}, {Howell}, {Lester}, {Scott}, {Feliz}, {Lund}, {Siverd}, {Stevens},
  {Narita}, {Fukui}, {Murgas}, {Palle}, {Sutton}, {Stassun}, {Bouma}, {Vezie},
  {Villase{\~n}or}, {Quintana}, \& {Smith}}]{2021AJ....162...54H}
{Hedges}, C., {Hughes}, A., {Zhou}, G., {et~al.} 2021, \aj, 162, 54

\bibitem[{{Herschel}(1864)}]{1864RSPT..154....1H}
{Herschel}, J. F.~W. 1864, Philosophical Transactions of the Royal Society of
  London Series I, 154, 1

\bibitem[{{Huber} {et~al.}(2013){Huber}, {Chaplin}, {Christensen-Dalsgaard},
  {Gilliland}, {Kjeldsen}, {Buchhave}, {Fischer}, {Lissauer}, {Rowe},
  {Sanchis-Ojeda}, {Basu}, {Handberg}, {Hekker}, {Howard}, {Isaacson},
  {Karoff}, {Latham}, {Lund}, {Lundkvist}, {Marcy}, {Miglio}, {Silva Aguirre},
  {Stello}, {Arentoft}, {Barclay}, {Bedding}, {Burke}, {Christiansen},
  {Elsworth}, {Haas}, {Kawaler}, {Metcalfe}, {Mullally}, \&
  {Thompson}}]{2013ApJ...767..127H}
{Huber}, D., {Chaplin}, W.~J., {Christensen-Dalsgaard}, J., {et~al.} 2013,
  \apj, 767, 127

\bibitem[{{Huber} {et~al.}(2017){Huber}, {Zinn}, {Bojsen-Hansen},
  {Pinsonneault}, {Sahlholdt}, {Serenelli}, {Silva Aguirre}, {Stassun},
  {Stello}, {Tayar}, {Bastien}, {Bedding}, {Buchhave}, {Chaplin}, {Davies},
  {Garc{\'\i}a}, {Latham}, {Mathur}, {Mosser}, \&
  {Sharma}}]{2017ApJ...844..102H}
{Huber}, D., {Zinn}, J., {Bojsen-Hansen}, M., {et~al.} 2017, \apj, 844, 102

\bibitem[{Hunter {et~al.}(2007)}]{matplotlib}
Hunter, J.~D., {et~al.} 2007, Computing in science and engineering, 9, 90

\bibitem[{{Isern}(2019)}]{2019ApJ...878L..11I}
{Isern}, J. 2019, \apjl, 878, L11

\bibitem[{{Izidoro} {et~al.}(2017){Izidoro}, {Ogihara}, {Raymond},
  {Morbidelli}, {Pierens}, {Bitsch}, {Cossou}, \&
  {Hersant}}]{2017MNRAS.470.1750I}
{Izidoro}, A., {Ogihara}, M., {Raymond}, S.~N., {et~al.} 2017, \mnras, 470,
  1750

\bibitem[{{Jeffries} {et~al.}(2023){Jeffries}, {Jackson}, {Wright}, {Weaver},
  {Gilmore}, {Randich}, {Bragaglia}, {Korn}, {Smiljanic}, {Biazzo}, {Casey},
  {Frasca}, {Gonneau}, {Guiglion}, {Morbidelli}, {Prisinzano}, {Sacco},
  {Tautvai{\v{s}}ien{\.{e}}}, {Worley}, \& {Zaggia}}]{Jeffries_2023}
{Jeffries}, R.~D., {Jackson}, R.~J., {Wright}, N.~J., {et~al.} 2023, \mnras,
  523, 802

\bibitem[{{Jenkins} {et~al.}(2002){Jenkins}, {Caldwell}, \&
  {Borucki}}]{2002ApJ...564..495J}
{Jenkins}, J.~M., {Caldwell}, D.~A., \& {Borucki}, W.~J. 2002, \apj, 564, 495

\bibitem[{{Johnson} {et~al.}(2017){Johnson}, {Petigura}, {Fulton}, {Marcy},
  {Howard}, {Isaacson}, {Hebb}, {Cargile}, {Morton}, {Weiss}, {Winn}, {Rogers},
  {Sinukoff}, \& {Hirsch}}]{2017AJ....154..108J}
{Johnson}, J.~A., {Petigura}, E.~A., {Fulton}, B.~J., {et~al.} 2017, \aj, 154,
  108

\bibitem[{{Kawaler}(1989)}]{Kawaler_1989}
{Kawaler}, S.~D. 1989, \apjl, 343, L65

\bibitem[{{Keppler} {et~al.}(2018){Keppler}, {Benisty}, {M{\"u}ller},
  {Henning}, {van Boekel}, {Cantalloube}, {Ginski}, {van Holstein}, {Maire},
  {Pohl}, {Samland}, {Avenhaus}, {Baudino}, {Boccaletti}, {de Boer},
  {Bonnefoy}, {Chauvin}, {Desidera}, {Langlois}, {Lazzoni}, {Marleau},
  {Mordasini}, {Pawellek}, {Stolker}, {Vigan}, {Zurlo}, {Birnstiel},
  {Brandner}, {Feldt}, {Flock}, {Girard}, {Gratton}, {Hagelberg}, {Isella},
  {Janson}, {Juhasz}, {Kemmer}, {Kral}, {Lagrange}, {Launhardt}, {Matter},
  {M{\'e}nard}, {Milli}, {Molli{\`e}re}, {Olofsson}, {P{\'e}rez}, {Pinilla},
  {Pinte}, {Quanz}, {Schmidt}, {Udry}, {Wahhaj}, {Williams}, {Buenzli},
  {Cudel}, {Dominik}, {Galicher}, {Kasper}, {Lannier}, {Mesa}, {Mouillet},
  {Peretti}, {Perrot}, {Salter}, {Sissa}, {Wildi}, {Abe}, {Antichi},
  {Augereau}, {Baruffolo}, {Baudoz}, {Bazzon}, {Beuzit}, {Blanchard}, {Brems},
  {Buey}, {De Caprio}, {Carbillet}, {Carle}, {Cascone}, {Cheetham}, {Claudi},
  {Costille}, {Delboulb{\'e}}, {Dohlen}, {Fantinel}, {Feautrier}, {Fusco},
  {Giro}, {Gluck}, {Gry}, {Hubin}, {Hugot}, {Jaquet}, {Le Mignant}, {Llored},
  {Madec}, {Magnard}, {Martinez}, {Maurel}, {Meyer}, {M{\"o}ller-Nilsson},
  {Moulin}, {Mugnier}, {Orign{\'e}}, {Pavlov}, {Perret}, {Petit}, {Pragt},
  {Puget}, {Rabou}, {Ramos}, {Rigal}, {Rochat}, {Roelfsema}, {Rousset}, {Roux},
  {Salasnich}, {Sauvage}, {Sevin}, {Soenke}, {Stadler}, {Suarez}, {Turatto}, \&
  {Weber}}]{2018A&A...617A..44K}
{Keppler}, M., {Benisty}, M., {M{\"u}ller}, A., {et~al.} 2018, \aap, 617, A44

\bibitem[{{Kerr} {et~al.}(2024){Kerr}, {Kraus}, {Krolikowski}, {Bouma}, \&
  {Farias}}]{Kerr2024}
{Kerr}, R., {Kraus}, A.~L., {Krolikowski}, D., {Bouma}, L.~G., \& {Farias},
  J.~P. 2024, arXiv e-prints, arXiv:2406.19530

\bibitem[{{Kerr} {et~al.}(2021){Kerr}, {Rizzuto}, {Kraus}, \&
  {Offner}}]{Kerr_2021}
{Kerr}, R. M.~P., {Rizzuto}, A.~C., {Kraus}, A.~L., \& {Offner}, S. S.~R. 2021,
  \apj, 917, 23

\bibitem[{{Kipping} \& {Bakos}(2011)}]{2011ApJ...733...36K}
{Kipping}, D., \& {Bakos}, G. 2011, \apj, 733, 36

\bibitem[{{Kirk} {et~al.}(2016){Kirk}, {Conroy}, {Pr{\v{s}}a}, {Abdul-Masih},
  {Kochoska}, {Matijevi{\v{c}}}, {Hambleton}, {Barclay}, {Bloemen}, {Boyajian},
  {Doyle}, {Fulton}, {Hoekstra}, {Jek}, {Kane}, {Kostov}, {Latham}, {Mazeh},
  {Orosz}, {Pepper}, {Quarles}, {Ragozzine}, {Shporer}, {Southworth},
  {Stassun}, {Thompson}, {Welsh}, {Agol}, {Derekas}, {Devor}, {Fischer},
  {Green}, {Gropp}, {Jacobs}, {Johnston}, {LaCourse}, {Saetre}, {Schwengeler},
  {Toczyski}, {Werner}, {Garrett}, {Gore}, {Martinez}, {Spitzer}, {Stevick},
  {Thomadis}, {Vrijmoet}, {Yenawine}, {Batalha}, \&
  {Borucki}}]{2016AJ....151...68K}
{Kirk}, B., {Conroy}, K., {Pr{\v{s}}a}, A., {et~al.} 2016, \aj, 151, 68

\bibitem[{{Klein} {et~al.}(2022){Klein}, {Zicher}, {Kavanagh}, {Nielsen},
  {Aigrain}, {Vidotto}, {Barrag{\'a}n}, {Strugarek}, {Nicholson}, {Donati}, \&
  {Bouvier}}]{2022MNRAS.512.5067K}
{Klein}, B., {Zicher}, N., {Kavanagh}, R.~D., {et~al.} 2022, \mnras, 512, 5067

\bibitem[{{Kolbl} {et~al.}(2015){Kolbl}, {Marcy}, {Isaacson}, \&
  {Howard}}]{2015AJ....149...18K}
{Kolbl}, R., {Marcy}, G.~W., {Isaacson}, H., \& {Howard}, A.~W. 2015, \aj, 149,
  18

\bibitem[{{Kounkel} \& {Covey}(2019)}]{2019AJ....158..122K}
{Kounkel}, M., \& {Covey}, K. 2019, \aj, 158, 122

\bibitem[{{Kounkel} {et~al.}(2020){Kounkel}, {Covey}, \&
  {Stassun}}]{Kounkel_2020}
{Kounkel}, M., {Covey}, K., \& {Stassun}, K.~G. 2020, \aj, 160, 279

\bibitem[{{Livingston} {et~al.}(2018){Livingston}, {Dai}, {Hirano}, {Gandolfi},
  {Nowak}, {Endl}, {Velasco}, {Fukui}, {Narita}, {Prieto-Arranz}, {Barragan},
  {Cusano}, {Albrecht}, {Cabrera}, {Cochran}, {Csizmadia}, {Deeg},
  {Eigm{\"u}ller}, {Erikson}, {Fridlund}, {Grziwa}, {Guenther}, {Hatzes},
  {Kawauchi}, {Korth}, {Nespral}, {Palle}, {P{\"a}tzold}, {Persson}, {Rauer},
  {Smith}, {Tamura}, {Tanaka}, {Van Eylen}, {Watanabe}, \&
  {Winn}}]{Livingston_2018}
{Livingston}, J.~H., {Dai}, F., {Hirano}, T., {et~al.} 2018, \aj, 155, 115

\bibitem[{{Lorenzo-Oliveira} {et~al.}(2020){Lorenzo-Oliveira}, {Mel{\'e}ndez},
  {Ponte}, \& {Galarza}}]{LorenzoOliveira2020}
{Lorenzo-Oliveira}, D., {Mel{\'e}ndez}, J., {Ponte}, G., \& {Galarza}, J.~Y.
  2020, \mnras, 495, L61

\bibitem[{{Lorenzo-Oliveira} {et~al.}(2019){Lorenzo-Oliveira}, {Mel{\'e}ndez},
  {Yana Galarza}, {Ponte}, {dos Santos}, {Spina}, {Bedell}, {Ram{\'\i}rez},
  {Bean}, \& {Asplund}}]{LorenzoOliveira2019}
{Lorenzo-Oliveira}, D., {Mel{\'e}ndez}, J., {Yana Galarza}, J., {et~al.} 2019,
  \mnras, 485, L68

\bibitem[{{Lu} {et~al.}(2024){Lu}, {Angus}, {Foreman-Mackey}, \&
  {Hattori}}]{2024AJ....167..159L}
{Lu}, Y., {Angus}, R., {Foreman-Mackey}, D., \& {Hattori}, S. 2024, \aj, 167,
  159

\bibitem[{{Lu} {et~al.}(2021){Lu}, {Angus}, {Curtis}, {David}, \&
  {Kiman}}]{2021AJ....161..189L}
{Lu}, Y.~L., {Angus}, R., {Curtis}, J.~L., {David}, T.~J., \& {Kiman}, R. 2021,
  \aj, 161, 189

\bibitem[{{Lu} {et~al.}(2022){Lu}, {Curtis}, {Angus}, {David}, \&
  {Hattori}}]{Lu2022}
{Lu}, Y.~L., {Curtis}, J.~L., {Angus}, R., {David}, T.~J., \& {Hattori}, S.
  2022, \aj, 164, 251

\bibitem[{{Mamajek} \& {Hillenbrand}(2008)}]{Mamajek_2008}
{Mamajek}, E.~E., \& {Hillenbrand}, L.~A. 2008, \apj, 687, 1264

\bibitem[{Mann {et~al.}(2016)Mann, Gaidos, Mace, Johnson, Bowler, LaCourse,
  Jacobs, Vanderburg, Kraus, Kaplan, \& Jaffe}]{Mann_K2_25_2016}
Mann, A.~W., Gaidos, E., Mace, G.~N., {et~al.} 2016, ApJ, 818

\bibitem[{{Marois} {et~al.}(2008){Marois}, {Macintosh}, {Barman}, {Zuckerman},
  {Song}, {Patience}, {Lafreni{\`e}re}, \& {Doyon}}]{2008Sci...322.1348M}
{Marois}, C., {Macintosh}, B., {Barman}, T., {et~al.} 2008, Science, 322, 1348

\bibitem[{{Masuda}(2014)}]{2014ApJ...783...53M}
{Masuda}, K. 2014, \apj, 783, 53

\bibitem[{{Masuda}(2022{\natexlab{a}})}]{2022ApJ...937...94M}
---. 2022{\natexlab{a}}, \apj, 937, 94

\bibitem[{{Masuda}(2022{\natexlab{b}})}]{Masuda2022_amplevoln}
---. 2022{\natexlab{b}}, \apj, 933, 195

\bibitem[{{Masuda} {et~al.}(2022){Masuda}, {Petigura}, \&
  {Hall}}]{Masuda2022infer}
{Masuda}, K., {Petigura}, E.~A., \& {Hall}, O.~J. 2022, \mnras, 510, 5623

\bibitem[{{Mathur} {et~al.}(2017){Mathur}, {Huber}, {Batalha}, {Ciardi},
  {Bastien}, {Bieryla}, {Buchhave}, {Cochran}, {Endl}, {Esquerdo}, {Furlan},
  {Howard}, {Howell}, {Isaacson}, {Latham}, {MacQueen}, \&
  {Silva}}]{Mathur_2017}
{Mathur}, S., {Huber}, D., {Batalha}, N.~M., {et~al.} 2017, \apjs, 229, 30

\bibitem[{{Mathur} {et~al.}(2023){Mathur}, {Claytor}, {Santos}, {Garc{\'\i}a},
  {Amard}, {Bugnet}, {Corsaro}, {Bonanno}, {Breton}, {Godoy-Rivera},
  {Pinsonneault}, \& {van Saders}}]{2023ApJ...952..131M}
{Mathur}, S., {Claytor}, Z.~R., {Santos}, {\^A}. R.~G., {et~al.} 2023, \apj,
  952, 131

\bibitem[{{Matt} {et~al.}(2015){Matt}, {Brun}, {Baraffe}, {Bouvier}, \&
  {Chabrier}}]{Matt_2015}
{Matt}, S.~P., {Brun}, A.~S., {Baraffe}, I., {Bouvier}, J., \& {Chabrier}, G.
  2015, \apjl, 799, L23

\bibitem[{{Mazeh} {et~al.}(2015){Mazeh}, {Perets}, {McQuillan}, \&
  {Goldstein}}]{Mazeh_2015}
{Mazeh}, T., {Perets}, H.~B., {McQuillan}, A., \& {Goldstein}, E.~S. 2015,
  \apj, 801, 3

\bibitem[{{Mazzi} {et~al.}(2024){Mazzi}, {Girardi}, {Trabucchi}, {Dalcanton},
  {Luger}, {Marigo}, {Miglio}, {Costa}, {Chen}, {Pastorelli}, {Fouesneau},
  {Zaggia}, {Bressan}, \& {Dal Tio}}]{Mazzi2024}
{Mazzi}, A., {Girardi}, L., {Trabucchi}, M., {et~al.} 2024, \mnras, 527, 583

\bibitem[{{McQuillan} {et~al.}(2013){McQuillan}, {Mazeh}, \&
  {Aigrain}}]{McQuillan_2013}
{McQuillan}, A., {Mazeh}, T., \& {Aigrain}, S. 2013, \apjl, 775, L11

\bibitem[{{McQuillan} {et~al.}(2014){McQuillan}, {Mazeh}, \&
  {Aigrain}}]{McQuillan_2014}
---. 2014, \apjs, 211, 24

\bibitem[{{Meibom} {et~al.}(2015){Meibom}, {Barnes}, {Platais}, {Gilliland},
  {Latham}, \& {Mathieu}}]{Meibom_2015}
{Meibom}, S., {Barnes}, S.~A., {Platais}, I., {et~al.} 2015, \nat, 517, 589

\bibitem[{{Meibom} {et~al.}(2007){Meibom}, {Mathieu}, \&
  {Stassun}}]{Meibom_2007}
{Meibom}, S., {Mathieu}, R.~D., \& {Stassun}, K.~G. 2007, \apjl, 665, L155

\bibitem[{{Meibom} {et~al.}(2013){Meibom}, {Torres}, {Fressin}, {Latham},
  {Rowe}, {Ciardi}, {Bryson}, {Rogers}, {Henze}, {Janes}, {Barnes}, {Marcy},
  {Isaacson}, {Fischer}, {Howell}, {Horch}, {Jenkins}, {Schuler}, \&
  {Crepp}}]{Meibom_2013}
{Meibom}, S., {Torres}, G., {Fressin}, F., {et~al.} 2013, \nat, 499, 55

\bibitem[{{Mills} \& {Fabrycky}(2017)}]{2017AJ....153...45M}
{Mills}, S.~M., \& {Fabrycky}, D.~C. 2017, \aj, 153, 45

\bibitem[{{Miyazaki} \& {Masuda}(2023)}]{Miyazaki2023}
{Miyazaki}, S., \& {Masuda}, K. 2023, \aj, 166, 209

\bibitem[{{Moe} \& {Kratter}(2021)}]{Moe2021}
{Moe}, M., \& {Kratter}, K.~M. 2021, \mnras, 507, 3593

\bibitem[{{Moe} {et~al.}(2019){Moe}, {Kratter}, \& {Badenes}}]{Moe2019}
{Moe}, M., {Kratter}, K.~M., \& {Badenes}, C. 2019, \apj, 875, 61

\bibitem[{{Mor} {et~al.}(2019){Mor}, {Robin}, {Figueras}, {Roca-F{\`a}brega},
  \& {Luri}}]{2019A&A...624L...1M}
{Mor}, R., {Robin}, A.~C., {Figueras}, F., {Roca-F{\`a}brega}, S., \& {Luri},
  X. 2019, \aap, 624, L1

\bibitem[{{Morton} {et~al.}(2016){Morton}, {Bryson}, {Coughlin}, {Rowe},
  {Ravichandran}, {Petigura}, {Haas}, \& {Batalha}}]{2016ApJ...822...86M}
{Morton}, T.~D., {Bryson}, S.~T., {Coughlin}, J.~L., {et~al.} 2016, \apj, 822,
  86

\bibitem[{{Nardiello} {et~al.}(2022){Nardiello}, {Malavolta}, {Desidera},
  {Baratella}, {D'Orazi}, {Messina}, {Biazzo}, {Benatti}, {Damasso}, {Rajpaul},
  {Bonomo}, {Capuzzo Dolcetta}, {Mallonn}, {Cale}, {Plavchan}, {El Mufti},
  {Bignamini}, {Borsa}, {Carleo}, {Claudi}, {Covino}, {Lanza}, {Maldonado},
  {Mancini}, {Micela}, {Molinari}, {Pinamonti}, {Piotto}, {Poretti},
  {Scandariato}, {Sozzetti}, {Andreuzzi}, {Boschin}, {Cosentino}, {Fiorenzano},
  {Harutyunyan}, {Knapic}, {Pedani}, {Affer}, {Maggio}, \&
  {Rainer}}]{Nardiello_2022}
{Nardiello}, D., {Malavolta}, L., {Desidera}, S., {et~al.} 2022, \aap, 664,
  A163

\bibitem[{{Newton} {et~al.}(2016){Newton}, {Irwin}, {Charbonneau},
  {Berta-Thompson}, {Dittmann}, \& {West}}]{2016ApJ...821...93N}
{Newton}, E.~R., {Irwin}, J., {Charbonneau}, D., {et~al.} 2016, \apj, 821, 93

\bibitem[{{Newton} {et~al.}(2021){Newton}, {Mann}, {Kraus}, {Livingston},
  {Vanderburg}, {Curtis}, {Thao}, {Hawkins}, {Wood}, {Rizzuto}, {Soubkiou},
  {Tofflemire}, {Zhou}, {Crossfield}, {Pearce}, {Collins}, {Conti}, {Tan},
  {Villeneuva}, {Spencer}, {Dragomir}, {Quinn}, {Jensen}, {Collins},
  {Stockdale}, {Cloutier}, {Hellier}, {Benkhaldoun}, {Ziegler}, {Brice{\~n}o},
  {Law}, {Benneke}, {Christiansen}, {Gorjian}, {Kane}, {Kreidberg}, {Morales},
  {Werner}, {Twicken}, {Levine}, {Ciardi}, {Guerrero}, {Hesse}, {Quintana},
  {Shiao}, {Smith}, {Torres}, {Ricker}, {Vanderspek}, {Seager}, {Winn},
  {Jenkins}, \& {Latham}}]{Newton_2021}
{Newton}, E.~R., {Mann}, A.~W., {Kraus}, A.~L., {et~al.} 2021, \aj, 161, 65

\bibitem[{{Nordstr{\"o}m} {et~al.}(2004){Nordstr{\"o}m}, {Mayor}, {Andersen},
  {Holmberg}, {Pont}, {J{\o}rgensen}, {Olsen}, {Udry}, \&
  {Mowlavi}}]{Nordstrom_2004}
{Nordstr{\"o}m}, B., {Mayor}, M., {Andersen}, J., {et~al.} 2004, \aap, 418, 989

\bibitem[{{Noyes} {et~al.}(1984){Noyes}, {Hartmann}, {Baliunas}, {Duncan}, \&
  {Vaughan}}]{Noyes_1984}
{Noyes}, R.~W., {Hartmann}, L.~W., {Baliunas}, S.~L., {Duncan}, D.~K., \&
  {Vaughan}, A.~H. 1984, \apj, 279, 763

\bibitem[{{O'Donovan} {et~al.}(2006){O'Donovan}, {Charbonneau}, {Mandushev},
  {Dunham}, {Latham}, {Torres}, {Sozzetti}, {Brown}, {Trauger}, {Belmonte},
  {Rabus}, {Almenara}, {Alonso}, {Deeg}, {Esquerdo}, {Falco}, {Hillenbrand},
  {Roussanova}, {Stefanik}, \& {Winn}}]{2006ApJ...651L..61O}
{O'Donovan}, F.~T., {Charbonneau}, D., {Mandushev}, G., {et~al.} 2006, \apjl,
  651, L61

\bibitem[{{Owen}(2019)}]{2019AREPS..47...67O}
{Owen}, J.~E. 2019, Annual Review of Earth and Planetary Sciences, 47, 67

\bibitem[{{Owen} \& {Lai}(2018)}]{Owen2018}
{Owen}, J.~E., \& {Lai}, D. 2018, \mnras, 479, 5012

\bibitem[{{Pecaut} \& {Mamajek}(2013)}]{Pecaut_2013}
{Pecaut}, M.~J., \& {Mamajek}, E.~E. 2013, \apjs, 208, 9

\bibitem[{{Petigura} {et~al.}(2017){Petigura}, {Howard}, {Marcy}, {Johnson},
  {Isaacson}, {Cargile}, {Hebb}, {Fulton}, {Weiss}, {Morton}, {Winn}, {Rogers},
  {Sinukoff}, {Hirsch}, \& {Crossfield}}]{2017AJ....154..107P}
{Petigura}, E.~A., {Howard}, A.~W., {Marcy}, G.~W., {et~al.} 2017, \aj, 154,
  107

\bibitem[{{Petigura} {et~al.}(2018){Petigura}, {Marcy}, {Winn}, {Weiss},
  {Fulton}, {Howard}, {Sinukoff}, {Isaacson}, {Morton}, \&
  {Johnson}}]{Petigura_2018}
{Petigura}, E.~A., {Marcy}, G.~W., {Winn}, J.~N., {et~al.} 2018, \aj, 155, 89

\bibitem[{{Petigura} {et~al.}(2022){Petigura}, {Rogers}, {Isaacson}, {Owen},
  {Kraus}, {Winn}, {MacDougall}, {Howard}, {Fulton}, {Kosiarek}, {Weiss},
  {Behmard}, \& {Blunt}}]{Petigura_2022}
{Petigura}, E.~A., {Rogers}, J.~G., {Isaacson}, H., {et~al.} 2022, \aj, 163,
  179

\bibitem[{{Plavchan} {et~al.}(2020){Plavchan}, {Barclay}, {Gagn{\'e}}, {Gao},
  {Cale}, {Matzko}, {Dragomir}, {Quinn}, {Feliz}, {Stassun}, {Crossfield},
  {Berardo}, {Latham}, {Tieu}, {Anglada-Escud{\'e}}, {Ricker}, {Vanderspek},
  {Seager}, {Winn}, {Jenkins}, {Rinehart}, {Krishnamurthy}, {Dynes}, {Doty},
  {Adams}, {Afanasev}, {Beichman}, {Bottom}, {Bowler}, {Brinkworth}, {Brown},
  {Cancino}, {Ciardi}, {Clampin}, {Clark}, {Collins}, {Davison},
  {Foreman-Mackey}, {Furlan}, {Gaidos}, {Geneser}, {Giddens}, {Gilbert},
  {Hall}, {Hellier}, {Henry}, {Horner}, {Howard}, {Huang}, {Huber}, {Kane},
  {Kenworthy}, {Kielkopf}, {Kipping}, {Klenke}, {Kruse}, {Latouf}, {Lowrance},
  {Mennesson}, {Mengel}, {Mills}, {Morton}, {Narita}, {Newton}, {Nishimoto},
  {Okumura}, {Palle}, {Pepper}, {Quintana}, {Roberge}, {Roccatagliata},
  {Schlieder}, {Tanner}, {Teske}, {Tinney}, {Vanderburg}, {von Braun}, {Walp},
  {Wang}, {Wang}, {Weigand}, {White}, {Wittenmyer}, {Wright}, {Youngblood},
  {Zhang}, \& {Zilberman}}]{Plavchan_2020}
{Plavchan}, P., {Barclay}, T., {Gagn{\'e}}, J., {et~al.} 2020, \nat, 582, 497

\bibitem[{{Quinn} {et~al.}(2012){Quinn}, {White}, {Latham}, {Buchhave},
  {Cantrell}, {Dahm}, {F{\H{u}}r{\'e}sz}, {Szentgyorgyi}, {Geary}, {Torres},
  {Bieryla}, {Berlind}, {Calkins}, {Esquerdo}, \&
  {Stefanik}}]{2012ApJ...756L..33Q}
{Quinn}, S.~N., {White}, R.~J., {Latham}, D.~W., {et~al.} 2012, \apjl, 756, L33

\bibitem[{{Raghavan} {et~al.}(2010){Raghavan}, {McAlister}, {Henry}, {Latham},
  {Marcy}, {Mason}, {Gies}, {White}, \& {ten Brummelaar}}]{Raghavan2010}
{Raghavan}, D., {McAlister}, H.~A., {Henry}, T.~J., {et~al.} 2010, \apjs, 190,
  1

\bibitem[{{Rampalli} {et~al.}(2021){Rampalli}, {Ag{\"u}eros}, {Curtis},
  {Douglas}, {N{\'u}{\~n}ez}, {Cargile}, {Covey}, {Gosnell}, {Kraus}, {Law}, \&
  {Mann}}]{Rampalli_2021}
{Rampalli}, R., {Ag{\"u}eros}, M.~A., {Curtis}, J.~L., {et~al.} 2021, \apj,
  921, 167

\bibitem[{{Rana}(1991)}]{Rana1991}
{Rana}, N.~C. 1991, \araa, 29, 129

\bibitem[{{Rauer} {et~al.}(2014){Rauer}, {Catala}, {Aerts}, {Appourchaux},
  {Benz}, {Brandeker}, {Christensen-Dalsgaard}, {Deleuil}, {Gizon}, {Goupil},
  {G{\"u}del}, {Janot-Pacheco}, {Mas-Hesse}, {Pagano}, {Piotto}, {Pollacco},
  {Santos}, {Smith}, {Su{\'a}rez}, {Szab{\'o}}, {Udry}, {Adibekyan}, {Alibert},
  {Almenara}, {Amaro-Seoane}, {Eiff}, {Asplund}, {Antonello}, {Barnes},
  {Baudin}, {Belkacem}, {Bergemann}, {Bihain}, {Birch}, {Bonfils}, {Boisse},
  {Bonomo}, {Borsa}, {Brand{\~a}o}, {Brocato}, {Brun}, {Burleigh}, {Burston},
  {Cabrera}, {Cassisi}, {Chaplin}, {Charpinet}, {Chiappini}, {Church},
  {Csizmadia}, {Cunha}, {Damasso}, {Davies}, {Deeg}, {D{\'{\i}}az}, {Dreizler},
  {Dreyer}, {Eggenberger}, {Ehrenreich}, {Eigm{\"u}ller}, {Erikson}, {Farmer},
  {Feltzing}, {de Oliveira Fialho}, {Figueira}, {Forveille}, {Fridlund},
  {Garc{\'{\i}}a}, {Giommi}, {Giuffrida}, {Godolt}, {Gomes da Silva},
  {Granzer}, {Grenfell}, {Grotsch-Noels}, {G{\"u}nther}, {Haswell}, {Hatzes},
  {H{\'e}brard}, {Hekker}, {Helled}, {Heng}, {Jenkins}, {Johansen},
  {Khodachenko}, {Kislyakova}, {Kley}, {Kolb}, {Krivova}, {Kupka}, {Lammer},
  {Lanza}, {Lebreton}, {Magrin}, {Marcos-Arenal}, {Marrese}, {Marques},
  {Martins}, {Mathis}, {Mathur}, {Messina}, {Miglio}, {Montalban}, {Montalto},
  {Monteiro}, {Moradi}, {Moravveji}, {Mordasini}, {Morel}, {Mortier},
  {Nascimbeni}, {Nelson}, {Nielsen}, {Noack}, {Norton}, {Ofir}, {Oshagh},
  {Ouazzani}, {P{\'a}pics}, {Parro}, {Petit}, {Plez}, {Poretti}, {Quirrenbach},
  {Ragazzoni}, {Raimondo}, {Rainer}, {Reese}, {Redmer}, {Reffert},
  {Rojas-Ayala}, {Roxburgh}, {Salmon}, {Santerne}, {Schneider}, {Schou},
  {Schuh}, {Schunker}, {Silva-Valio}, {Silvotti}, {Skillen}, {Snellen}, {Sohl},
  {Sousa}, {Sozzetti}, {Stello}, {Strassmeier}, {{\v S}vanda}, {Szab{\'o}},
  {Tkachenko}, {Valencia}, {Van Grootel}, {Vauclair}, {Ventura}, {Wagner},
  {Walton}, {Weingrill}, {Werner}, {Wheatley}, \& {Zwintz}}]{Rauer14}
{Rauer}, H., {Catala}, C., {Aerts}, C., {et~al.} 2014, Experimental Astronomy,
  38, 249

\bibitem[{{Raymond} {et~al.}(2014){Raymond}, {Kokubo}, {Morbidelli},
  {Morishima}, \& {Walsh}}]{2014prpl.conf..595R}
{Raymond}, S.~N., {Kokubo}, E., {Morbidelli}, A., {Morishima}, R., \& {Walsh},
  K.~J. 2014, in Protostars and Planets VI, ed. H.~{Beuther}, R.~S. {Klessen},
  C.~P. {Dullemond}, \& T.~{Henning}, 595--618

\bibitem[{{Rebull} {et~al.}(2022){Rebull}, {Stauffer}, {Hillenbrand}, {Cody},
  {Kruse}, \& {Powell}}]{Rebull_2022}
{Rebull}, L.~M., {Stauffer}, J.~R., {Hillenbrand}, L.~A., {et~al.} 2022, \aj,
  164, 80

\bibitem[{{Rebull} {et~al.}(2016){Rebull}, {Stauffer}, {Bouvier}, {Cody},
  {Hillenbrand}, {Soderblom}, {Valenti}, {Barrado}, {Bouy}, {Ciardi},
  {Pinsonneault}, {Stassun}, {Micela}, {Aigrain}, {Vrba}, {Somers}, {Gillen},
  \& {Collier Cameron}}]{Rebull_2016}
{Rebull}, L.~M., {Stauffer}, J.~R., {Bouvier}, J., {et~al.} 2016, \aj, 152, 114

\bibitem[{{Reinhold} {et~al.}(2019){Reinhold}, {Bell}, {Kuszlewicz}, {Hekker},
  \& {Shapiro}}]{Reinhold2019}
{Reinhold}, T., {Bell}, K.~J., {Kuszlewicz}, J., {Hekker}, S., \& {Shapiro},
  A.~I. 2019, \aap, 621, A21

\bibitem[{{Reinhold} \& {Gizon}(2015)}]{Reinhold_2015}
{Reinhold}, T., \& {Gizon}, L. 2015, \aap, 583, A65

\bibitem[{{Reinhold} {et~al.}(2023){Reinhold}, {Shapiro}, {Solanki}, \&
  {Basri}}]{Reinhold2023}
{Reinhold}, T., {Shapiro}, A.~I., {Solanki}, S.~K., \& {Basri}, G. 2023, \aap,
  678, A24

\bibitem[{{Ricker} {et~al.}(2015){Ricker}, {Winn}, {Vanderspek}, {Latham},
  {Bakos}, {Bean}, {Berta-Thompson}, {Brown}, {Buchhave}, {Butler}, {Butler},
  {Chaplin}, {Charbonneau}, {Christensen-Dalsgaard}, {Clampin}, {Deming},
  {Doty}, {De Lee}, {Dressing}, {Dunham}, {Endl}, {Fressin}, {Ge}, {Henning},
  {Holman}, {Howard}, {Ida}, {Jenkins}, {Jernigan}, {Johnson}, {Kaltenegger},
  {Kawai}, {Kjeldsen}, {Laughlin}, {Levine}, {Lin}, {Lissauer}, {MacQueen},
  {Marcy}, {McCullough}, {Morton}, {Narita}, {Paegert}, {Palle}, {Pepe},
  {Pepper}, {Quirrenbach}, {Rinehart}, {Sasselov}, {Sato}, {Seager},
  {Sozzetti}, {Stassun}, {Sullivan}, {Szentgyorgyi}, {Torres}, {Udry}, \&
  {Villasenor}}]{2015JATIS...1a4003R}
{Ricker}, G.~R., {Winn}, J.~N., {Vanderspek}, R., {et~al.} 2015, Journal of
  Astronomical Telescopes, Instruments, and Systems, 1, 014003

\bibitem[{{Rizzuto} {et~al.}(2020){Rizzuto}, {Newton}, {Mann}, {Tofflemire},
  {Vanderburg}, {Kraus}, {Wood}, {Quinn}, {Zhou}, {Thao}, {Law}, {Ziegler}, \&
  {Brice{\~n}o}}]{Rizzuto_2020}
{Rizzuto}, A.~C., {Newton}, E.~R., {Mann}, A.~W., {et~al.} 2020, \aj, 160, 33

\bibitem[{{Rogers} \& {Owen}(2021)}]{Rogers_2021}
{Rogers}, J.~G., \& {Owen}, J.~E. 2021, \mnras, 503, 1526

\bibitem[{{Ruiz-Lara} {et~al.}(2020){Ruiz-Lara}, {Gallart}, {Bernard}, \&
  {Cassisi}}]{2020NatAs...4..965R}
{Ruiz-Lara}, T., {Gallart}, C., {Bernard}, E.~J., \& {Cassisi}, S. 2020, Nature
  Astronomy, 4, 965

\bibitem[{{Sandoval} {et~al.}(2021){Sandoval}, {Contardo}, \&
  {David}}]{Sandoval_2021}
{Sandoval}, A., {Contardo}, G., \& {David}, T.~J. 2021, \apj, 911, 117

\bibitem[{{Santos} {et~al.}(2021){Santos}, {Breton}, {Mathur}, \&
  {Garc{\'\i}a}}]{Santos_2021}
{Santos}, A.~R.~G., {Breton}, S.~N., {Mathur}, S., \& {Garc{\'\i}a}, R.~A.
  2021, \apjs, 255, 17

\bibitem[{{Santos} {et~al.}(2019){Santos}, {Garc{\'\i}a}, {Mathur}, {Bugnet},
  {van Saders}, {Metcalfe}, {Simonian}, \& {Pinsonneault}}]{Santos_2019}
{Santos}, A.~R.~G., {Garc{\'\i}a}, R.~A., {Mathur}, S., {et~al.} 2019, \apjs,
  244, 21

\bibitem[{{Saunders} {et~al.}(2024){Saunders}, {van Saders}, {Lyttle},
  {Metcalfe}, {Li}, {Davies}, {Hall}, {Ball}, {Townsend}, {Creevey}, \&
  {Dodds}}]{2024ApJ...962..138S}
{Saunders}, N., {van Saders}, J.~L., {Lyttle}, A.~J., {et~al.} 2024, \apj, 962,
  138

\bibitem[{{Schmidt} {et~al.}(2024){Schmidt}, {Schlaufman}, \&
  {Hamer}}]{Schmidt2024}
{Schmidt}, S.~P., {Schlaufman}, K.~C., \& {Hamer}, J.~H. 2024, \aj, 168, 109

\bibitem[{{See} {et~al.}(2024){See}, {Lu}, {Amard}, \& {Roquette}}]{See2024}
{See}, V., {Lu}, Y.~L., {Amard}, L., \& {Roquette}, J. 2024, \mnras, 533, 1290

\bibitem[{{Sestito} \& {Randich}(2005)}]{2005A&A...442..615S}
{Sestito}, P., \& {Randich}, S. 2005, \aap, 442, 615

\bibitem[{{Silva Aguirre} {et~al.}(2018){Silva Aguirre}, {Bojsen-Hansen},
  {Slumstrup}, {Casagrande}, {Kawata}, {Ciuc{\v{a}}}, {Handberg}, {Lund},
  {Mosumgaard}, {Huber}, {Johnson}, {Pinsonneault}, {Serenelli}, {Stello},
  {Tayar}, {Bird}, {Cassisi}, {Hon}, {Martig}, {Nissen}, {Rix},
  {Sch{\"o}nrich}, {Sahlholdt}, {Trick}, \& {Yu}}]{SilvaAguirre2018}
{Silva Aguirre}, V., {Bojsen-Hansen}, M., {Slumstrup}, D., {et~al.} 2018,
  \mnras, 475, 5487

\bibitem[{{Simonian} {et~al.}(2019){Simonian}, {Pinsonneault}, \&
  {Terndrup}}]{2019ApJ...871..174S}
{Simonian}, G. V.~A., {Pinsonneault}, M.~H., \& {Terndrup}, D.~M. 2019, \apj,
  871, 174

\bibitem[{{Skrutskie} {et~al.}(2006){Skrutskie}, {Cutri}, {Stiening},
  {Weinberg}, {Schneider}, {Carpenter}, {Beichman}, {Capps}, {Chester},
  {Elias}, {Huchra}, {Liebert}, {Lonsdale}, {Monet}, {Price}, {Seitzer},
  {Jarrett}, {Kirkpatrick}, {Gizis}, {Howard}, {Evans}, {Fowler}, {Fullmer},
  {Hurt}, {Light}, {Kopan}, {Marsh}, {McCallon}, {Tam}, {Van Dyk}, \&
  {Wheelock}}]{Skrutskie06}
{Skrutskie}, M.~F., {Cutri}, R.~M., {Stiening}, R., {et~al.} 2006, \aj, 131,
  1163

\bibitem[{{Skumanich}(1972)}]{Skumanich_1972}
{Skumanich}, A. 1972, \apj, 171, 565

\bibitem[{{Snodgrass} \& {Ulrich}(1990)}]{Snodgrass1990}
{Snodgrass}, H.~B., \& {Ulrich}, R.~K. 1990, \apj, 351, 309

\bibitem[{{Soderblom}(2010)}]{Soderblom_2010}
{Soderblom}, D.~R. 2010, \araa, 48, 581

\bibitem[{{Southworth}(2011)}]{2011MNRAS.417.2166S}
{Southworth}, J. 2011, \mnras, 417, 2166

\bibitem[{{Sozzetti} {et~al.}(2007){Sozzetti}, {Torres}, {Charbonneau},
  {Latham}, {Holman}, {Winn}, {Laird}, \& {O'Donovan}}]{2007ApJ...664.1190S}
{Sozzetti}, A., {Torres}, G., {Charbonneau}, D., {et~al.} 2007, \apj, 664, 1190

\bibitem[{{Spada} \& {Lanzafame}(2020)}]{Spada_2020}
{Spada}, F., \& {Lanzafame}, A.~C. 2020, \aap, 636, A76

\bibitem[{{Thao} {et~al.}(2024){Thao}, {Mann}, {Feinstein}, {Gao}, {Thorngren},
  {Rotman}, {Welbanks}, {Brown}, {Duvvuri}, {France}, {Longo}, {Sandoval},
  {Schneider}, {Wilson}, {Youngblood}, {Vanderburg}, {Barber}, {Wood},
  {Batalha}, {Kraus}, {Murray}, {Newton}, {Rizzuto}, {Tofflemire}, {Tsai},
  {Bean}, {Berta-Thompson}, {Evans-Soma}, {Froning}, {Kempton}, {Miguel}, \&
  {Pineda}}]{Thao2024}
{Thao}, P.~C., {Mann}, A.~W., {Feinstein}, A.~D., {et~al.} 2024, arXiv
  e-prints, arXiv:2409.16355

\bibitem[{{Thompson} {et~al.}(2018){Thompson}, {Coughlin}, {Hoffman},
  {Mullally}, {Christiansen}, {Burke}, {Bryson}, {Batalha}, {Haas},
  {Catanzarite}, {Rowe}, {Barentsen}, {Caldwell}, {Clarke}, {Jenkins}, {Li},
  {Latham}, {Lissauer}, {Mathur}, {Morris}, {Seader}, {Smith}, {Klaus},
  {Twicken}, {Van Cleve}, {Wohler}, {Akeson}, {Ciardi}, {Cochran}, {Henze},
  {Howell}, {Huber}, {Pr{\v{s}}a}, {Ram{\'\i}rez}, {Morton}, {Barclay},
  {Campbell}, {Chaplin}, {Charbonneau}, {Christensen-Dalsgaard}, {Dotson},
  {Doyle}, {Dunham}, {Dupree}, {Ford}, {Geary}, {Girouard}, {Isaacson},
  {Kjeldsen}, {Quintana}, {Ragozzine}, {Shabram}, {Shporer}, {Silva Aguirre},
  {Steffen}, {Still}, {Tenenbaum}, {Welsh}, {Wolfgang}, {Zamudio}, {Koch}, \&
  {Borucki}}]{Thompson_2018}
{Thompson}, S.~E., {Coughlin}, J.~L., {Hoffman}, K., {et~al.} 2018, \apjs, 235,
  38

\bibitem[{{Tofflemire} {et~al.}(2021){Tofflemire}, {Rizzuto}, {Newton},
  {Kraus}, {Mann}, {Vanderburg}, {Nelson}, {Hawkins}, {Wood}, {Zhou}, {Quinn},
  {Howell}, {Collins}, {Schwarz}, {Stassun}, {Bouma}, {Essack}, {Osborn},
  {Boyd}, {F{\H{u}}r{\'e}sz}, {Glidden}, {Twicken}, {Wohler}, {McLean},
  {Ricker}, {Vanderspek}, {Latham}, {Seager}, {Winn}, \&
  {Jenkins}}]{Tofflemire_2021}
{Tofflemire}, B.~M., {Rizzuto}, A.~C., {Newton}, E.~R., {et~al.} 2021, \aj,
  161, 171

\bibitem[{{Tran} {et~al.}(2024){Tran}, {Bowler}, {Cochran}, {Halverson},
  {Mahadevan}, {Ninan}, {Robertson}, {Stef{\'a}nsson}, \&
  {Terrien}}]{2024AJ....167..193T}
{Tran}, Q.~H., {Bowler}, B.~P., {Cochran}, W.~D., {et~al.} 2024, \aj, 167, 193

\bibitem[{{Vach} {et~al.}(2024){Vach}, {Zhou}, {Huang}, {Rogers}, {Bouma},
  {Douglas}, {Kunimoto}, {Mann}, {Barber}, {Quinn}, {Latham}, {Bieryla}, \&
  {Collins}}]{Vach2024}
{Vach}, S., {Zhou}, G., {Huang}, C.~X., {et~al.} 2024, \aj, 167, 210

\bibitem[{Van Der~Walt {et~al.}(2011)Van Der~Walt, Colbert, \&
  Varoquaux}]{numpy}
Van Der~Walt, S., Colbert, S.~C., \& Varoquaux, G. 2011, Computing in Science
  \& Engineering, 13, 22

\bibitem[{{Van Eylen} {et~al.}(2018){Van Eylen}, {Agentoft}, {Lundkvist},
  {Kjeldsen}, {Owen}, {Fulton}, {Petigura}, \& {Snellen}}]{2018MNRAS.479.4786V}
{Van Eylen}, V., {Agentoft}, C., {Lundkvist}, M.~S., {et~al.} 2018, \mnras,
  479, 4786

\bibitem[{{van Saders} {et~al.}(2016){van Saders}, {Ceillier}, {Metcalfe},
  {Silva Aguirre}, {Pinsonneault}, {Garc{\'\i}a}, {Mathur}, \&
  {Davies}}]{vanSaders_2016}
{van Saders}, J.~L., {Ceillier}, T., {Metcalfe}, T.~S., {et~al.} 2016, \nat,
  529, 181

\bibitem[{{Villumsen}(1983)}]{Villumsen1983}
{Villumsen}, J.~V. 1983, \apj, 274, 632

\bibitem[{Virtanen {et~al.}(2020)Virtanen, Gommers, Oliphant, Haberland, Reddy,
  Cournapeau, Burovski, Peterson, Weckesser, Bright, {van der Walt}, Brett,
  Wilson, Millman, Mayorov, Nelson, Jones, Kern, Larson, Carey, Polat, Feng,
  Moore, {VanderPlas}, Laxalde, Perktold, Cimrman, Henriksen, Quintero, Harris,
  Archibald, Ribeiro, Pedregosa, {van Mulbregt}, \& {SciPy 1.0
  Contributors}}]{scipy}
Virtanen, P., Gommers, R., Oliphant, T.~E., {et~al.} 2020, Nature Methods, 17,
  261

\bibitem[{{Vogt} {et~al.}(1994){Vogt}, {Allen}, {Bigelow}, {Bresee}, {Brown},
  {Cantrall}, {Conrad}, {Couture}, {Delaney}, {Epps}, {Hilyard}, {Hilyard},
  {Horn}, {Jern}, {Kanto}, {Keane}, {Kibrick}, {Lewis}, {Osborne},
  {Pardeilhan}, {Pfister}, {Ricketts}, {Robinson}, {Stover}, {Tucker}, {Ward},
  \& {Wei}}]{vogt_hires_1994}
{Vogt}, S.~S., {Allen}, S.~L., {Bigelow}, B.~C., {et~al.} 1994, {SPIE
  Conference Series}, ed. D.~L. {Crawford} \& E.~R. {Craine}, Vol. 2198

\bibitem[{{Walkowicz} \& {Basri}(2013)}]{Walkowicz_2013}
{Walkowicz}, L.~M., \& {Basri}, G.~S. 2013, \mnras, 436, 1883

\bibitem[{{Wilson} {et~al.}(2023){Wilson}, {Barclay}, {Powell}, {Schlieder},
  {Hedges}, {Montet}, {Quintana}, {Mcdonald}, {Penny}, {Espinoza}, \&
  {Kerins}}]{Wilson2023}
{Wilson}, R.~F., {Barclay}, T., {Powell}, B.~P., {et~al.} 2023, \apjs, 269, 5

\bibitem[{{Winn} {et~al.}(2008){Winn}, {Johnson}, {Narita}, {Suto}, {Turner},
  {Fischer}, {Butler}, {Vogt}, {O'Donovan}, \& {Gaudi}}]{2008ApJ...682.1283W}
{Winn}, J.~N., {Johnson}, J.~A., {Narita}, N., {et~al.} 2008, \apj, 682, 1283

\bibitem[{{Wood} {et~al.}(2023){Wood}, {Mann}, {Barber}, {Bush}, {Kraus},
  {Tofflemire}, {Vanderburg}, {Newton}, {Feiden}, {Zhou}, {Bouma}, {Quinn},
  {Armstrong}, {Osborn}, {Adibekyan}, {Mena}, {Sousa}, {Gagn{\'e}}, {Fields},
  {Milburn}, {Thao}, {Schmidt}, {Gnilka}, {Howell}, {Law}, {Ziegler},
  {Brice{\~n}o}, {Ricker}, {Vanderspek}, {Latham}, {Seager}, {Winn}, {Jenkins},
  {Schlieder}, {Osborn}, {Twicken}, {Ciardi}, \& {Huang}}]{Wood_2023}
{Wood}, M.~L., {Mann}, A.~W., {Barber}, M.~G., {et~al.} 2023, \aj, 165, 85

\bibitem[{{Xiang} \& {Rix}(2022)}]{2022Natur.603..599X}
{Xiang}, M., \& {Rix}, H.-W. 2022, \nat, 603, 599

\bibitem[{{York} {et~al.}(2000){York}, {Adelman}, {Anderson}, {Anderson},
  {Annis}, {Bahcall}, {Bakken}, {Barkhouser}, {Bastian}, {Berman}, {Boroski},
  {Bracker}, {Briegel}, {Briggs}, {Brinkmann}, {Brunner}, {Burles}, {Carey},
  {Carr}, {Castander}, {Chen}, {Colestock}, {Connolly}, {Crocker}, {Csabai},
  {Czarapata}, {Davis}, {Doi}, {Dombeck}, {Eisenstein}, {Ellman}, {Elms},
  {Evans}, {Fan}, {Federwitz}, {Fiscelli}, {Friedman}, {Frieman}, {Fukugita},
  {Gillespie}, {Gunn}, {Gurbani}, {de Haas}, {Haldeman}, {Harris}, {Hayes},
  {Heckman}, {Hennessy}, {Hindsley}, {Holm}, {Holmgren}, {Huang}, {Hull},
  {Husby}, {Ichikawa}, {Ichikawa}, {Ivezi{\'c}}, {Kent}, {Kim}, {Kinney},
  {Klaene}, {Kleinman}, {Kleinman}, {Knapp}, {Korienek}, {Kron}, {Kunszt},
  {Lamb}, {Lee}, {Leger}, {Limmongkol}, {Lindenmeyer}, {Long}, {Loomis},
  {Loveday}, {Lucinio}, {Lupton}, {MacKinnon}, {Mannery}, {Mantsch}, {Margon},
  {McGehee}, {McKay}, {Meiksin}, {Merelli}, {Monet}, {Munn}, {Narayanan},
  {Nash}, {Neilsen}, {Neswold}, {Newberg}, {Nichol}, {Nicinski}, {Nonino},
  {Okada}, {Okamura}, {Ostriker}, {Owen}, {Pauls}, {Peoples}, {Peterson},
  {Petravick}, {Pier}, {Pope}, {Pordes}, {Prosapio}, {Rechenmacher}, {Quinn},
  {Richards}, {Richmond}, {Rivetta}, {Rockosi}, {Ruthmansdorfer}, {Sandford},
  {Schlegel}, {Schneider}, {Sekiguchi}, {Sergey}, {Shimasaku}, {Siegmund},
  {Smee}, {Smith}, {Snedden}, {Stone}, {Stoughton}, {Strauss}, {Stubbs},
  {SubbaRao}, {Szalay}, {Szapudi}, {Szokoly}, {Thakar}, {Tremonti}, {Tucker},
  {Uomoto}, {Vanden Berk}, {Vogeley}, {Waddell}, {Wang}, {Watanabe},
  {Weinberg}, {Yanny}, {Yasuda}, \& {SDSS Collaboration}}]{2000AJ....120.1579Y}
{York}, D.~G., {Adelman}, J., {Anderson}, John~E., J., {et~al.} 2000, \aj, 120,
  1579

\bibitem[{{Zakhozhay} {et~al.}(2022){Zakhozhay}, {Launhardt}, {Trifonov},
  {K{\"u}rster}, {Reffert}, {Henning}, {Brahm}, {Vin{\'e}s}, {Marleau}, \&
  {Patel}}]{Zakhozhay_2022}
{Zakhozhay}, O.~V., {Launhardt}, R., {Trifonov}, T., {et~al.} 2022, \aap, 667,
  L14

\bibitem[{{Zari} {et~al.}(2018){Zari}, {Hashemi}, {Brown}, {Jardine}, \& {de
  Zeeuw}}]{Zari_2018}
{Zari}, E., {Hashemi}, H., {Brown}, A.~G.~A., {Jardine}, K., \& {de Zeeuw},
  P.~T. 2018, \aap, 620, A172

\bibitem[{{Zhou} {et~al.}(2022){Zhou}, {Wirth}, {Huang}, {Venner}, {Franson},
  {Quinn}, {Bouma}, {Kraus}, {Mann}, {Newton}, {Dragomir}, {Heitzmann},
  {Lowson}, {Douglas}, {Battley}, {Gillen}, {Triaud}, {Latham}, {Howell},
  {Hartman}, {Tofflemire}, {Wittenmyer}, {Bowler}, {Horner}, {Kane},
  {Kielkopf}, {Plavchan}, {Wright}, {Addison}, {Mengel}, {Okumura}, {Ricker},
  {Vanderspek}, {Seager}, {Jenkins}, {Winn}, {Daylan}, {Fausnaugh}, \&
  {Kunimoto}}]{Zhou_2022}
{Zhou}, G., {Wirth}, C.~P., {Huang}, C.~X., {et~al.} 2022, \aj, 163, 289

\bibitem[{{Ziegler} {et~al.}(2017){Ziegler}, {Law}, {Morton}, {Baranec},
  {Riddle}, {Atkinson}, {Baker}, {Roberts}, \& {Ciardi}}]{2017AJ....153...66Z}
{Ziegler}, C., {Law}, N.~M., {Morton}, T., {et~al.} 2017, \aj, 153, 66

\bibitem[{{Zong} {et~al.}(2018){Zong}, {Fu}, {De Cat}, {Shi}, {Luo}, {Zhang},
  {Frasca}, {Corbally}, {Molenda-{\.Z}akowicz}, {Catanzaro}, {Gray}, {Wang},
  {Pan}, {Ren}, {Zhang}, {Jin}, {Wu}, {Dong}, {Xie}, {Zhang}, {Hou}, \&
  {LAMOST-Kepler Collaboration}}]{Zong2018}
{Zong}, W., {Fu}, J.-N., {De Cat}, P., {et~al.} 2018, \apjs, 238, 30

\end{thebibliography}

\appendix

\section{Rotation Period Catalog Comparison}
\label{app:reinhold23}

\begin{figure*}[!b]
  \begin{center}
    \leavevmode
      \includegraphics[width=0.48\textwidth]{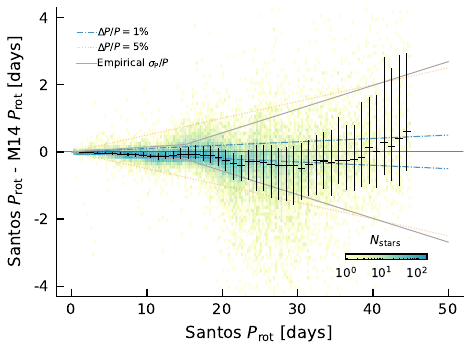}
      \includegraphics[width=0.48\textwidth]{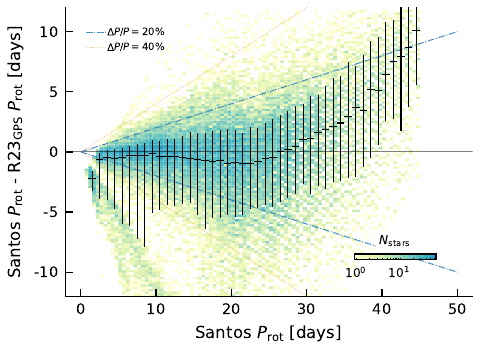}
  \end{center}
  \vspace{-0.6cm}
  \caption{
    {\bf Comparison of reported literature stellar rotation periods}.
    ``Santos'' refers to the concatenation of \citet{Santos_2019} and
    \citet{Santos_2021}.  ``M14'' refers to \citet{McQuillan_2014}.
    ``R23$_{\rm GPS}$'' refers to the Gradient of Power Spectrum
    periods from \citet{Reinhold2023}.  The background is a 2-D
    histogram with a logarithmic color stretch that counts overlapping
    stars between these studies.  The black errorbars show the median
    and $\pm$1$\sigma$ range of period the differences in 1-day bins.
    The
    solid gray ``empirical $\sigma_{\rm P}/P$'' line in the left panel
    shows the expected $\pm 1\sigma$ range that would be spanned if
    the independent $P_{\rm rot}$ measurements had gaussian
    uncertainties drawn following the empirical estimate described in
    Section~\ref{subsec:rotsel}.
    Note the different vertical scales.
    \label{fig:perioddiff}
  }
\end{figure*}

While we opted for the \citetalias{Santos_2019} and
\citetalias{Santos_2021} rotation period catalogs, the \deleted{recent }analysis 
by \citet{Reinhold2023} bears comment.  \citet{Reinhold2023}
used a similar selection function as \citeauthor{Santos_2021}
(all Kepler stars), and considered a novel period measurement
approach based on the gradient of the power spectrum (GPS).  Their method
likely provides greater completeness, due to its greater sensitivity
to signals that are only weakly periodic.  

Figure~\ref{fig:perioddiff} shows histograms of the differences
between reported periods from these various studies.  The left panel
compares overlapping stars from \citet{McQuillan_2014} and
the \citeauthor{Santos_2021} studies.  The periods agree at a precision of
$\lesssim$0.01$P_{\rm rot}$ for $P_{\rm rot}\lesssim$15\,days, and at
$\lesssim$0.03$P_{\rm rot}$ for $P_{\rm rot}\approx$30\,days.  
A small bias exists at $P_{\rm rot}$$\approx$10\,days,
in the sense that the Santos periods tend to be $\approx$1\% faster
for such stars.

The right panel compares overlapping periods from the
\citet{Reinhold2023} GPS method and the \citeauthor{Santos_2021} studies.  The
periods agree at a precision of $\lesssim$0.20$P_{\rm rot}$ for
$P_{\rm rot}\lesssim$15\,days, and at $\lesssim$0.30$P_{\rm rot}$ for
$P_{\rm rot}\approx$30\,days.  A  bias develops past $P_{\rm
rot}\gtrsim$30\,days, in the sense that the \citeauthor{Santos_2021}
periods tend to be 10-20\% longer than the GPS periods.  The origin of
this larger scatter could be connected to the need for the GPS method
to determine the ``$\alpha$ calibration factor'', which is known only
at a statistical level for large stellar populations
\citep[see][]{Reinhold2023}.  While we considered deriving independent
rotation-based ages using the \citet{Reinhold2023} GPS periods with an
inflated empirical uncertainty, we ultimately opted against this path.
A rotation period difference of 20-30\% away from the ``true period''
could sufficiently skew a star's age that we prefer to restrict our
attention to the stars for which precision is a possible
outcome.

\section{Defining An ``Isochronally Far From The Main Sequence'' Boundary}
\label{app:linemethod}

The (arbitrary) method used to construct the orange line in 
the lower panel of Figure~\ref{fig:stellarprops} was as follows.
The line is	defined as
the maximum of two numerically-determined functions, $f_0(T_{\rm eff})$ and $f_1(T_{\rm eff})$.
We defined $f_0$ by fitting an
$N^{\rm th}$-order polynomial to the stars with rotation periods for which $t_{\rm
	B20,iso}/(+\sigma_{t,{\rm B20,iso}})\approx 3$
and $T_{\rm eff}\in[3700,5900]$\,K, where
$t_{\rm B20,iso}$ was the isochronal age reported by
\citet{Berger_2020a_catalog}.  We let the order of the fit vary from
$N$=1 to 10, minimized the Bayesian Information Criterion,
and found $N$=6.  
By eye, this yielded a plausible locus for $T_{\rm eff}$$\lesssim$5300\,K.
However, for F and G dwarfs, the resulting locus
allowed for stars that were ``too evolved''; the density of KOIs 
was anomalously low in this region of the $\log g$ vs.\ $T$$_{\rm eff}$ plane.
We therefore visually selected KOIs that appeared
to be near the main sequence -- i.e.~most of the yellow
points in Figure~\ref{fig:stellarprops} -- and used them to fit a separate
polynomial $f_1$ through a similar BIC-minimization procedure,
which yielded $N$=3.
This polynomial, $f_1$, is shown in very faint opacity in Figure~\ref{fig:stellarprops}.
The portion of the orange locus from $\approx$5300--6200\,K
is set as $f_1 + c$, for $c$ a constant offset that we defined to be
0.1\,dex above the ``KOI main sequence''.
The exact break-point in temperature is automatically set by ${\rm max}(f_0, f_1+c)$.
While we do report rotation-based ages for stars
above this orange locus,
they are flagged as not being near the main sequence in the $\log g$ vs.~$T_{\rm eff}$ plane.

\section{Cases With ``Inconsistent'' Rotation and Lithium-Based Ages}
\label{app:inconsistent}

In this section we discuss the systems with confirmed planets that
have nominally discrepant rotation and lithium-based ages.  We find
that only one of them, Kepler-786, is interesting.

\paragraph{Kepler-1}---TrES-2/Kepler-1 \citep{2006ApJ...651L..61O} is
a $\approx$2.5\,day near-grazing hot Jupiter orbiting a G0 V primary
which has been studied in detail by multiple investigators
\citep[e.g.][]{2007ApJ...664.1190S,2008ApJ...682.1283W,2011ApJ...733...36K,2011MNRAS.417.2166S}.
Our lithium-based age of this system (\trestwotli\,Myr), derived from
EW$_{\rm Li}^{\ast}$=81$\pm$3\,m\AA, qualitatively agrees with the
previously noted lithium abundance \citep{2007ApJ...664.1190S}.
However, we detect no photometric rotation signal, and the
spectroscopic $v\sin i$ is low.  The lithium age is strongly
asymmetric because of the scatter in EW$_{\rm Li}$ vs.~$T_{\rm eff}$
over ages $t$$\gtrsim$1\,Gyr for early G dwarfs.  We calculate a
$2\sigma$ upper limit on $t_{\rm Li}$ of 2.8\,Gyr, and a $3\sigma$
upper limit of 6.1\,Gyr.  The non-detection of rotation, while mildly
surprising, is not shocking if the system's true age is in this long
tail.

\paragraph{Kepler-101}---This $\approx$5500\,K star has EW$_{\rm
Li}^{\ast}$=100$\pm$5\,m\AA, nominally implying $t_{\rm
Li}$$\approx$200-800\,Myr, but no rotation detection.  Our automated
flags noted that this star has a low surface gravity, high luminosity,
and is far from the main sequence.  \citet{2014A&A...572A...2B}
concur: this star is a sub-giant slightly more massive than the Sun;
no rotation detection is expected; the lithium likely simply survived
over the star's main sequence lifetime.

\paragraph{Kepler-108}---This system of two mutually inclined giant
planets \citep{2017AJ....153...45M} has a $T_{\rm
eff}$$\approx$5600\,K host star with a strong lithium detection, and
no rotation detection.  The host star is entering the red giant
branch; this has been previously noted through asteroseismology
\citep{2013ApJ...767..127H}, and our analysis of the Keck/HIRES
spectrum with \texttt{SpecMatch-Synth} \citep{2017AJ....154..107P}
concurs, yielding $T_{\rm eff}$=5668$\pm$100\,K, $\log g$=3.8$\pm$0.1,
[Fe/H]=0.38$\pm$0.06, and $v\sin i$=3.2$\pm$1.0\kms.  Given the star's
high mass, the high lithium content and lack of rotation detection is
not surprising.


\paragraph{Kepler-786}---This K3 V star has a surprisingly high
lithium content.  The spectrum is single-lined, with
\texttt{SpecMatch-Synth} derived parameters of $T_{\rm
eff}$=4769$\pm$100\,K, $\log g$=4.5$\pm$0.1, [Fe/H]=0.11$\pm$0.06, and
$v\sin i$=2.4$\pm$0.6\kms.  Yet with EW$_{\rm
Li}^{\ast}$=83$\pm$6\,m\AA, the lithium age of \kepseveneightsix\,Myr
would predict \replaced{an obvious}{a short-period} rotation signal.  \replaced{None is present in the
Kepler light curve}{The actual signal that is present has an amplitude of $\approx$0.1-0.3\% and a period of $\approx$33\,days}, consistent with the low $v\sin i$.  Out of all
``apparently discrepant'' lithium and rotation measurements discussed
in this appendix, this is the only one that seems to remain discrepant
after scrutiny.  The  \ion{Ca}{2} doublet is in emission, with
$R'_{\rm HK}=-4.7\pm0.5$, which suggests an age at least as old as the
Hyades \citep{Mamajek_2008}.  The Balmer lines are in absorption, and
display no obvious signatures of youth.

\paragraph{Kepler-1644}---The rotational age of this system
(\kepsixteenfourfour\,Myr) is nominally much lower than the lithium
limit ($>$767\,Myr).  The Kepler light curve shows a $\approx$1\%
amplitude 1.4\,day rotation signal with many flares.  However the
automated quality flags note that the star has low (photometric)
surface gravity, a high RUWE (${\rm RUWE}_{\rm DR3}$=7.0), and that it
is far from the main sequence.  The Keck/HIRES spectrum also shows
visually narrow lines, with a \texttt{SpecMatch-Synth} $v \sin i \leq
2$\kms.  However, we performed a cross-correlation between the HIRES
spectrum and the nearest matches in the Keck/HIRES template library
\citep{2015AJ....149...18K}, and found that on 31 July 2022 (UT) the
system displayed a broad CCF with a blended second component at
$\approx$$+26$\,\kms\ relative to the primary, with a best-fit
flux-ratio of $\approx$3.4\%, and a preferred $T_{\rm
eff,B}$$\approx$4400\,K.  The spectrum and astrometric excess noise
therefore point to this system being an unresolved binary, which calls
the reliability of the rotational age into question.  The
non-detection of the companion from Robo-AO imaging at Palomar
\citep{2017AJ....153...66Z} suggests that the companion(s) are likely
within $\rho$$\lesssim$0.3$''$.

\paragraph{Kepler-1699}---This system is in a similar qualitative
regime as Kepler-1644, with an apparently young $t_{\rm
gyro}$=\kepsixteenninenine\,Myr derived from a $\approx$2\% amplitude
4.2\,day rotation signal, and no evidence for lithium.  The system has
${\rm RUWE}_{\rm DR3}$=19.1.  From the same style of CCF analysis from
the Keck/HIRES spectrum acquired on 31 Aug 2022 (UT), we also find a
double-peaked CCF, in this case with a secondary component at
$\approx$$-16$\,\kms\ relative to the primary with $T_{\rm
eff,B}$$\approx$4900\,K.  This putative companion is similarly not
detected in high-resolution imaging \citep{2017AJ....153...66Z}.  The
rotational age is questionable, given that we do not know which
source the rotation signal is from, or whether the stars have
interacted.

\paragraph{Kepler-1943}---This system nominally has a
\kepnineteenfourthree\,Myr lithium age, and no reported rotation
detection.  However, the star is flagged as being over-luminous,
low-surface gravity, and far from the main sequence.  In other words,
it is a subgiant.

\paragraph{Kepler-639, Kepler-320, Kepler-1719, Kepler-1876,
Kepler-1072, Kepler-1743, Kepler-1929, Kepler-1488}---These eight
systems all have nominally two-sided lithium age posteriors between 1
and 3\,Gyr, and yet lack rotation period detections.  All are
subgiants, flagged with $Q_{\rm star}$ under various combinations of
bits 1, 2, and 9.

%

\section{Age Diagnostics}
\label{app:age_diagnostic}

\begin{figure*}[!t]
  \begin{center}
    \leavevmode
        \includegraphics[width=0.48\textwidth]{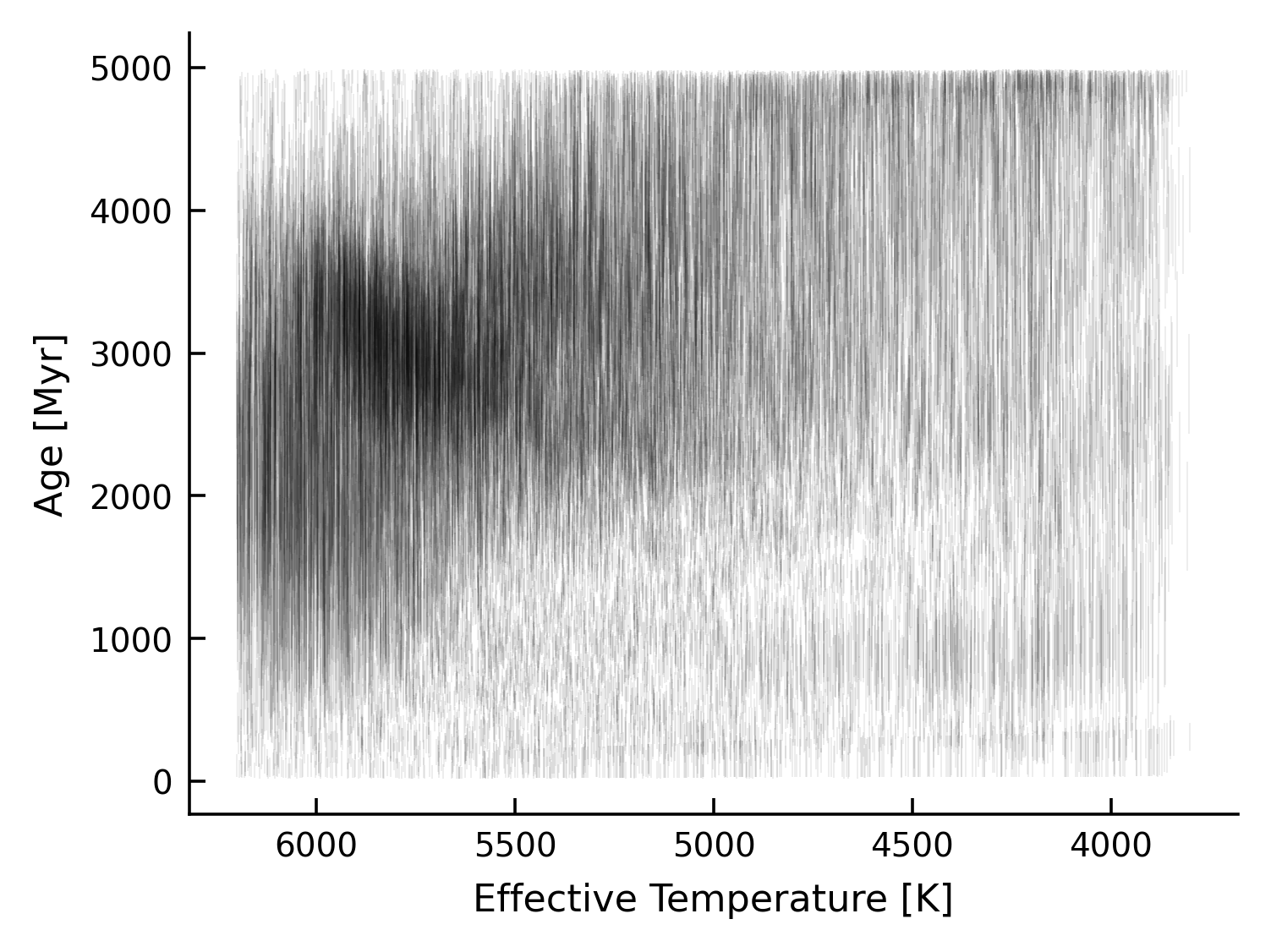}
    \includegraphics[width=0.48\textwidth]{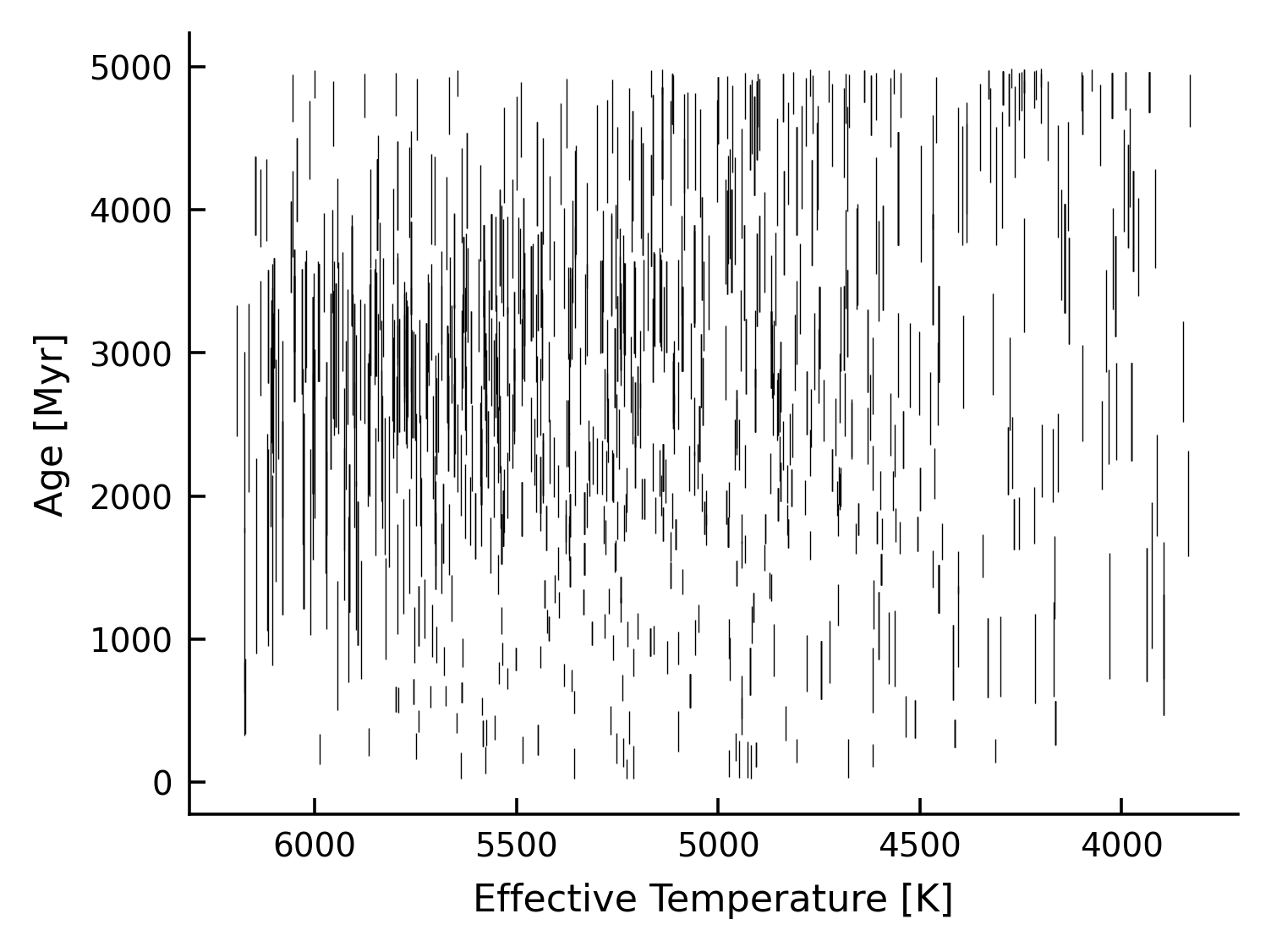}
  \end{center}
  \vspace{-0.6cm}
  \caption{
    {\bf Rotation-based age vs.~effective temperature}.
    The entire Kepler sample is shown on the left.
    KOI hosts are shown on the right.
    Each bar shows the $\pm$1$\sigma$ uncertainty for a star's age as
    derived from its rotation period.  
    Notable features are discussed in
    Appendix~\ref{app:age_diagnostic}.
    \label{fig:gyroage_vs_teff}
  }
\end{figure*}

Figure~\ref{fig:gyroage_vs_teff} is a visualization of $t_{\rm gyro}$
vs.~$T_{\rm eff}$ for our rotation-based age catalog. 
Each bar denotes the $\pm$1$\sigma$ uncertainty for a star's
\deleted{rotation-based }age:
this plot can be viewed as the transformation of
the left panel of Figure~\ref{fig:prot_vs_teff}.
The overdensity of $P_{\rm rot}$$\approx$20\,day G dwarfs corresponds
to the large overdensity at $t_{\rm gyro}$$\approx$3\,Gyr.
The (non-physical) deficit of $\sim$22\,day rotators appears as a
deficit between 5500--6000\,K.
A second deficit is also visible from 3800--5000\,K;
it is associated with the ``intermediate period gap'', which may be
associated with a transition from spot-dominated to faculae-dominated
light curves \citep[e.g.][]{Reinhold2019}\added{, or with rapid spin-down at Rossby numbers near 0.5 \citep{Lu2022}}.
At $\approx$1\,Gyr, stalled spin-down yields larger uncertainties for
K dwarfs.
The pile-up near $\approx$5\,Gyr is imposed by our choice of prior;
given that \texttt{gyro-interp} returns non-calibrated ages in this
regime, we truncated our analysis at 4\,Gyr.

\section{What if we only considered McQuillan's periods?}
\label{app:mcqonly}

\begin{figure*}[!t]
	\begin{center}
		\subfloat{
			\includegraphics[width=0.44\textwidth]{prot_teff_Santos19_Santos21_dquality.pdf}
		}
		\subfloat{
			\includegraphics[width=0.44\textwidth]{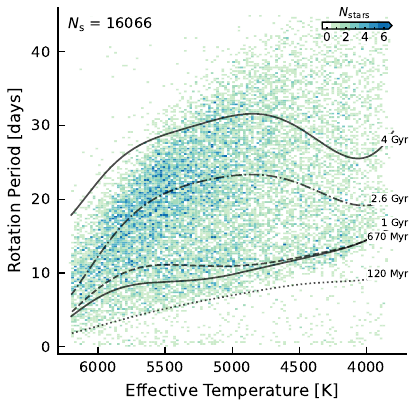}
		}
	\end{center}
	\vspace{-0.6cm}
	\caption{
    {\bf Variant of Figure~\ref{fig:prot_vs_teff}}, comparing the default
    rotation periods ({\it left}) from \citet{Santos_2019,Santos_2021}
    against the \citet{McQuillan_2014} rotation periods ({\it right}).
    All stars are required to be amenable to rotation-based age dating
    (i.e.~none of bit flags zero through \replaced{nine}{ten} raised).
	}
	\label{fig:prot_vs_teff_mcq}
\end{figure*}

\begin{figure*}[!th]
  \begin{center}
    \leavevmode
    \subfloat{
        \includegraphics[width=.85\textwidth]{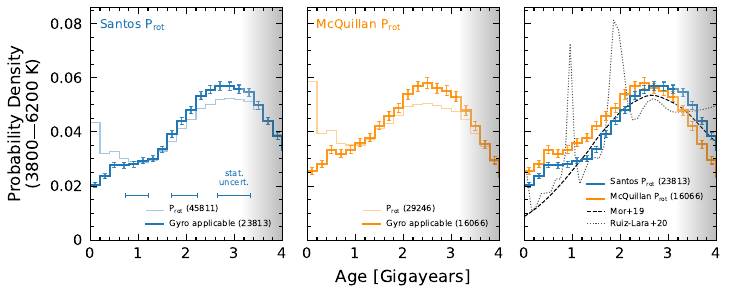}
    }

	\vspace{-0.35cm}
    \subfloat{
        \includegraphics[width=.85\textwidth]{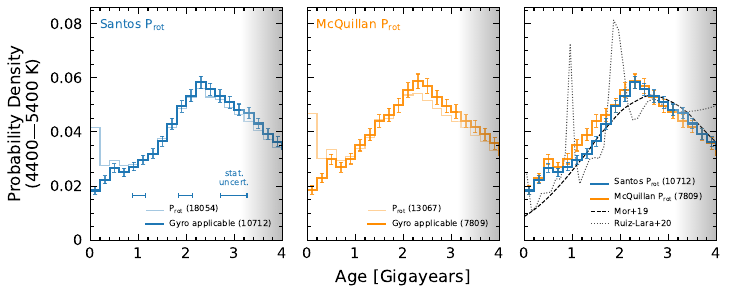}
    }
  \end{center}
  \vspace{-0.66cm}
  \caption{{\bf Variant of Figure~\ref{fig:hist_tgyro},} comparing
  our default adopted age distribution (left) based on the \citet{Santos_2019,Santos_2021}
  rotation periods against the age distribution (middle) based only on rotation periods
  reported by \citet{McQuillan_2014}.
  \label{fig:hist_tgyro_mcq}
  }
\end{figure*}

We argued in Section~\ref{subsec:rotsel} that adopting
the periods from \citet{Santos_2019,Santos_2021} yielded the best
possible balance between homogeneity and sensitivity for both KOIs and
for the broader Kepler sample.  Nonetheless, it is interesting to
consider what would happen if we were to analyze only the rotation
periods from \citet{McQuillan_2014}.
Figures~\ref{fig:prot_vs_teff_mcq} and~\ref{fig:hist_tgyro_mcq} repeat
the analysis, but make this alternate choice.
Figure~\ref{fig:prot_vs_teff_mcq} shows that while the overall
contours of the $P_{\rm rot}$ vs.~$T_{\rm eff}$ distributions are
similar, the \citeauthor{Santos_2019} distribution includes more
stars, particularly at longer rotation periods.  This is connected to
a stronger noise-dependent cutoff in the \citet{McQuillan_2014} sample
than in the \citeauthor{Santos_2019} samples (see
\citealt{Masuda2022_amplevoln} Figures~10 and 15).
Figure~\ref{fig:hist_tgyro_mcq} shows the impact on the inferred
rotation-based age histograms: while the \citeauthor{Santos_2019}
distribution peaks at 2.8-3.0\,Gyr, the \citeauthor{McQuillan_2014}
distribution peaks at 2.4-2.6\,Gyr.  A number of metrics shift.  For
instance, the ratio of the star formation rate 3\,Gyr age to today is
\ratiosfr$\pm$\uncratiosfr\ assuming the \citeauthor{Santos_2019}
periods, and \mcquillanonlyratiosfr$\pm$\mcquillanonlyuncratiosfr\
assuming the \citeauthor{McQuillan_2014} periods.  Similarly, the
ratio of ``old'' (2-3\,Gyr) to ``young'' (0-1\,Gyr) stars shifts from
\ratioobtoybstars\ in the \citeauthor{Santos_2019} case to
\mcquillanonlyratioobtoybstars\ in the \citeauthor{McQuillan_2014}
case.  The K dwarf age distributions (Figure~\ref{fig:hist_tgyro_mcq}
lower panel) however appear quite similar.  Generally, these plots
support the conclusion that the $t_{\rm gyro}$ age distribution in the
Kepler field is non-uniform, with a peak in the \replaced{2.5}{2.3}-3\,Gyr range.
Nonetheless, some of the details do shift depending on the choice of
rotation detection pipeline.

\clearpage
\listofchanges

\end{document}